# Ion beam treatment of thick polystyrene films

Alexey Kondyurin

School of Physics, University of Sydney, Camperdown, NSW 2050, Australia
Ewingar Scientific, Ewingar, NSW 2469, Australia

**Abstract**
Spincoated polystyrene films of different thickness from 78 nm to 1.3 μm on silicon wafers were treated by nitrogen ions with an energy of 20 keV. Ellipsometric measurements and FTIR spectra showed modification of the surface layer corresponding to the depth of ion penetration into the polymer (about 70 nm). However, washing of the deep layers and subsequent measurements showed that free radicals formed in the thin modified layer migrated into the bulk layer and caused a number of changes in the chemical structure of the deep polystyrene layer. Thus, despite the small depth of ion penetration into the polystyrene film, the entire film is modified to a depth much greater than the projective range of ions. Therefore, the ion beam treatment of a polymer is a surface modification method only conditionally. It is necessary to take into account free-radical reactions and the possibility of their migration into the deep layers of the polymer.

## 1. Introduction

Ion beam treatment is a powerful method for deep modification of polymer structure [1]. In this method, only a thin surface layer of the polymer is modified and deeper layers of the polymer are preserved. This corresponds to the theory of scattering of particles flying into the target on electrons and atoms, according to the results of the first works of Thompson and Rutherford. Therefore, in the literature, only a thin surface layer of the polymer is considered, corresponding to the penetration depth of the bombarding ions, and deeper layers are not considered.
The effects of structure changing in a thin surface layer are associated with carbonization, oxidation and activation of the polymer surface, which has practical significance [2]. The effect of the bombarding ions with energies higher than the binding energy in the macromolecule is based on the transfer of kinetic energy to the atoms and electrons of the macromolecule during collisions with the bombarding ion. As a result, atoms, fragments of macromolecules and electrons can be knocked out. The first two lead to the appearance of free radicals - molecules and atoms with broken chemical bonds, that is, with unpaired electrons in the valency orbit. Such free radicals are extremely active and trigger chain free-radical reactions in the place where they were created [1, 3].
Therefore, the main changes in the polymer structure as a result of ion treatment occur in a thin surface layer, the thickness of which corresponds to the projective range of the ion in this material. Experimental studies of the thickness of the changed layer well confirm the correctness of this model for different materials including polymers. However, as has been shown in a number of studies [4-7], the effect of structure changes is a polymer material is observed deeper than the projective range of an ion. In particular, measuring the gel fraction of a film of polyethylene, polyisoprene and polystyrene showed that the thickness of the cross-linked layer exceeds the depth of the ion range [1].
Since a change in the polymer structure at depths exceeding the projective ion range may affect the properties of treated polymer, this paper presents a more detailed study of the polymer structure treated by high energy ions. A thin spin-coated polystyrene film on a silicon wafer was used in the experiment. An ellipsometry and FTIR spectroscopy were used to measure the film properties, including after washing off the non-crosslinked part of the film, the so-called gel-fraction. The choice of polystyrene was due to the fundamental classical question: does the polymer structure change at depths exceeding the projective ion range? As well as the ability to create films of different thicknesses and study them in detail using structural analysis methods was also a reason. The scheme of experiment is shown in Fig. 1. A polystyrene film from a toluene solution was spincoated onto a silicon wafer. Then the wafer was treated with an ion beam. It was assumed that a carbonized layer appears on the surface of the polystyrene film, corresponding in thickness to the projective range of nitrogen ions. Under the carbonized layer there should be a layer of pristine polystyrene not modified by nitrogen ions. Then the wafer is washed in toluene so that the pristine layer of polystyrene completely dissolves and only the carbonized layer remains on the wafer. Measurements of the ellipsometry and FTIR spectra at each stage of processing should characterise these layers.

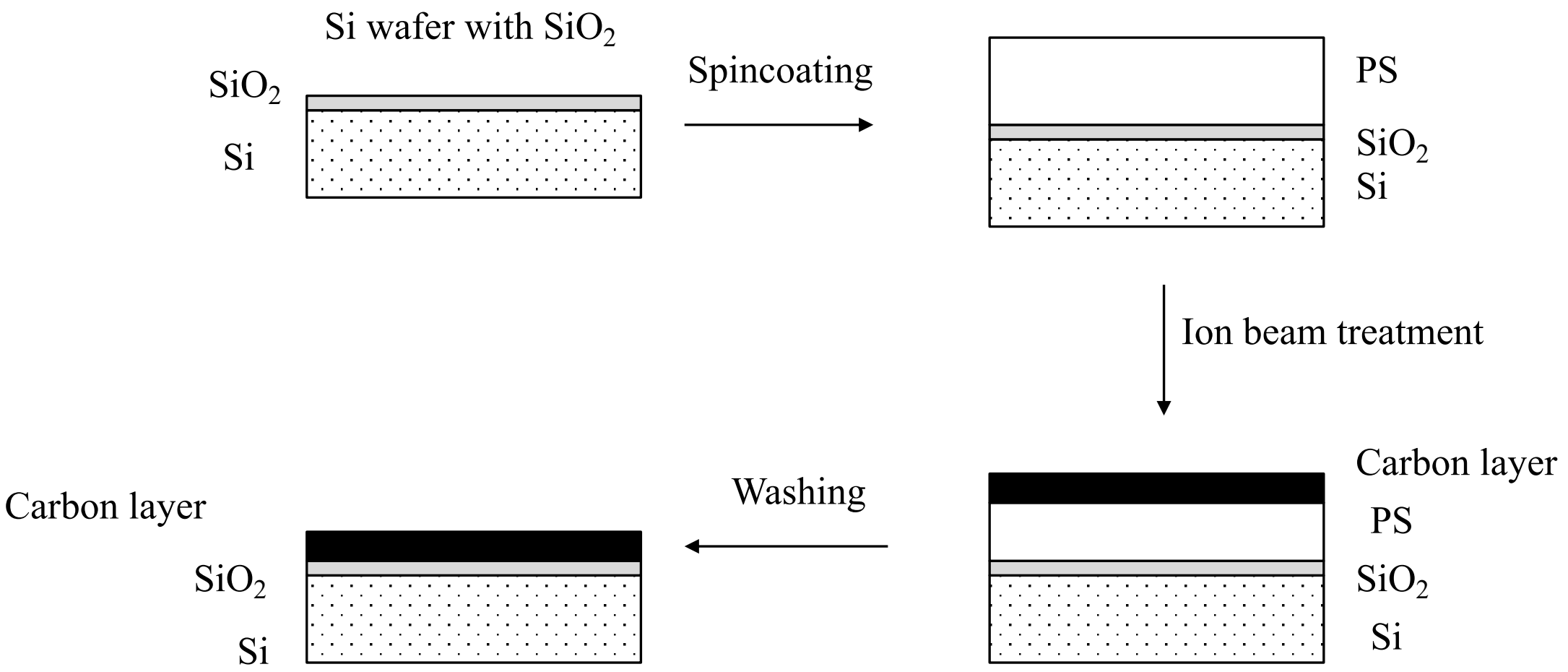

Fig.1. The scheme of experiment.

## 2. Experiment

Polystyrene (PS) films were prepared by spincoating on silicon wafer. Silicon wafer was low doped and had low electrical conductivity, which ensured low FTIR absorption of silicon wafer. The backside of the wafer was not polished to exclude interference in FTIR spectra. The silicon wafer was cut to 10 x 10 mm in size. The spin coater speed was 2000 rpm. An SCS G3P-8 spin coater was used. Austrex 400 polystyrene in granules was obtained from Polystyrene Australia Pty. Ltd. The granules were dissolved in ultra-high purity toluene obtained from Sigma Aldrich, Australia (Product No. 34866, purity >99.9%). Solutions of different concentrations were prepared and films of different thickness were obtained at the same spin coater speed. The presence of toluene was monitored by FTIR spectra.
Polystyrene films were treated by nitrogen ions with an energy of 20 keV. The samples were fixed on the stainless steel holder of 150 mm diameter and covered by Faraday cap with stainless steel mesh. The Faraday cap with mesh were electrically connected to the holder. The distance between mesh and the sample surface was 50 mm. The mesh was made of 0.05 mm wire with 2 mm cells. The maximum pressure of the residual atmosphere in the vacuum chamber was $10^{-6}$ Torr ($1.3\cdot10^{-4}$ Pa) provided by scroll pump (Edwards, XDS-10) and turbomolecular pump (nEXT). The nitrogen pressure during the treatment was $2\cdot10^{-3}$ Torr ($2.7\cdot10^{-1}$ Pa) provided by MKS flow controller and measured by Pfeiffer vacuumeter. The vacuum chamber was made of aluminium cylinder. The two pairs of copper coils were fixed outside of the chamber and used for constant axial magnetic field of 5 mT inside the chamber. The radio-frequency power of 13.75 MHz and 100 W (12 W reverse power) was generated by ACG-10 ENI Power Systems, matched with Comdel CPM-2000 matching network and applied via inductive antenna of one turn with cooler to the glass part of the vacuum chamber. The ion beam treatment was carried out in a pulsed mode. The negative 20 kV high voltage pulse of squared shape was generated by ANSTO PI$^3$ generator based on Eimac 8960 power tetrode with air-cooling and applied to the sample holder. The pulse duration was 20 microseconds. The pulse repetition frequency was 50 Hz. The ion voltage and current were monitored by digital oscilloscope as well as an average ion current was measured by ANSTO PI$^3$ generator measurement unit. The sample holder was grounded between the high voltage pulses. According to preliminary data, this mode did not lead to overheating of the polystyrene samples above 40 °C. The treatment time was 40, 80, 800 and 1600 seconds, which corresponded to an ion fluence from $0.5\cdot10^{15}$ ions/cm$^2$ to $20\cdot10^{15}$ ions/cm$^2$.
For control, the PS samples were treated by rf-plasma without applying high voltage. In this case, the samples were placed on a high-voltage electrode and covered with a mesh, as in the case of the ion beam treatment. Since the size of the mesh cell was much smaller than the length of the electromagnetic wave of the plasma, the plasma intensity under the mesh was very weak and cannot be compared with the plasma above the mesh. Such plasma treatment was used only to compare the conditions of plasma effect on the samples between the applied high-voltage pulses.
For the measurements of the polystyrene gel-fraction, the silicon wafers with the polystyrene film were washed in a sufficient amount of pure toluene (50 ml for each sample) for 10 minutes. The samples were then dried on open air.
The thickness of polystyrene films and their optical parameters were measured using spectral ellipsometry. A Woollam M2000V ellipsometer was used. The spectra were recorded at beam incidence angles of 65, 70 and 75 degrees in the range from 350 nm to 800 nm of light wavelength. A multilayer optical model of polystyrene was applied. FTIR transmission spectra were recorded on a Bomem spectrometer with a spectral resolution of 4 cm$^{-1}$. Number of spectra collected for each sample was 2000 to get sufficient signal-to-noise ratio. The spectra of the silicon wafer were subtracted. The spectra of water vapor in the spectrometer were also subtracted. TRIM-95 and SRIM-2003 software were used to simulate the nitrogen ion penetration into polystyrene target.

## 3. Results

### 3.1. Untreated films

The ellipsometry results for the untreated PS film of 80 nm thickness are shown in Fig. 2.

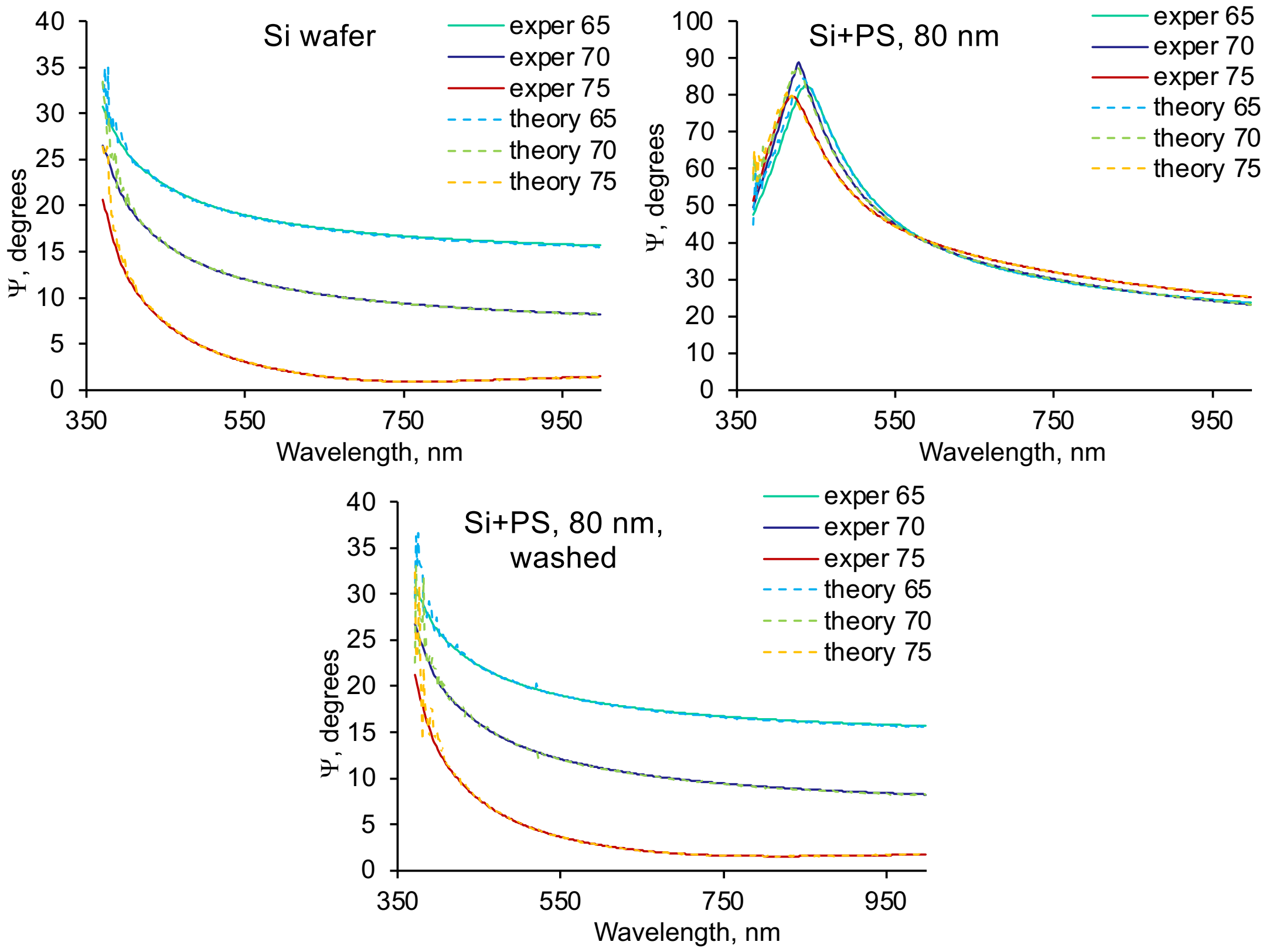

Fig.2. Ellipsometric Ψ function of a silicon wafer coated with polystyrene: original silicon wafer with an oxide layer, a silicon wafer coated with an 80 nm layer of polystyrene, and a silicon wafer coated with polystyrene and then washed in toluene. Solid lines are experimental. Dashed lines are theoretical calculations. Beam incidence angles are given in degrees.

The Ψ function graphs of the original silicon wafer are fitted with the optical model consisting of bulk silicon with 1 nm thick surface layer of silicon oxide. When a polystyrene layer is applied, the Ψ function shows a maximum in the region of 400 nm. The optical model of such a sample corresponds to a PS layer of 80 nm thick. The optical parameters refractive index and extinction coefficient of untreated PS were determined and used in further measurements.

The Ψ function and optical parameters of the silicon wafer with the applied and then washed off PS layer showed the absence of the PS layer on the surface. The error of measurements for the thickness of the residual PS on the silicon wafer surface is about 0.1 nm, which corresponds to the accuracy of the ellipsometric measurements limited by the noise level of the spectrum. Based on this, the accuracy of measuring for the PS residuals is less than 0.2% by volume.

The Ψ function of ellipsometric measurements for PS films with different thicknesses up to 1330 nm are shown in Figures 3 and 4. With increasing PS film thickness, the maxima and minima of the Ψ function shift to the region of long wavelengths and the number of function extremes increases. For the film of 1330 nm thickness, the number of function extremes approaches the number of experimental points of the spectrum and the measurement accuracy decreases. Therefore, this value of the PS film thickness was accepted as the maximal thickness for this type of measurements.

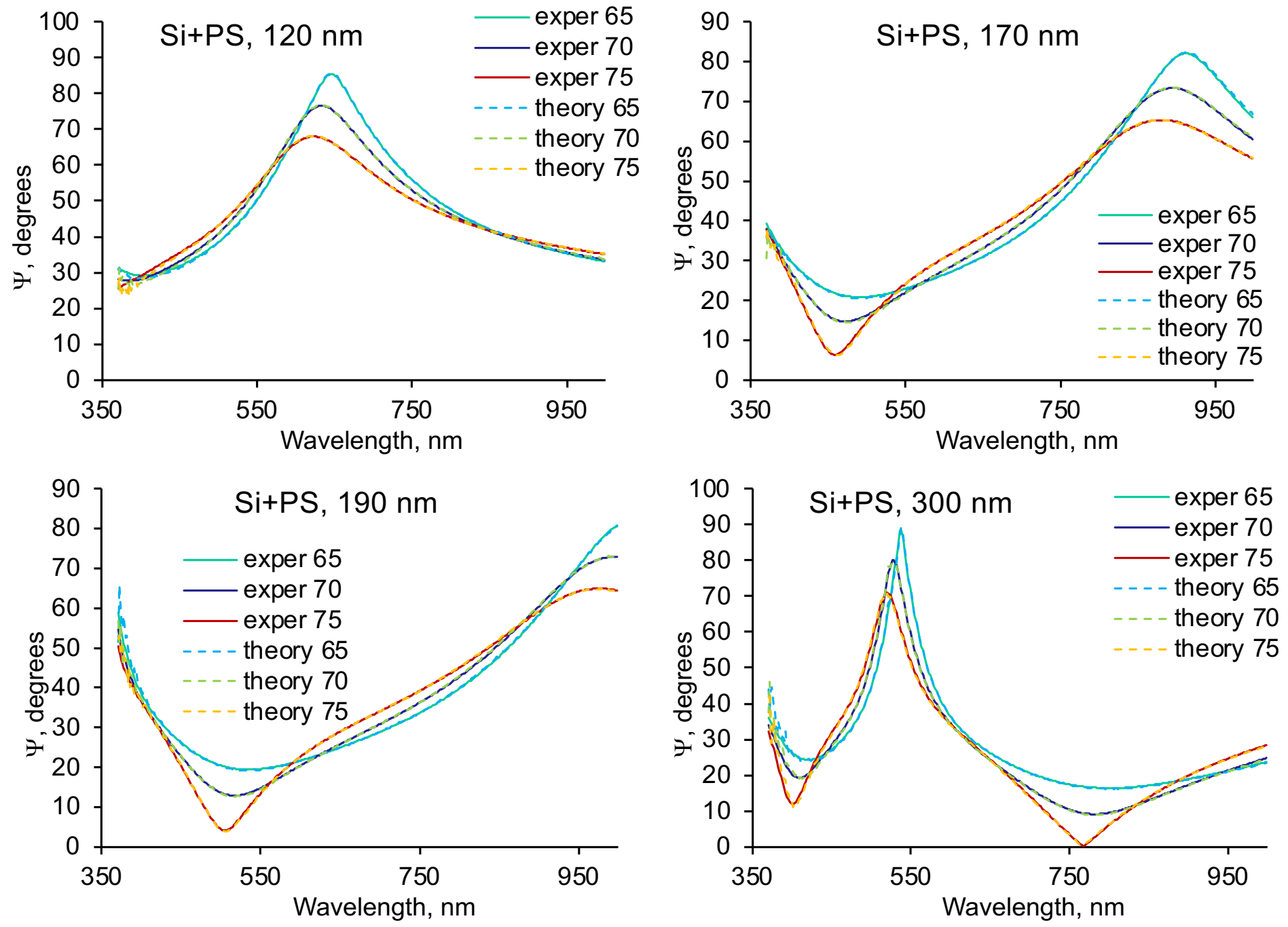

Fig.3. Ellipsometric Ψ function of a silicon wafer coated with polystyrene of 120, 170, 190 and 300 nm. Solid lines are experimental. Dashed lines are theoretical calculations. Beam incidence angles are given in degrees.

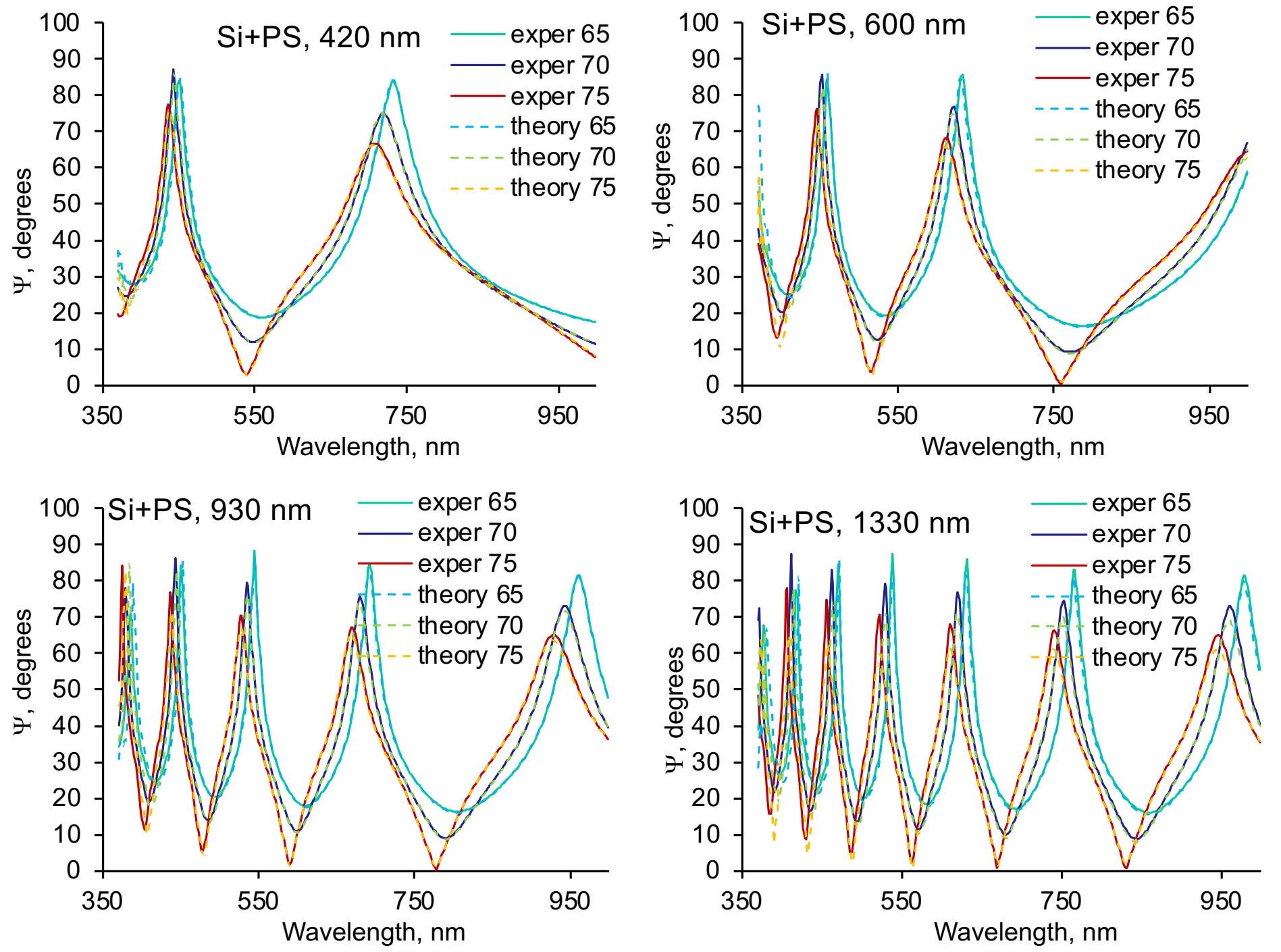

Fig.4. Ellipsometric Ψ function of a silicon wafer coated with polystyrene of 420, 600, 930 and 1330 nm. Solid lines are experimental. Dashed lines are theoretical calculations. Beam incidence angles are given in degrees.

It should also be noted that with increasing film thickness, the discrepancies between the experimental and theoretical spectrum curves increase. That is, the accuracy of the film thickness measurements decreases. This is well visible in the position of the maxima and minimum of the Ψ function, where the discrepancies in the amplitude of the Ψ function are observed more clearly. Taking into account that the penetration depth of nitrogen ions during ion-beam treatment was about 70 nm, the error in measuring the PS layer thickness for a thick film increases and becomes significant compared to the thickness of the carbonized layer. Therefore, a thickness of 1330 nm was designated as a limit for these measurements. For thicker films, if it would be required, other methods of the gel fraction measuring should apparently be used.

### 3.2. Films after ion beam treatment

The ellipsometry results of the PS films with different ion beam treatment times, which correspond to different ion fluences, are shown in Fig. 5. With an increase of the treatment time, the position of the Ψ function maximum shifts to the region of short wavelengths, which corresponds to a decrease in the thickness of the PS film.

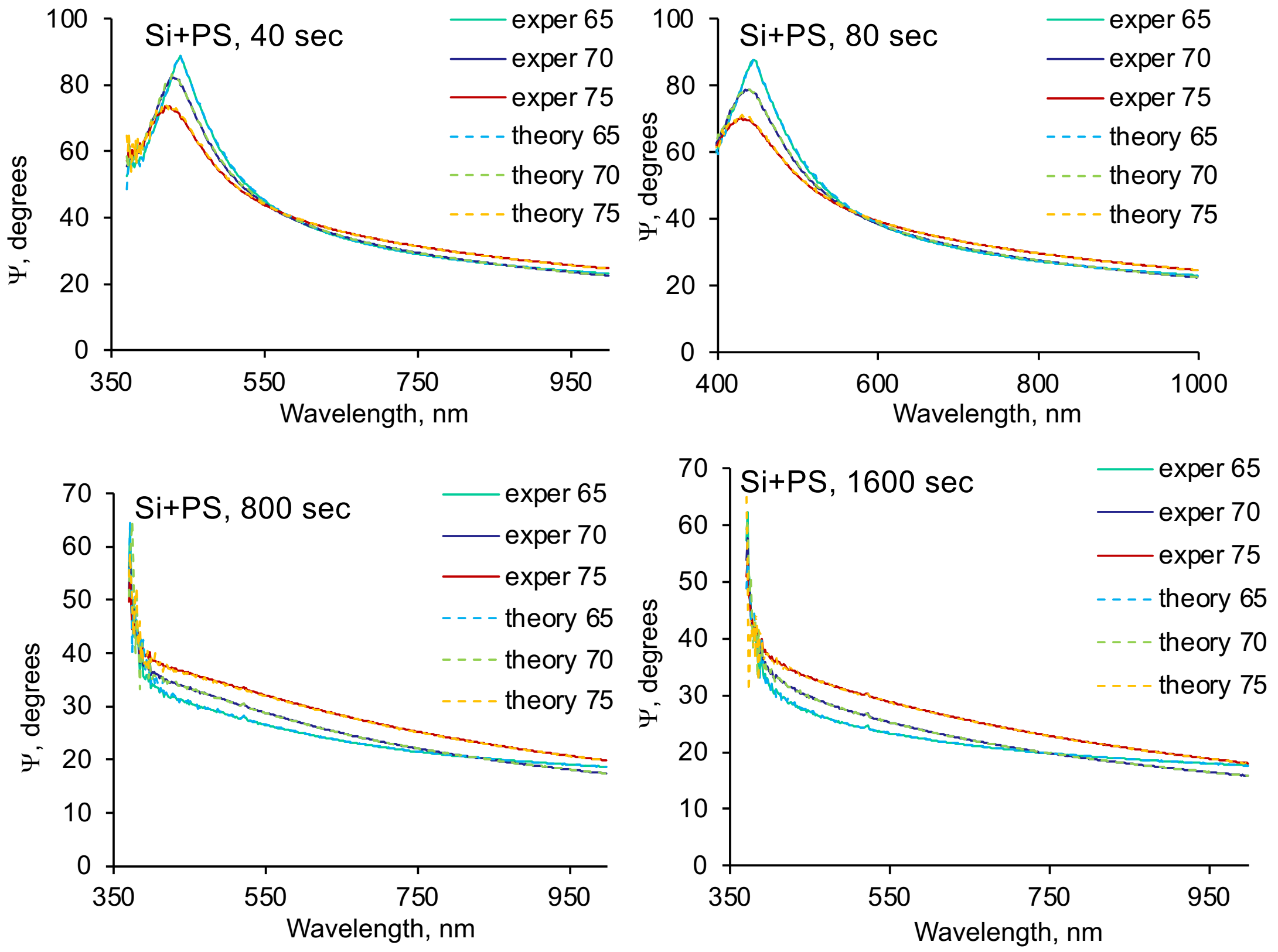

Fig.5. Ellipsometric Ψ function of a silicon wafer coated with polystyrene of 78 nm thickness and treated by nitrogen ions of 20 keV energy during 40, 80, 800 и 1600 seconds. Solid lines are experimental. Dashed lines are theoretical calculations. Beam incidence angles are given in degrees.

The experimental ellipsometric Ψ and δ functions were used to fit an optical model of the samples. For this purpose, the extinction coefficient and refractive index of silicon and silicon oxide were used, the parameters of which were determined from the spectra of control measurements of the silicon wafer without PS. Then, a layer with Cauchy function was added to the model and the coefficients of the Cauchy function for the untreated PS were found. Then, the spectra of refractive index and extinction coefficient of the PS layer were calculated from the function values. The calculation results are shown in Fig. 6. Ellipsometric data for other treatment times were analyzed similarly.

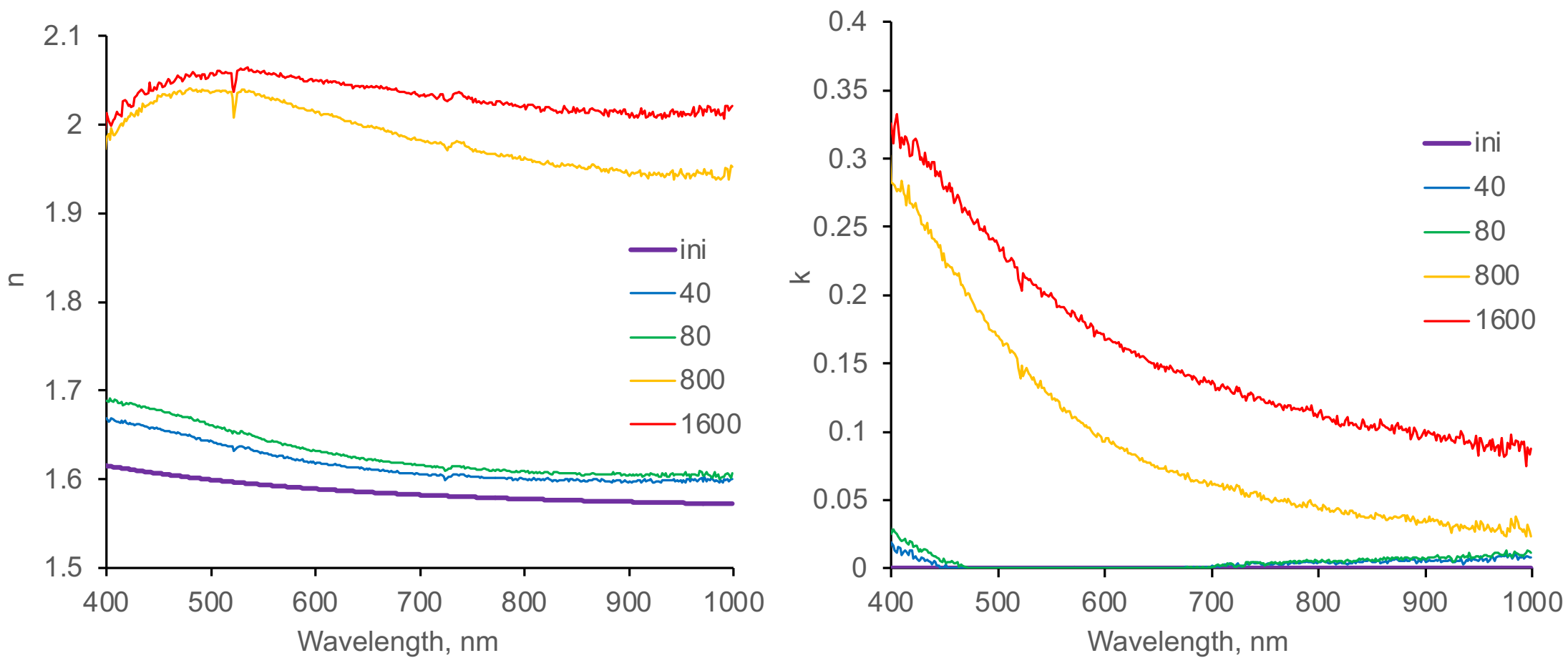

Fig.6. Spectra of refractive index n and extinction coefficient k for the polystyrene of 78 nm thickness and treated by nitrogen ions of 20 keV energy during 40, 80, 800 и 1600 seconds.

The results show that the refractive index of untreated PS corresponds to the literature data and lies in the range of values around 1.6 in the present wavelength range. The extinction coefficient of untreated PS is zero, which also corresponds to the literature data.

The refractive index of the PS layer treated with an ion beam increases with increasing treatment time (fluence). The dependence of the refractive index is nonlinear and saturation of growth occurs when treated for more than 800 seconds. The maximum values of the refractive index in the entire measured range of wavelengths lie in the range of 2-2.1. Such high values of the refractive index correspond to the refractive index of carbonized structures such as graphite, graphene and diamond.

The extinction coefficient increases with the increase in fluence. The frequency dependence of the extinction coefficient corresponds to aromatic condensed structures with different degrees of conjugation. These results are consistent with the literature data on polymer carbonization as a result of ion-beam treatment.

Taking into account the penetration depth of nitrogen ions with an energy of 20 keV into PS, which is about 70 nm, the PS film after 1600 sec of treatment is almost completely carbonized. Therefore, the optical properties of such a film were further used to determine the thickness of the carbonized layer for other films of treated PS with other thicknesses. Optical models of the carbonized layer for different times of ion-beam treatment are shown in Fig. 7.

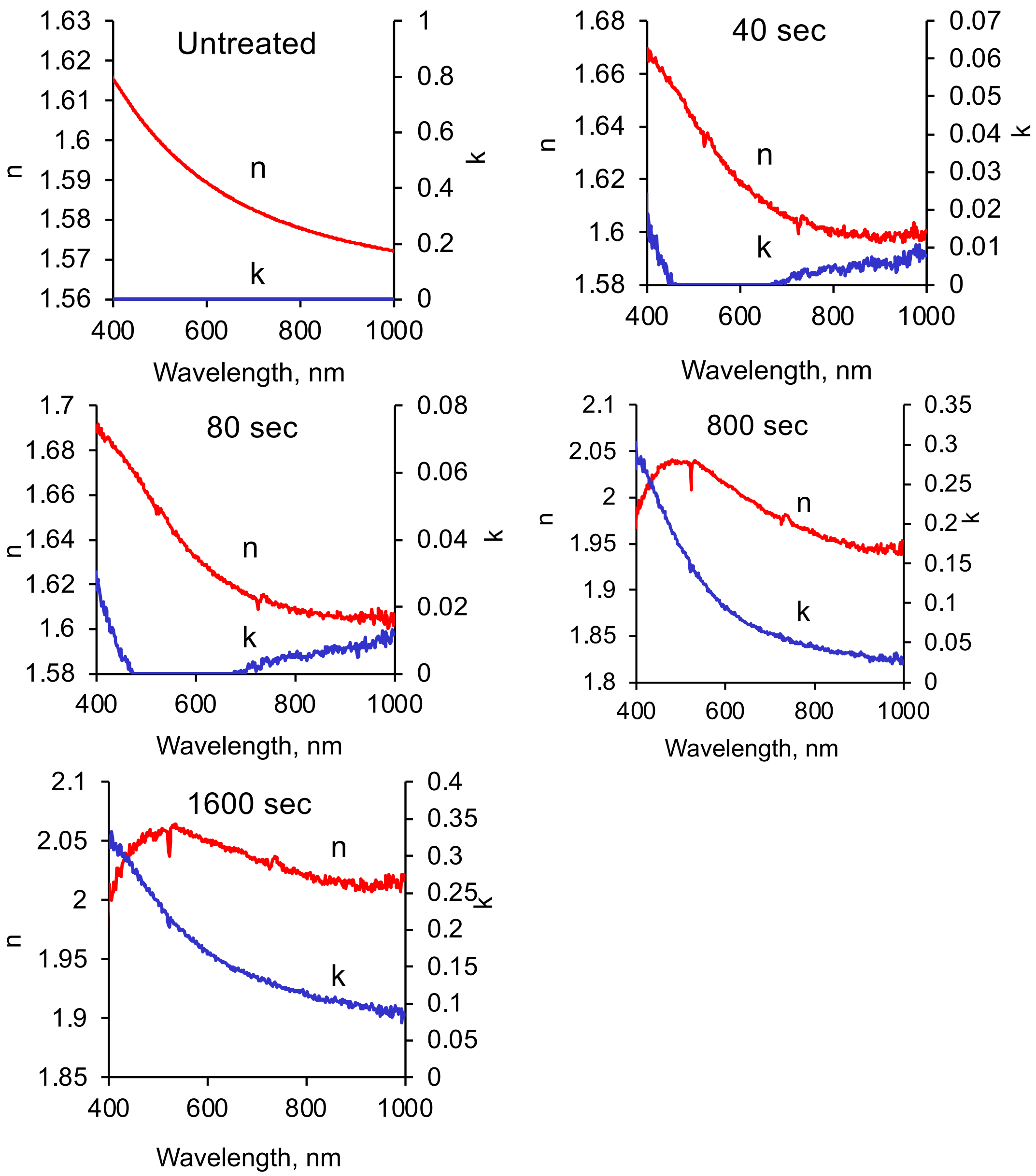

Fig.7. Optical parameters of the carbonized layer for different times of ion-beam treatment.

Using the optical parameters of the silicon wafer with the oxide layer, the optical parameters of the untreated polystyrene and the optical parameters of the carbonized layer with different degrees of carbonization caused by different times of ion-beam treatment, all other polystyrene samples with a thickness greater than the depth of nitrogen ion penetration in polystyrene were analysed.

Figure 8 shows the Ψ function of PS films with an initial nominal thickness of 120 nm treated with nitrogen ions with an energy of 20 keV. The optical model of these samples included a bulk silicon layer, a silicon oxide layer, a layer of untreated PS, and a layer of carbonized PS with the corresponding time of ion-beam treatment. In this case, the thickness of the silicon oxide was fixed. The thicknesses of the carbonized layer and the unchanged PS layer were fitted. The results of fitting the ellipsometric functions showed good agreement with the experimental spectra.

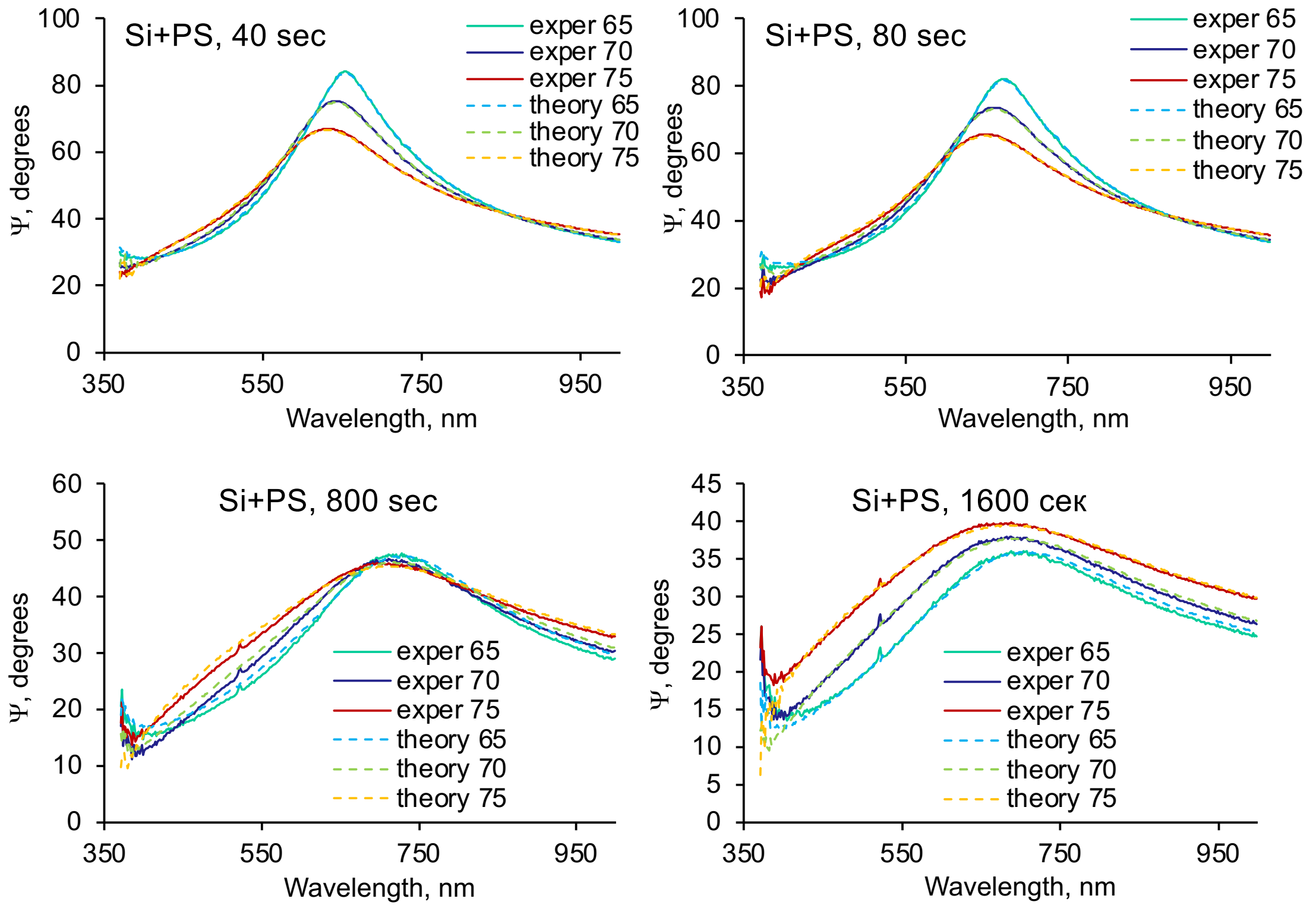

Fig.8. Ellipsometric Ψ function of a silicon wafer coated with polystyrene of 120 nm thickness and treated by nitrogen ions of 20 keV energy during 40, 80, 800 и 1600 seconds. Solid lines are experimental. Dashed lines are theoretical calculations. Beam incidence angles are given in degrees.

Other variants of fitting the optical parameters, such as fitting the optical parameters of the carbonized layer, led to unrealistic results of the ellipsometric Ψ and δ functions and did not correspond to the experimental spectra. Attempts to detect changes in the optical properties in the PS layer under the carbonized layer also did not yield results. Apparently, the PS layer under the carbonized layer does not change its optical parameters or these changes cannot be determined by this method.

Other samples with different thickness and treatment time were also measured and the results are shown in Fig. 9-15. For all samples, a fairly good match between the theoretical and experimental curves is observed.

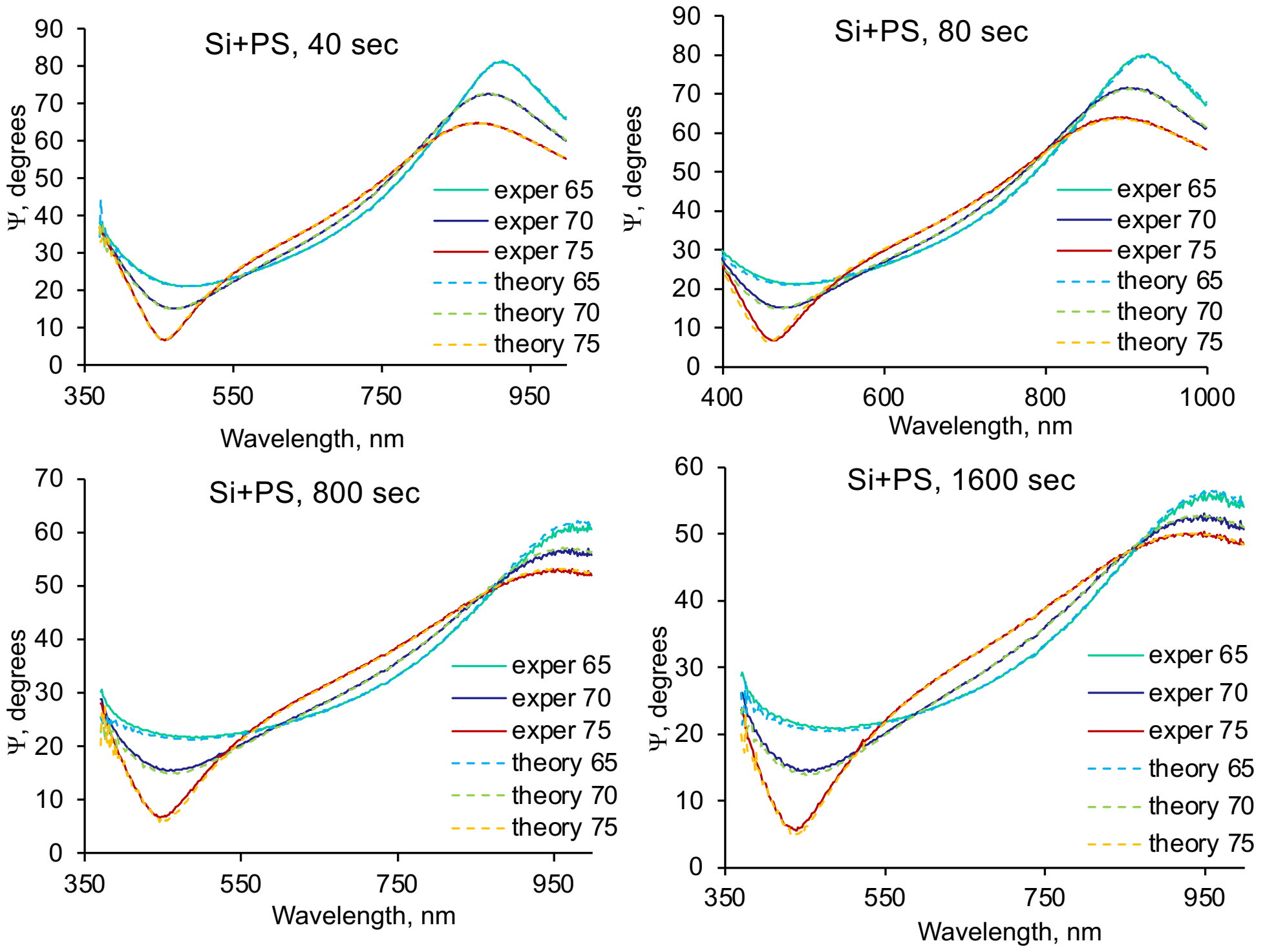

Fig.9. Ellipsometric Ψ function of a silicon wafer coated with polystyrene of 170 nm thickness and treated by nitrogen ions of 20 keV energy during 40, 80, 800 и 1600 seconds. Solid lines are experimental. Dashed lines are theoretical calculations. Beam incidence angles are given in degrees.

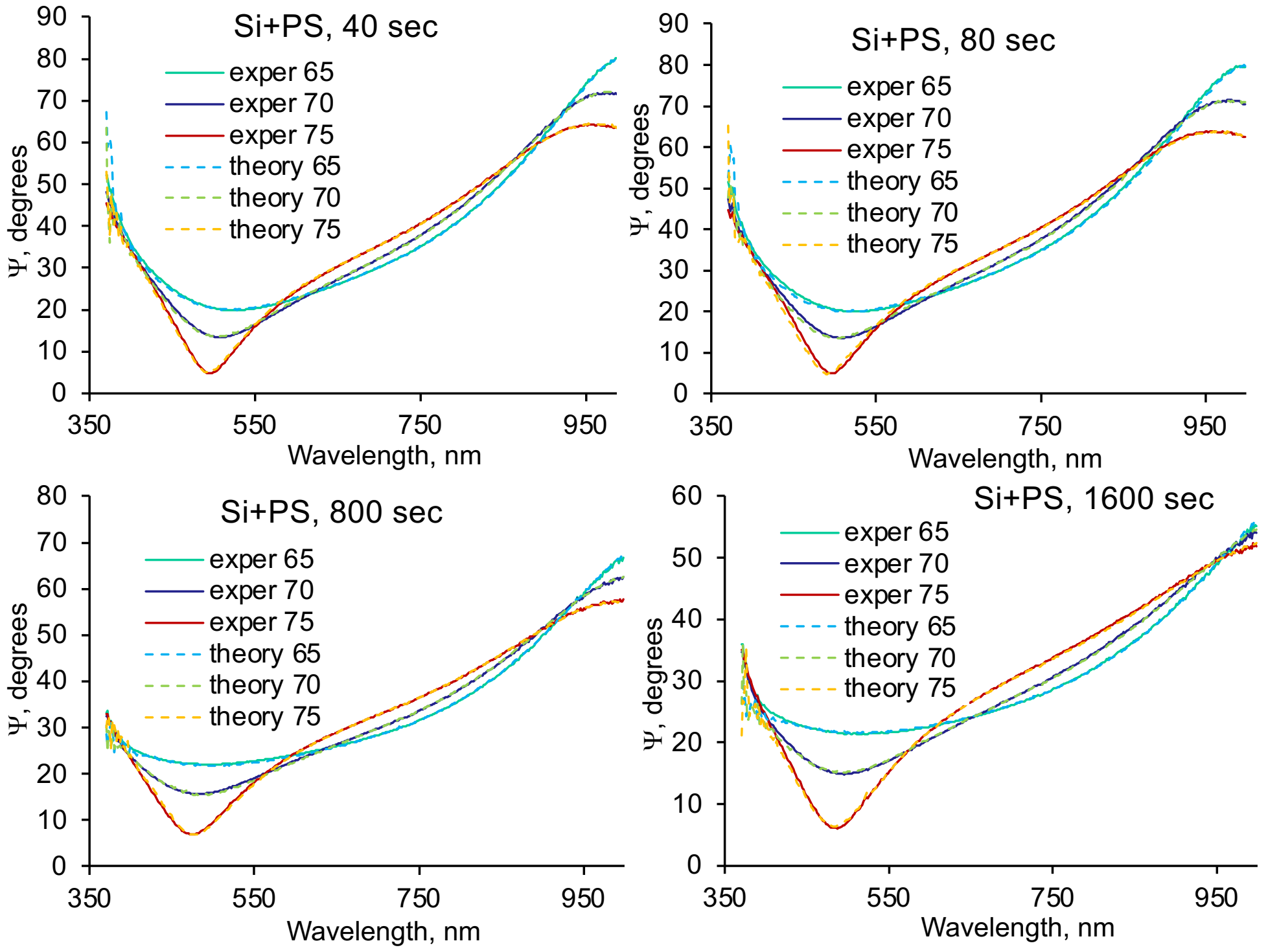

Fig.10. Ellipsometric Ψ function of a silicon wafer coated with polystyrene of 190 nm thickness and treated by nitrogen ions of 20 keV energy during 40, 80, 800 и 1600 seconds. Solid lines are experimental. Dashed lines are theoretical calculations. Beam incidence angles are given in degrees.

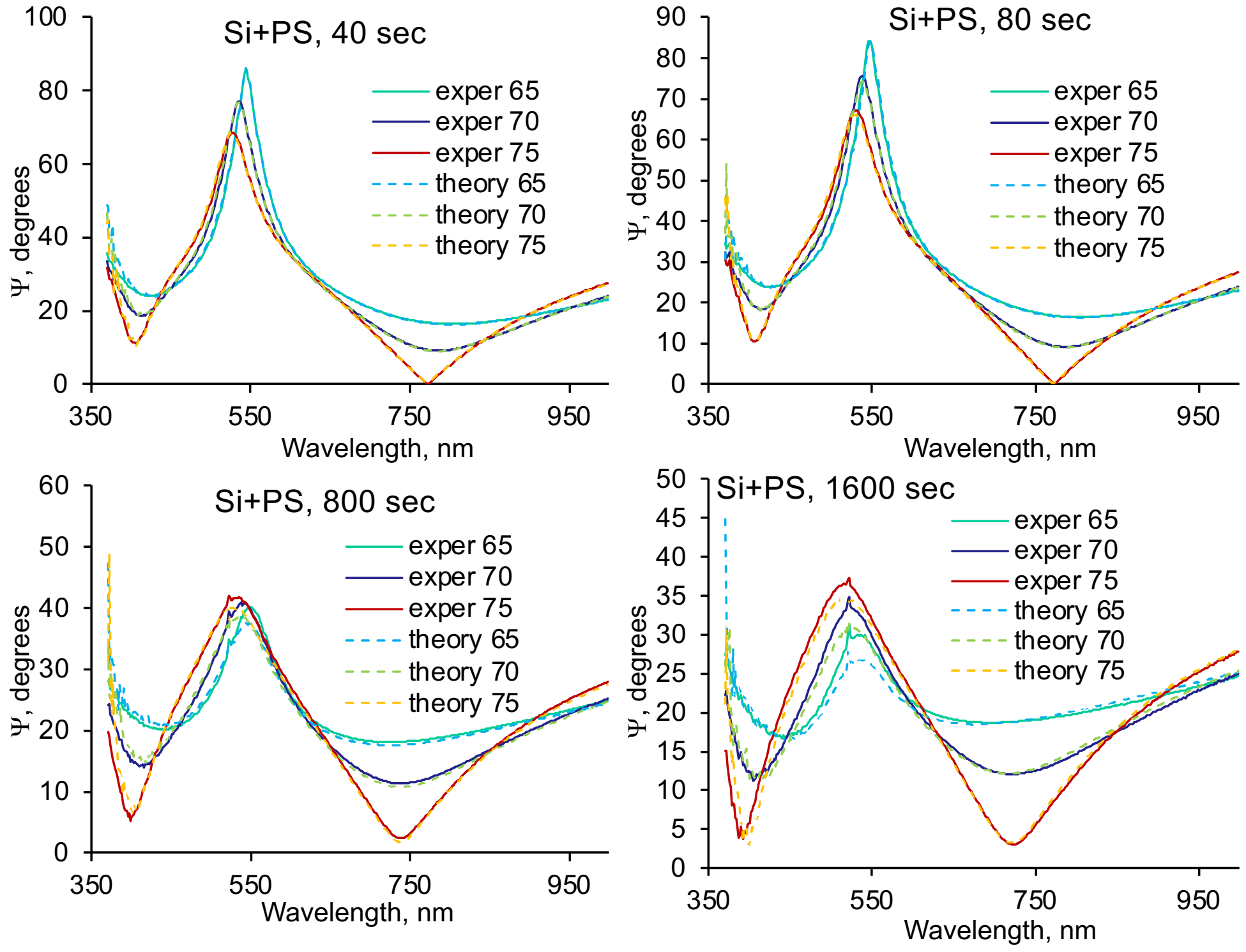

Fig.11. Ellipsometric Ψ function of a silicon wafer coated with polystyrene of 300 nm thickness and treated by nitrogen ions of 20 keV energy during 40, 80, 800 и 1600 seconds. Solid lines are experimental. Dashed lines are theoretical calculations. Beam incidence angles are given in degrees.

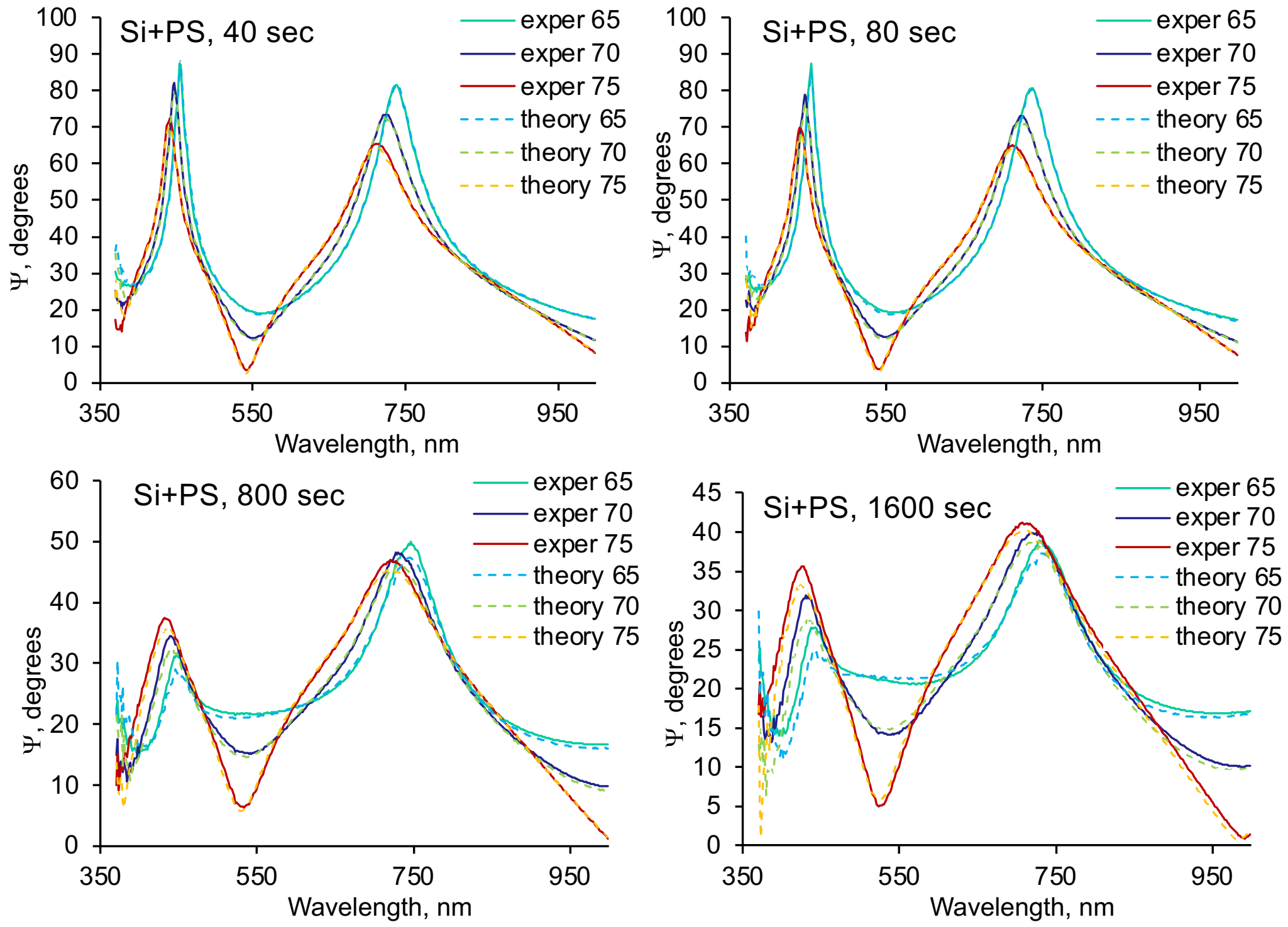

Fig.12. Ellipsometric Ψ function of a silicon wafer coated with polystyrene of 420 nm thickness and treated by nitrogen ions of 20 keV energy during 40, 80, 800 и 1600 seconds. Solid lines are experimental. Dashed lines are theoretical calculations. Beam incidence angles are given in degrees.

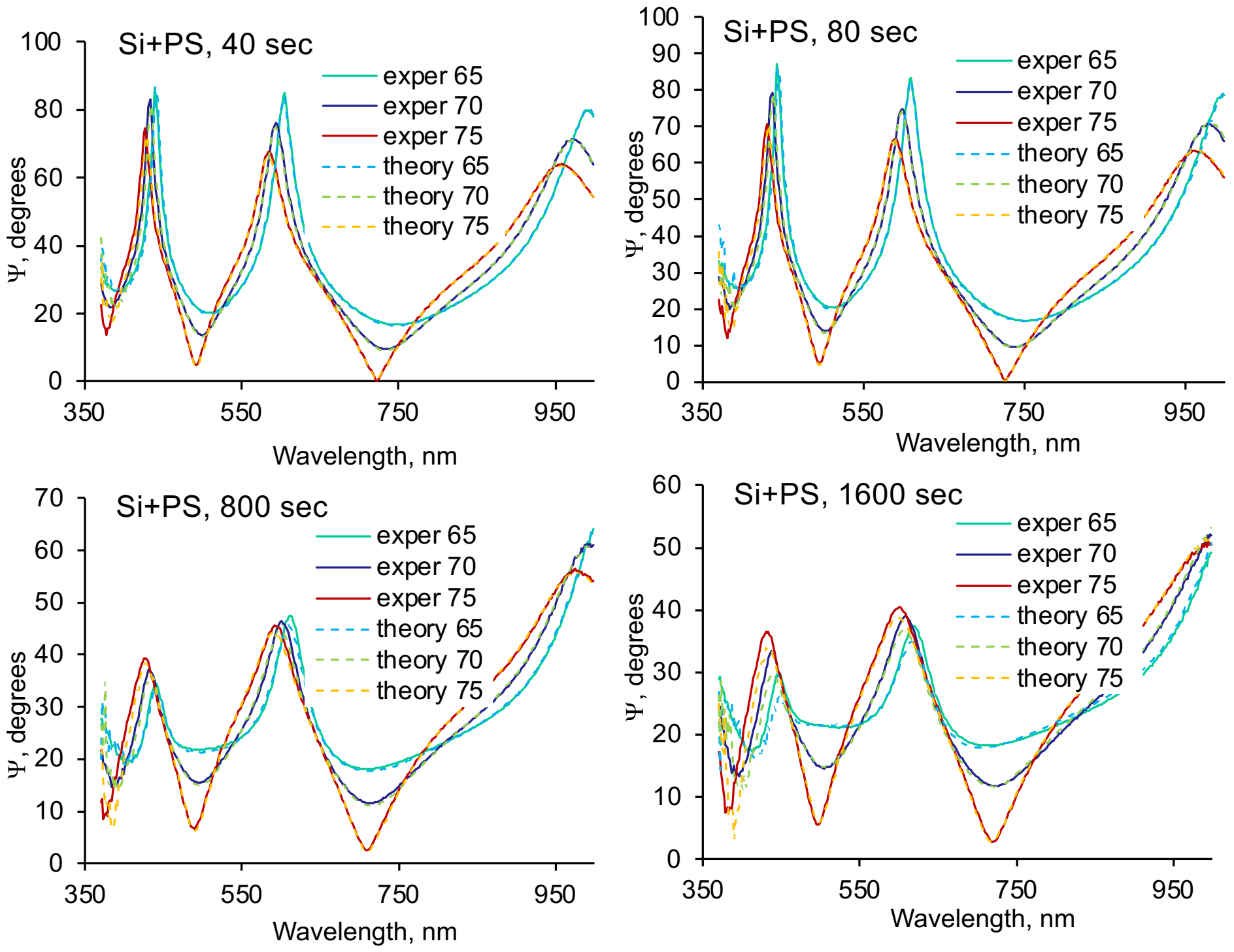

Fig.13. Ellipsometric Ψ function of a silicon wafer coated with polystyrene of 600 nm thickness and treated by nitrogen ions of 20 keV energy during 40, 80, 800 и 1600 seconds. Solid lines are experimental. Dashed lines are theoretical calculations. Beam incidence angles are given in degrees.

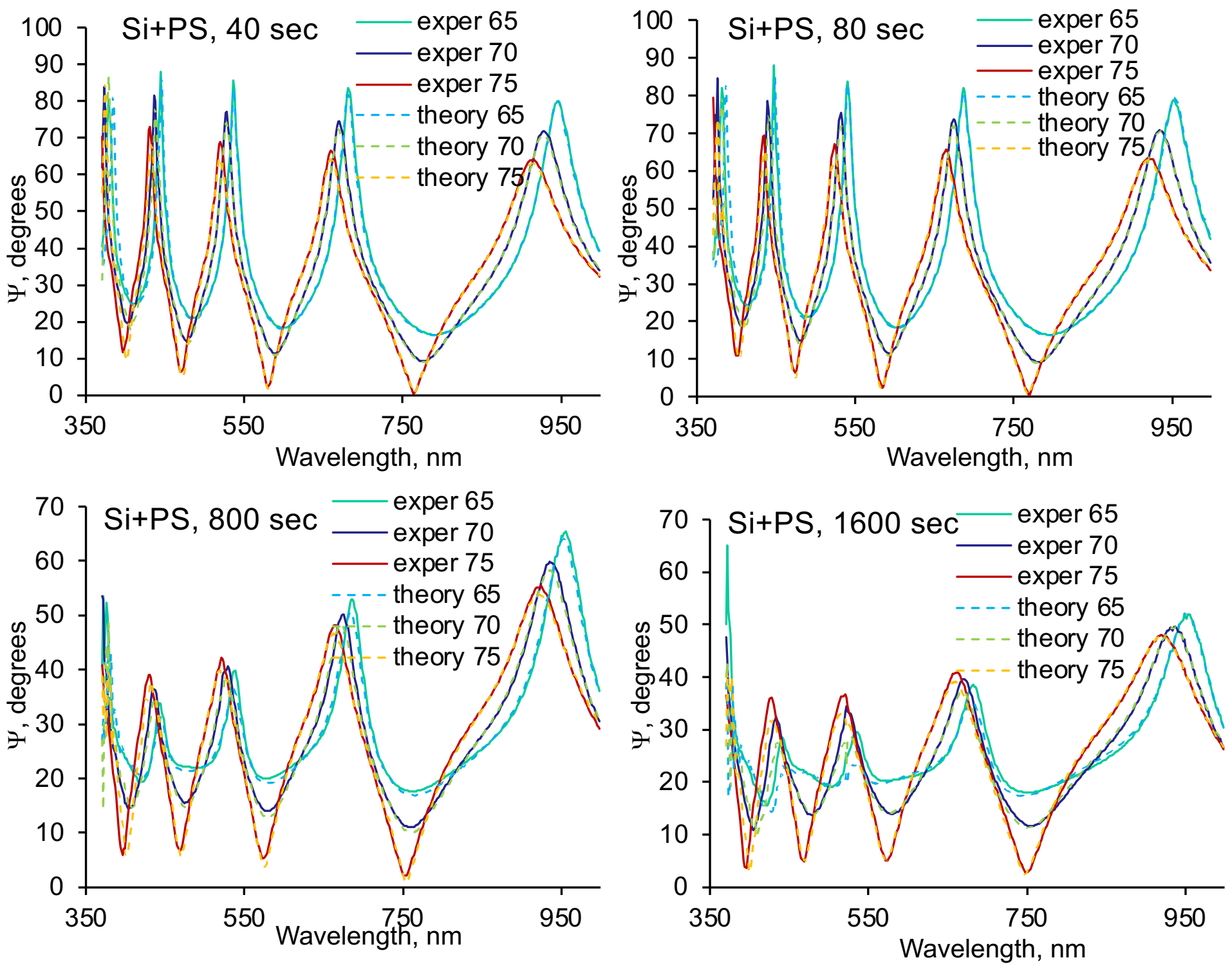

Fig.14. Ellipsometric Ψ function of a silicon wafer coated with polystyrene of 930 nm thickness and treated by nitrogen ions of 20 keV energy during 40, 80, 800 и 1600 seconds. Solid lines are experimental. Dashed lines are theoretical calculations. Beam incidence angles are given in degrees.

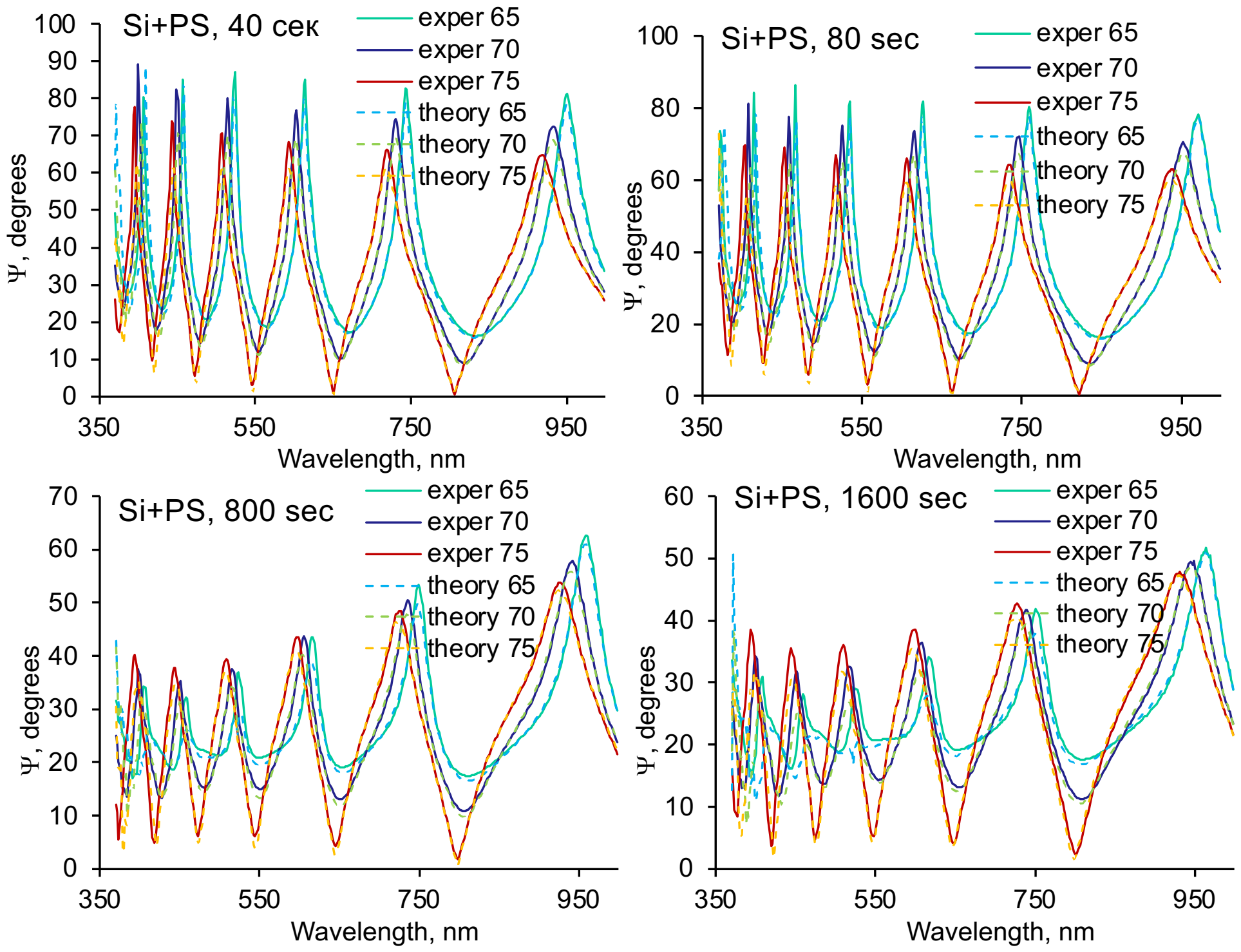

Fig.15. Ellipsometric Ψ function of a silicon wafer coated with polystyrene of 1330 nm thickness and treated by nitrogen ions of 20 keV energy during 40, 80, 800 и 1600 seconds. Solid lines are experimental. Dashed lines are theoretical calculations. Beam incidence angles are given in degrees.

For all samples with different thicknesses, it is clear observed that for the short treatment time the coincidence of theoretical and experimental curves is much better than for the long treatment time. This is especially clearly noticeable by the sharpness of the extremes in the maxima and minima of the curves. This feature of the ellipsometric curves is associated with the heterogeneity of the boundary between the carbonized surface layer and unchanged bulk layers of PS. Since the boundary between the layers is determined by chemical reactions of carbonization of PS macromolecules, which in turn are determined by the number of atoms knocked out by the incoming nitrogen ion, a sharp boundary between the layers should not be expected. Accordingly, the interference peaks of the light beam reflected from a somewhat blurred boundary between the layers have fuzzy transitions in brightness, which is observed as a blurring of the maxima and minima of the experimental curves. With the growth of the difference in the optical coefficients of the PS layers, which increases with the treatment time, the blurring of the extremes of the ellipsometric curves becomes more noticeable. However, we note that attempts to use models of the optical parameters of the PS layers with additional transition layers between them did not improve the agreement of the theoretical and experimental ellipsometric curves. Therefore, such assumptions remain unproven and, perhaps, require more detailed consideration and analysis.

These results showed that the thickness of the carbonized surface layer does not depend on the treatment time and the thickness of the initial PS film (Fig. 16).

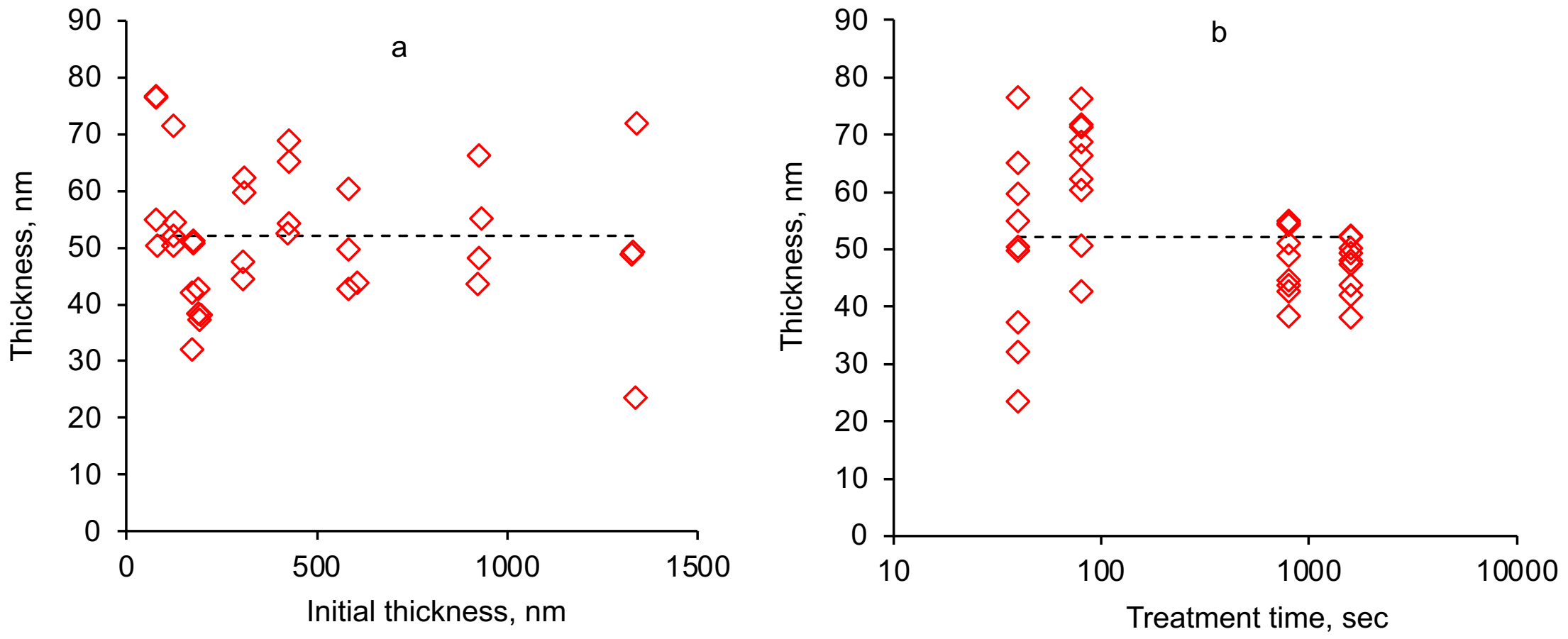

Fig.16. Thickness of the carbonized surface layer of PS after ion beam treatment (a) depending on the thickness of the initial PS layer and (b) depending on the treatment time. By the ellipsometric data.

The thickness of the carbonized layer was compared with the penetration depth of nitrogen ions in the PS (Fig. 17). For this purpose, the depth penetration for $N^+$ and $N_2^+$ ions was calculated using SRIM code. In the low-temperature nonequilibrium plasma from which the ions are accelerated, nitrogen is present in the form of molecules and individual atoms. Preliminary measurements showed that the plasma contains approximately equal amounts of molecular and atomic nitrogen ions. Therefore, the penetration depth of ions in the PS was calculated for these two types of ions. The carbonization of the PS surface layer is associated with the knocking out of carbon and hydrogen atoms from the PS macromolecules and the formation of new carbon-containing groups. Therefore, for comparison, the concentration profile of PS atom vacancies formed as a result of collisions with atomic and molecular nitrogen ions was analysed.

The average thickness of the carbonized layer for all films with different thickness and all treatment times is 52±12 nm. For short treatment time (low ion fluence), the standard deviation of the carbonized layer thickness from the average value is from 10 to 16 nm for all sample thicknesses. For long treatment times corresponding to complete carbonization of the surface layer, the standard deviation is in a range of 4-6 nm. For atomic ions, the penetration depth into the PS is 86 nm, and for molecular ions, it is 43 nm. Thus, the experimental value of the carbonized layer thickness lies between these two values. That is, the carbonized layer in this experiment is formed as a result of the sum of the carbonization processes from the atomic and molecular ions penetration.

The pristine PS with initial optical parameters lays under the carbonised layer and occupies the entire remaining film thickness. The thickness of such an unchanged layer decreases proportionally to the treatment time, which is evident from the vertical shift of the trend lines in Fig. 17(b).

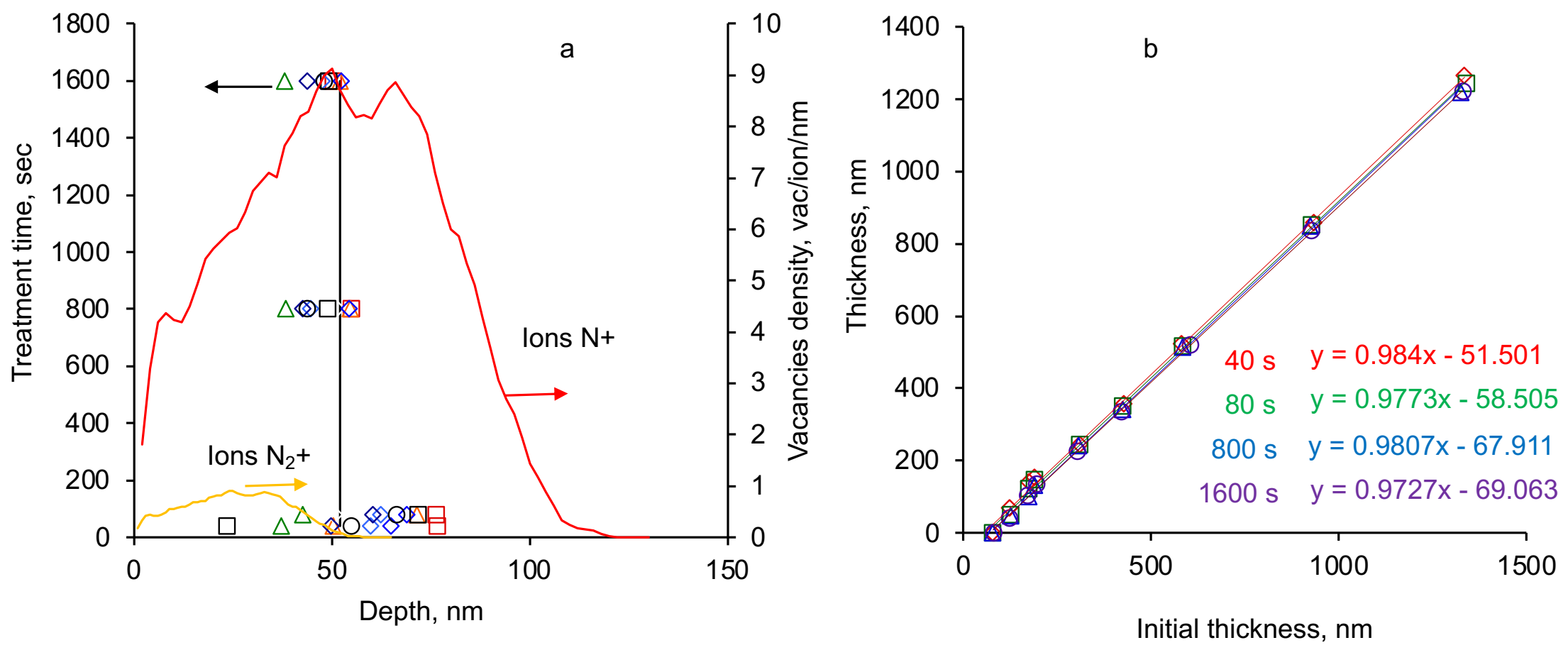

Fig. 17. (a) Thickness of the carbonized layer for all samples (horizontal axis) depending on the treatment time (left vertical axis). Different colours and signs correspond to different initial thicknesses of the samples (see Fig. 16). The black vertical line corresponds to the average values of the carbonized layer for all sample thicknesses. The density of vacancies in PS caused by the penetration of atomic and molecular ions calculated using the TRIM program are shown as yellow and red curves (right vertical axis). (b) Thickness of the unchanged bulk layer of the treated PS in dependence on the initial PS thickness. The fitting linear parameters are given for each treatment time. The treatment time is given in seconds.

The carbonized surface layer of the PS forms due to structural transformations of the original PS. The layer lying deeper the ion penetration depth remains unaffected by collision processes. This is determined by the optical properties of the underlying PS under the carbonized layer. However, the thickness of the underlying PS decreases with the time of ion beam treatment (Fig. 18).

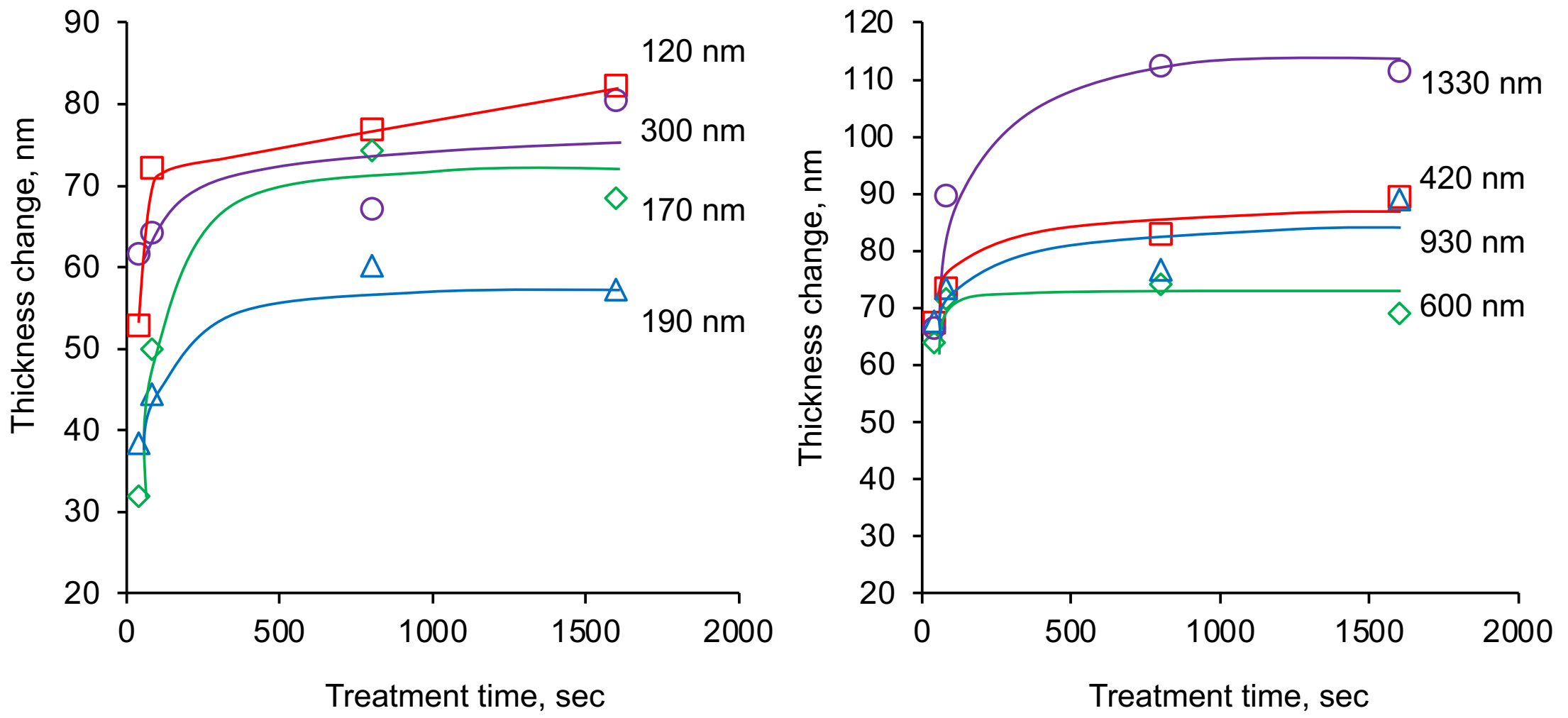

Fig. 18. Change in the thickness of the bulk unchanged layer lying under the carbonized layer depending on the time of ion beam treatment for samples of different thicknesses.

This is observed for all sample thicknesses. Note that the results presented show a decrease in the thickness of the unchanged underneath layer after the formation of the top surface carbonized layer and do not reflect structural transformations due to collision processes with penetrating ions from the plasma. The decrease in the unchanged layer is quite significant and for thick samples exceeds the value of the carbonized layer.
Based on the total thickness of the PS layer after treatment, the etching rate of the PS can be calculated. The etching rate of the PS during ion beam treatment is nonlinear (Fig. 19). These results coincide with the previously obtained results on the etching rate of thin PS films. The nonlinearity of the etching process is due to the formation of a carbonized surface layer, which has a much lower etching rate in plasma and ion beam. Therefore, the integral etching rate of the PS samples for the entire treatment time for long treatment time is much lower than for short treatment time.

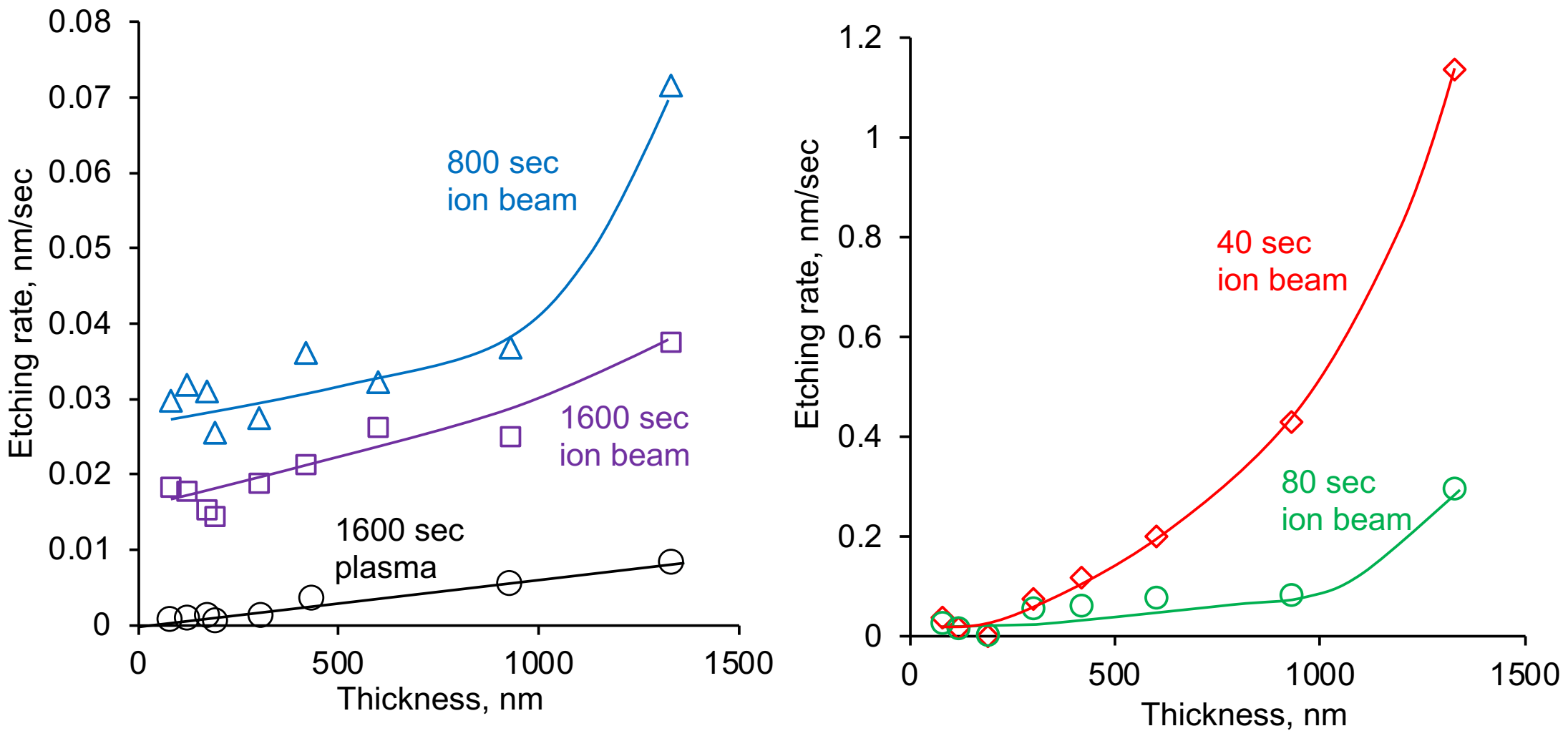

Fig. 19. Etching rate of PS samples with ion beam treatment and with only plasma treatment depending on the thickness of the samples at different treatment times.

It should be noted that the etching of a polymer film in the literature is always considered as a surface process. The etching is described as knocking out atoms or fragments of polymer molecules from the surface. In this case, the etching rate should not depend on the thickness of the sample. However, in the case of ion beam treatment, the etching rate increases with increasing sample thickness. That is, the etching process in this case occurs not only from the surface, but affects deeper layers of the polymer. That is, the etching process is volumetric. This means that the modification of the polymer structure occurs at depths greater than the depth of ion penetration. Therefore, the changes in the chemical structure of the bulk polymer layer should be observed.
The analysis of the PS structure after ion beam treatment was made using FTIR absorption spectra. The spectra were recorded in the transmission geometry. Therefore, the spectra intensity can be directly used for a quantitative analysis of the chemical group concentration. The silicon wafers were taken with weak doping, which gave low light absorption in the region of middle wavelengths of the FTIR spectra. The reverse side of the silicon wafers was not polished. Therefore, the spectra do not show interference inherent in the spectra of silicon wafers with double-sided polishing. The results of the spectra of silicon wafers with a spincoated PS layer are shown in Fig. 20.

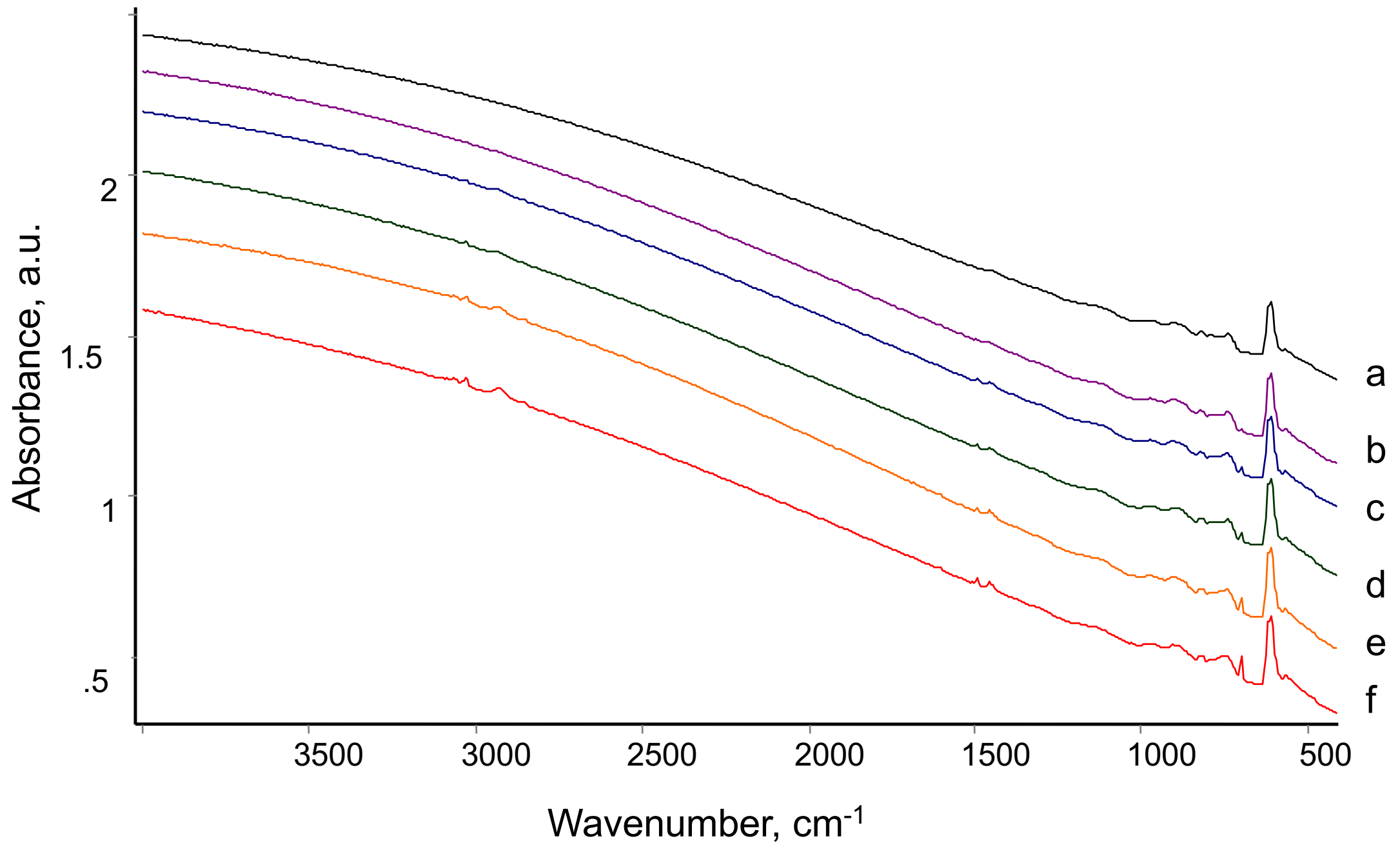

Fig.20. FTIR transmission spectra of silicon wafers with a spincoated PS nanolayer: (a) control wafer without PS, (b) 78 nm PS layer, (c) 172 nm PS layer, (d) 191 nm PS layer, (e) 304 nm PS layer, (f) 430 nm PS layer. The spectra are shifted vertically for better observation.

As can be seen from the presented results, the spectra show strong scattering of light on the inhomogeneities of the back side of the silicon wafer, and a number of relatively weak lines of silicon and silicon oxide are visible against its background. Analysis of the PS and its layers is impossible from such original spectra.

To analyze the PS spectra, the spectrum of the silicon wafer itself without the PS layer was subtracted from the all original spectra. The resulting spectra were treated using the baseline linearisation method and shifted along the absorption axis for better observation. The resulted spectra are shown in Fig. 21.

The spectra of all samples show lines at 3081, 3060, 3026, 1602 and 1493 $cm^{-1}$ of vibrations in the aromatic ring and lines at 2922, 2851 and 1452 $cm^{-1}$ of vibrations of the methyl group of the polystyrene macromolecule. These lines were used further for the analysis of the PS spectra. Other low-intensity lines could not be used due to the high noise level of the spectrum compared to their intensity.

Fig. 22 shows the spectra of a PS nanolayer with a nominal thickness of 78 nm treated with ion beam treatment for different treatment time. In the spectrum of untreated PS, only lines of PS macromolecules are observed. In the spectrum of samples treated for a short time, the PS lines are preserved, but their intensity is much lower. In the spectra of samples treated for a long time, the PS lines are not visible.

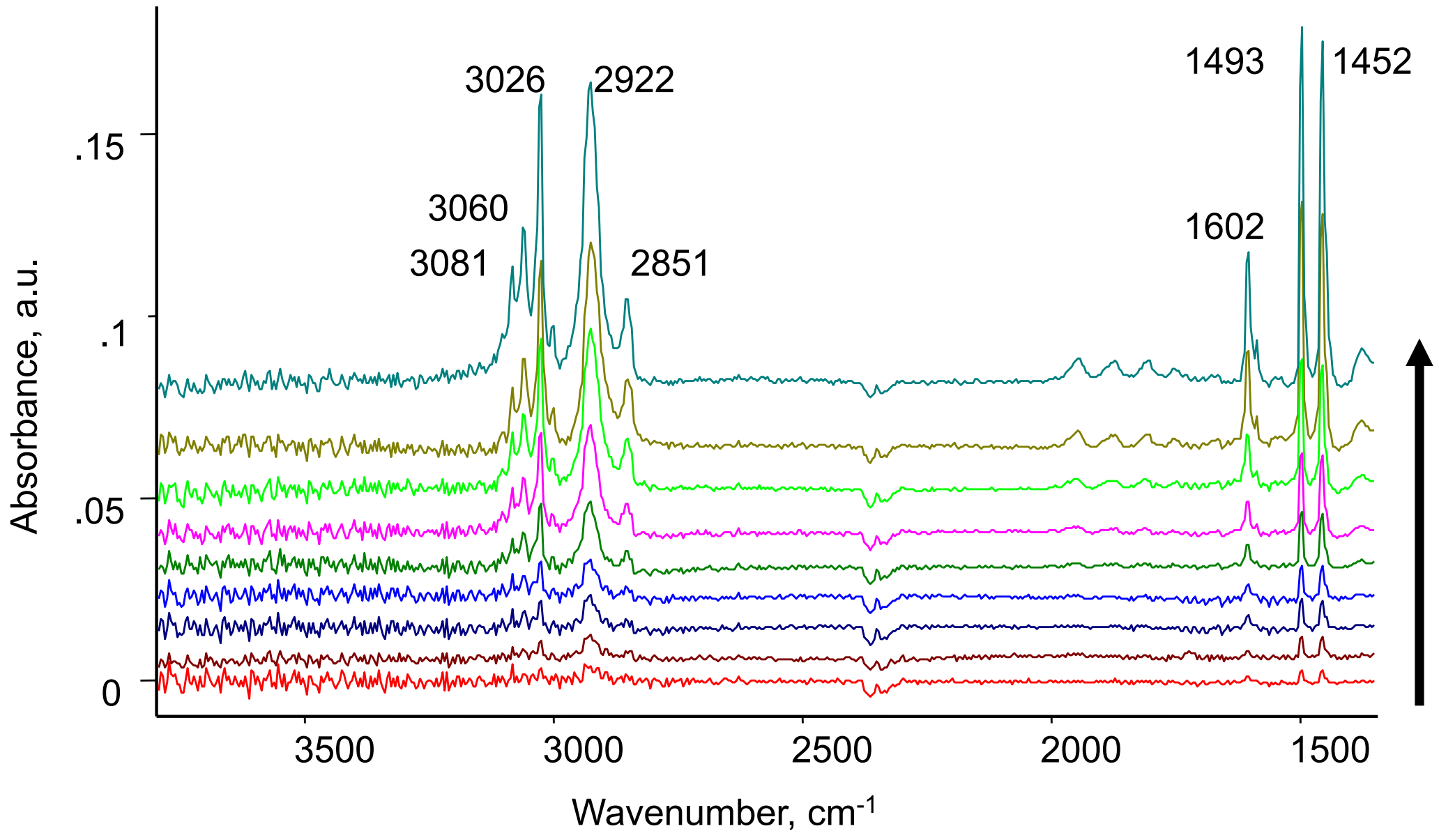

Fig.21. FTIR transmission spectra of PS nanolayers of different thicknesses treated with plasma for 1600 seconds. The order of spectra (by arrow) is as follows: 78 nm, 123 nm, 172 nm, 191 nm, 304 nm, 430 nm, 594 nm, 928 nm, 1331 nm. Spectra of silicon wafer are subtracted. Spectra are treated by baseline linearisation and shifted along the absorption axis for better observation.

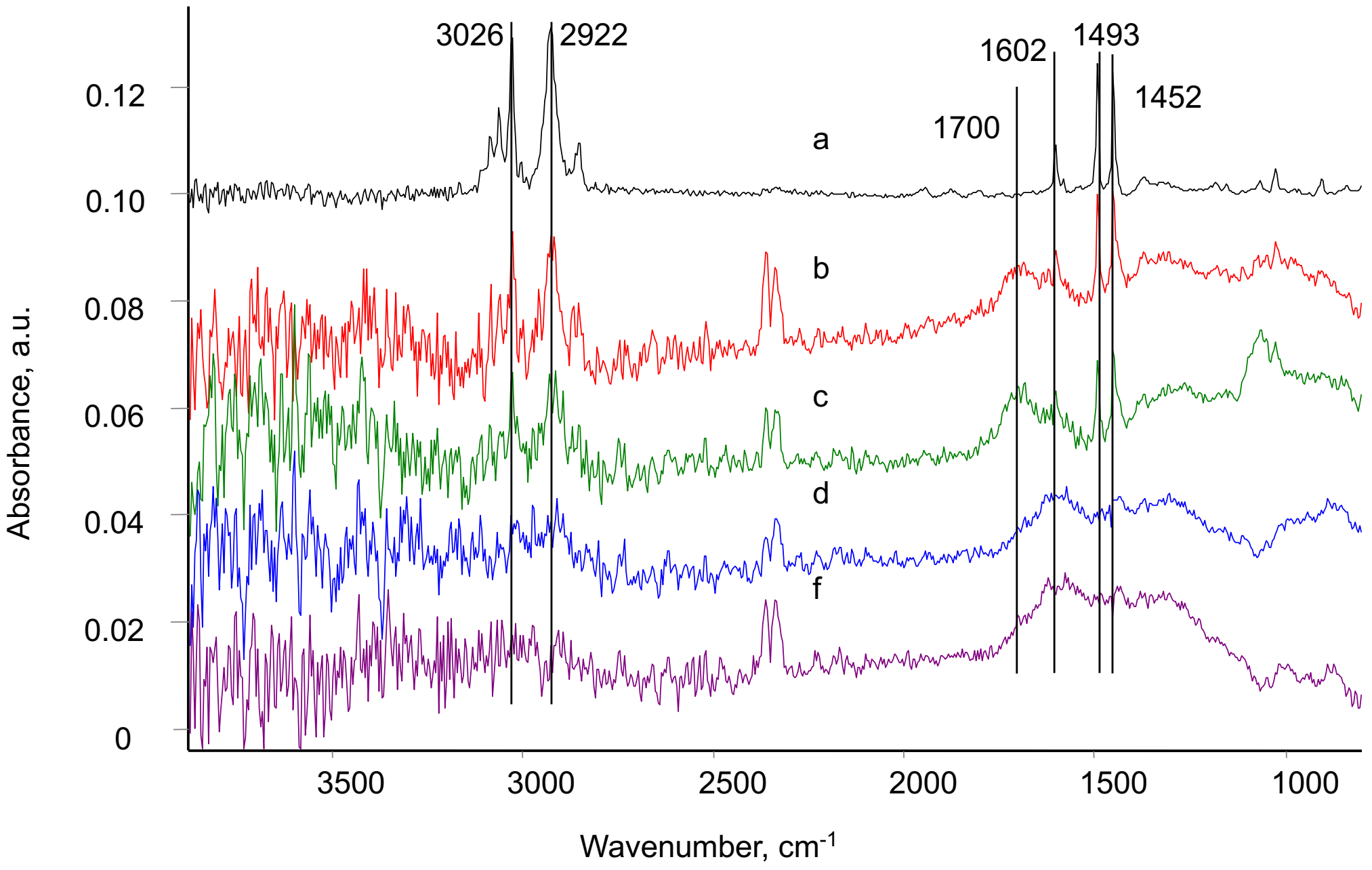

Fig.22. FTIR transmission spectra of a PS nanolayer with a nominal thickness of 78 nm treated with ion beam treatment: (a) untreated PS, (b) treated for 40 seconds, (c) treated for 80 seconds, (d) treated for 800 seconds and (f) treated for 1600 seconds. Spectra b-f are stretched along the absorption axis for better observation. Spectra of the silicon wafer are subtracted. Spectra are treated by baseline linearisation and shifted along the absorption axis for better observation. The marked lines are further used for quantitative analysis.

In addition to the PS lines, new lines are observed in the spectra of the treated samples in the region of 3500, 1800-1600 and 1300-1100 $cm^{-1}$. These broad lines are related to the vibrations of hydroxyl, carbonyl, unsaturated carbon-carbon groups in the carbonized PS layer. Since the thickness of this PS layer is close to the penetration depth of nitrogen ions during ion beam treatment, it can be stated that the spectrum of the sample processed for 1600 seconds completely corresponds to the carbonized layer. Fig. 23 shows the spectra of a PS nanolayer with a nominal thickness of 1330 nm treated with ion beam for different treatment times. In the spectrum of both untreated and treated PS, only lines of PS macromolecules are observed. The lines of the carbonized layer are not visible against the strong lines of the PS itself. Therefore, for thick samples for which the thickness of the PS layer exceeded the thickness of the carbonized layer. For further analysis, the spectra of the untreated PS were subtracted from the spectra of the treated samples.

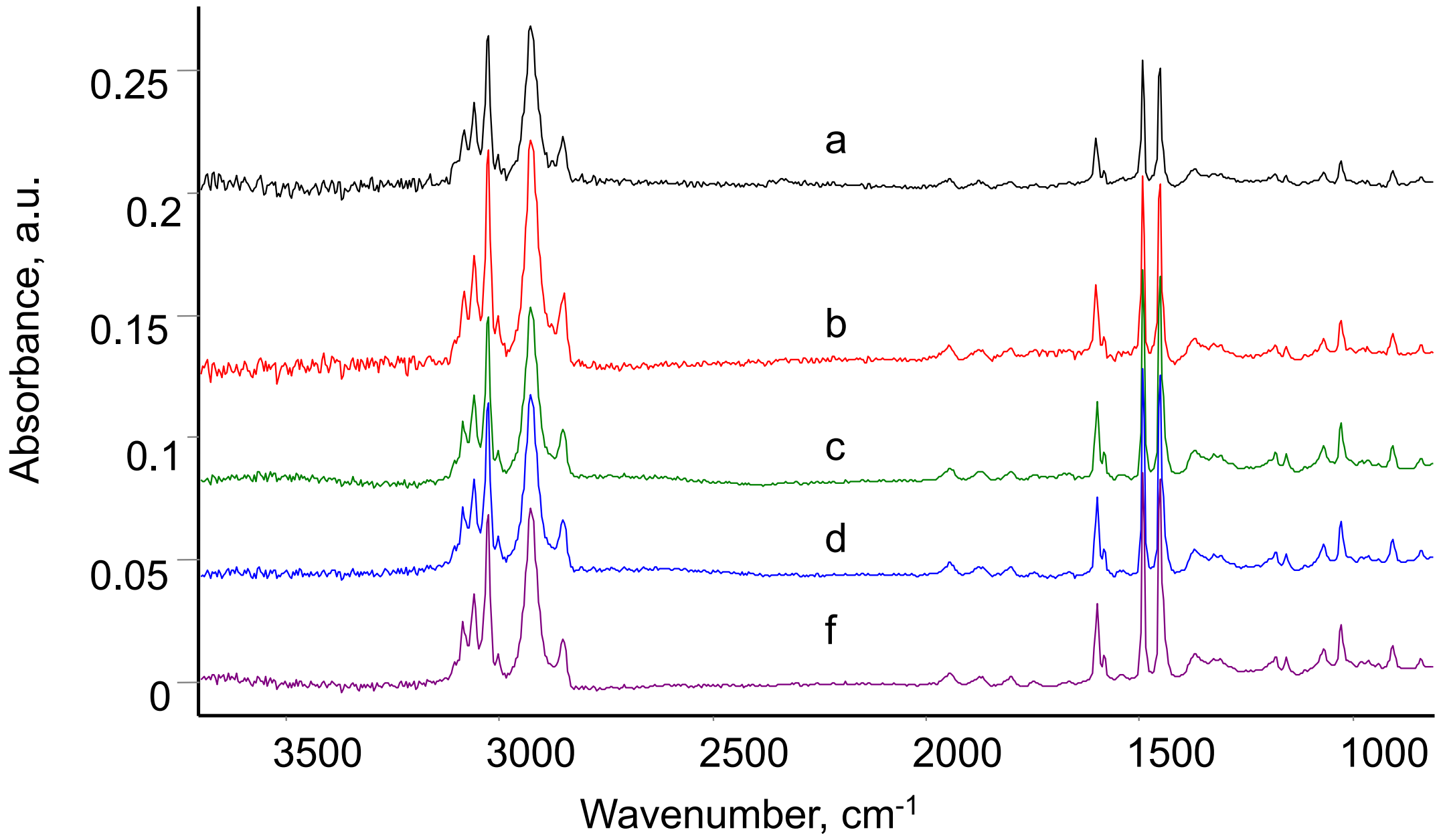

Fig.23. FTIR transmission spectra of a PS nanolayer with a nominal thickness of 1330 nm: (a) untreated PS, (b) treated for 40 seconds, (c) treated for 80 seconds, (d) treated for 800 seconds, and (f) treated for 1600 seconds. The spectra of the silicon wafer are subtracted. The spectra are treated by baseline linearisation and shifted along the absorption axis for better observation.

The results of spectrum subtracting of the untreated PS from the spectra of the treated samples are shown in Figs. 24–32. For subtraction, the intensity of the spectrum of untreated PS was multiplied by a fitting coefficient so that all lines of untreated PS were, as far as possible, completely subtracted from the spectra of the samples.

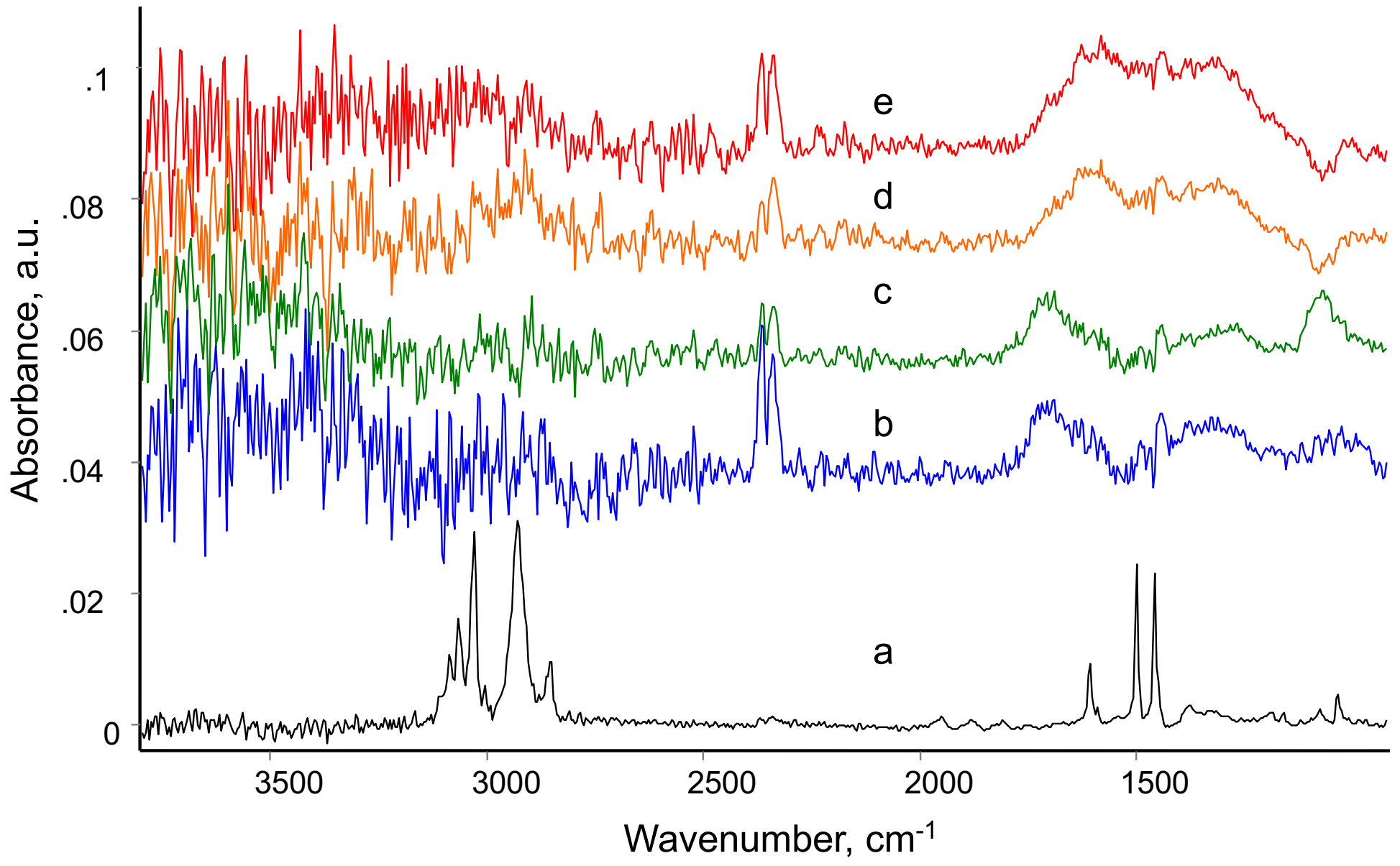

Fig. 24. FTIR transmission spectrum of untreated PS nanolayer with thickness of 78 nm (a) and differential spectra of PS nanolayer treated by ion-plasma treatment for (b) 40 seconds, (c) 80 seconds, (d) 800 seconds, (e) 1600 seconds. Spectra of silicon wafer are subtracted. Spectra are processed by baseline linearisation method and shifted along absorption axis for better observation. To obtain differential spectra (b-e) spectrum of untreated PS (a) is subtracted from corresponding spectra of treated samples and absorption scale is stretched for better observation.

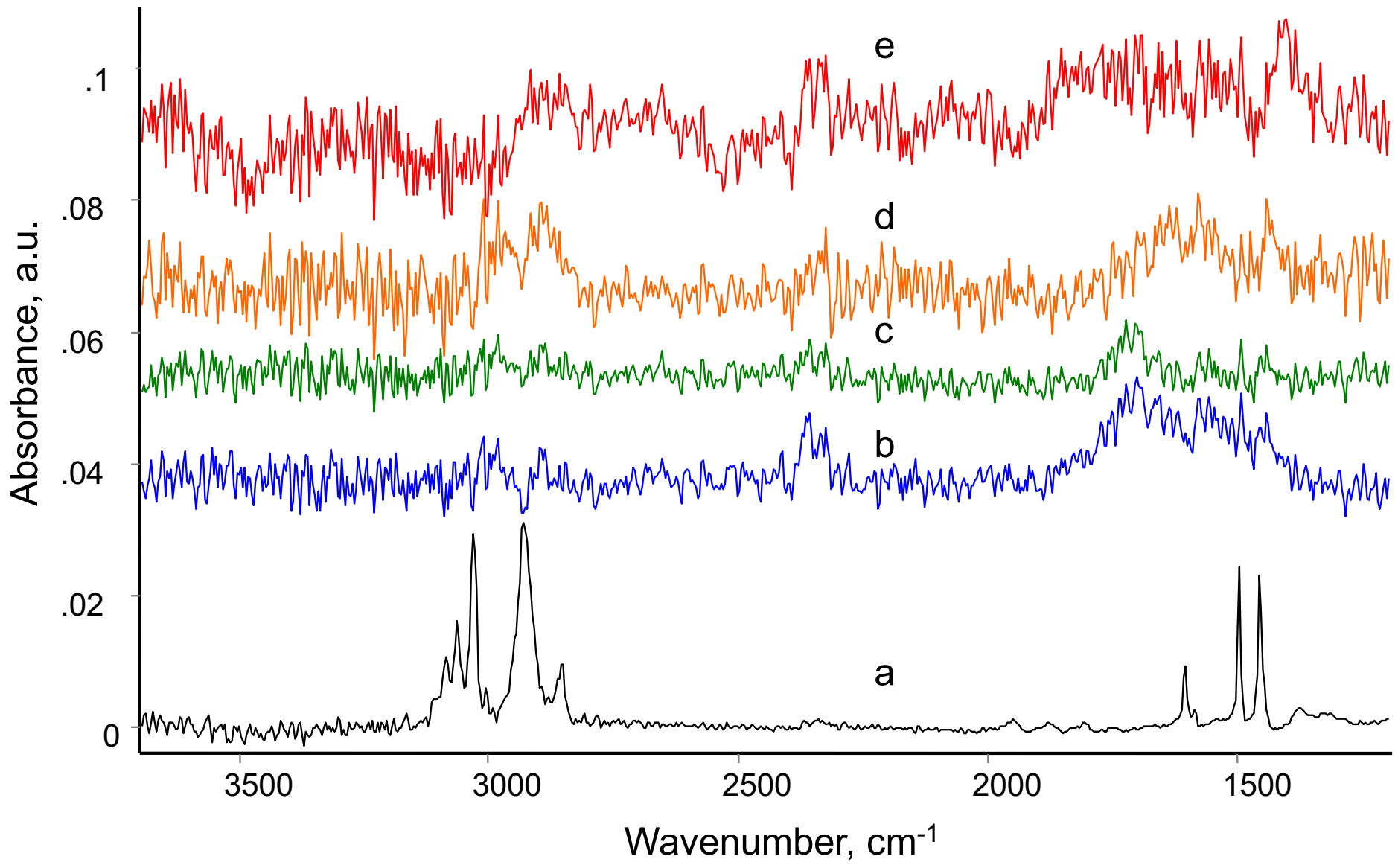

Fig. 25. FTIR transmission spectrum of untreated PS nanolayer with thickness of 122 nm (a) and differential spectra of PS nanolayer treated by ion-plasma treatment for (b) 40 seconds, (c) 80 seconds, (d) 800 seconds, (e) 1600 seconds. Spectra of silicon wafer are subtracted. Spectra are processed by baseline linearisation method and shifted along absorption axis for better observation. To obtain differential spectra (b-e) spectrum of untreated PS (a) is subtracted from corresponding spectra of treated samples and absorption scale is stretched for better observation.

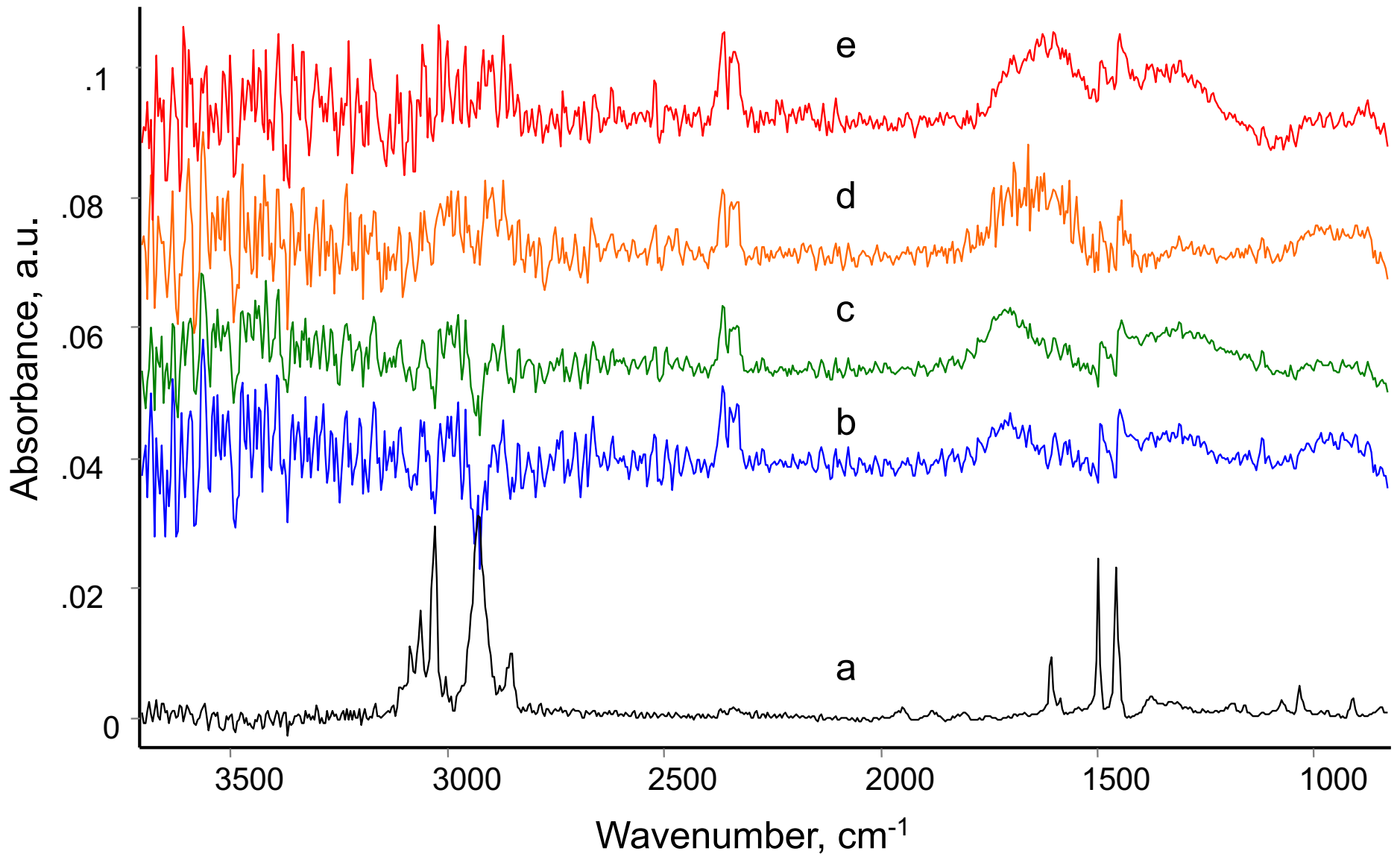

Fig. 26. FTIR transmission spectrum of untreated PS nanolayer with thickness of 170 nm (a) and differential spectra of PS nanolayer treated by ion-plasma treatment for (b) 40 seconds, (c) 80 seconds, (d) 800 seconds, (e) 1600 seconds. Spectra of silicon wafer are subtracted. Spectra are processed by baseline linearisation method and shifted along absorption axis for better observation. To obtain differential spectra (b-e) spectrum of untreated PS (a) is subtracted from corresponding spectra of treated samples and absorption scale is stretched for better observation.

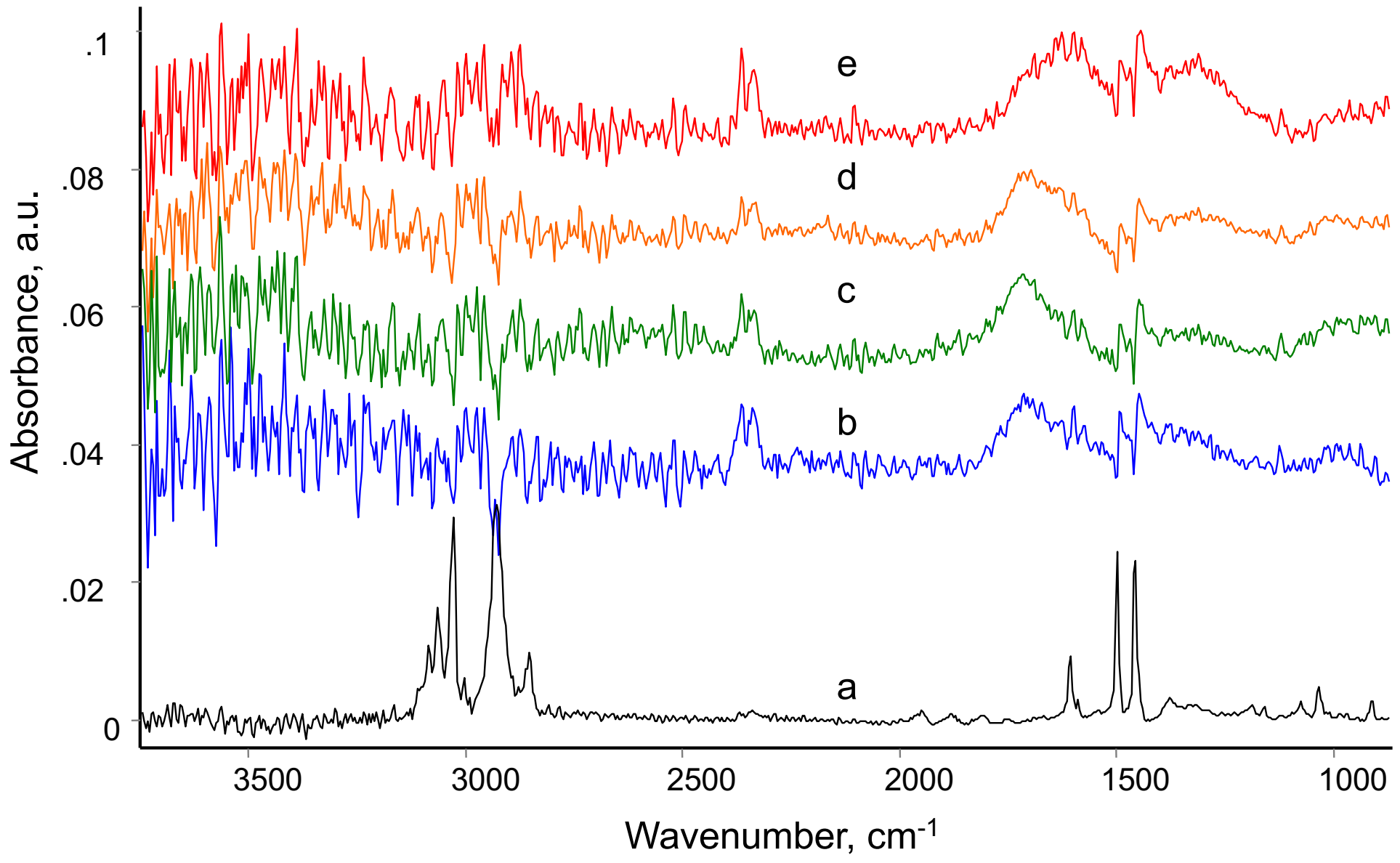

Fig. 27. FTIR transmission spectrum of untreated PS nanolayer with thickness of 190 nm (a) and differential spectra of PS nanolayer treated by ion-plasma treatment for (b) 40 seconds, (c) 80 seconds, (d) 800 seconds, (e) 1600 seconds. Spectra of silicon wafer are subtracted. Spectra are processed by baseline linearisation method and shifted along absorption axis for better observation. To obtain differential spectra (b-e) spectrum of untreated PS (a) is subtracted from corresponding spectra of treated samples and absorption scale is stretched for better observation.

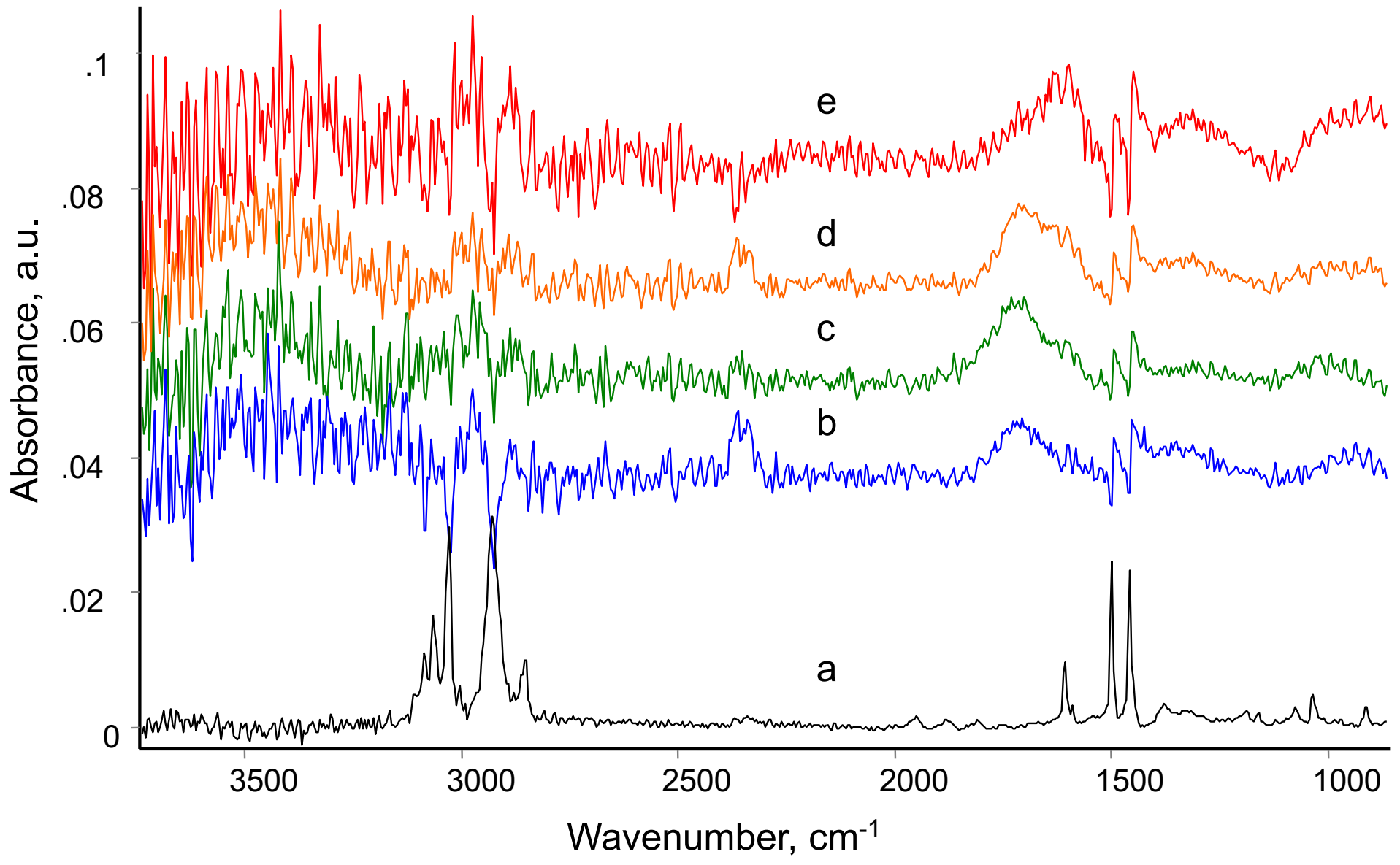

Fig. 28. FTIR transmission spectrum of untreated PS nanolayer with thickness of 310 nm (a) and differential spectra of PS nanolayer treated by ion-plasma treatment for (b) 40 seconds, (c) 80 seconds, (d) 800 seconds, (e) 1600 seconds. Spectra of silicon wafer are subtracted. Spectra are processed by baseline linearisation method and shifted along absorption axis for better observation. To obtain differential spectra (b-e) spectrum of untreated PS (a) is subtracted from corresponding spectra of treated samples and absorption scale is stretched for better observation.

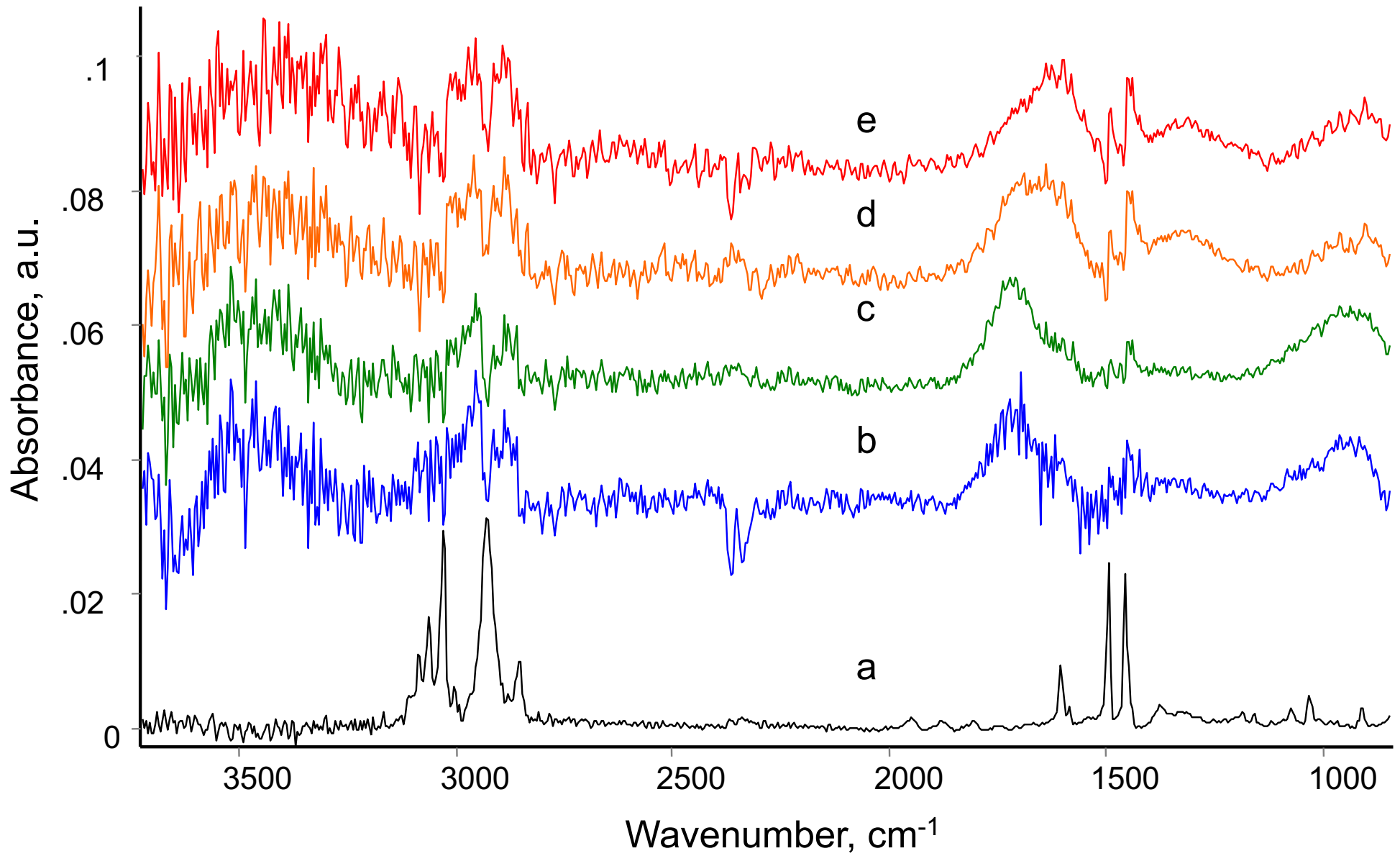

Fig. 29. FTIR transmission spectrum of untreated PS nanolayer with thickness of 420 nm (a) and differential spectra of PS nanolayer treated by ion-plasma treatment for (b) 40 seconds, (c) 80 seconds, (d) 800 seconds, (e) 1600 seconds. Spectra of silicon wafer are subtracted. Spectra are processed by baseline linearisation method and shifted along absorption axis for better observation. To obtain differential spectra (b-e) spectrum of untreated PS (a) is subtracted from corresponding spectra of treated samples and absorption scale is stretched for better observation.

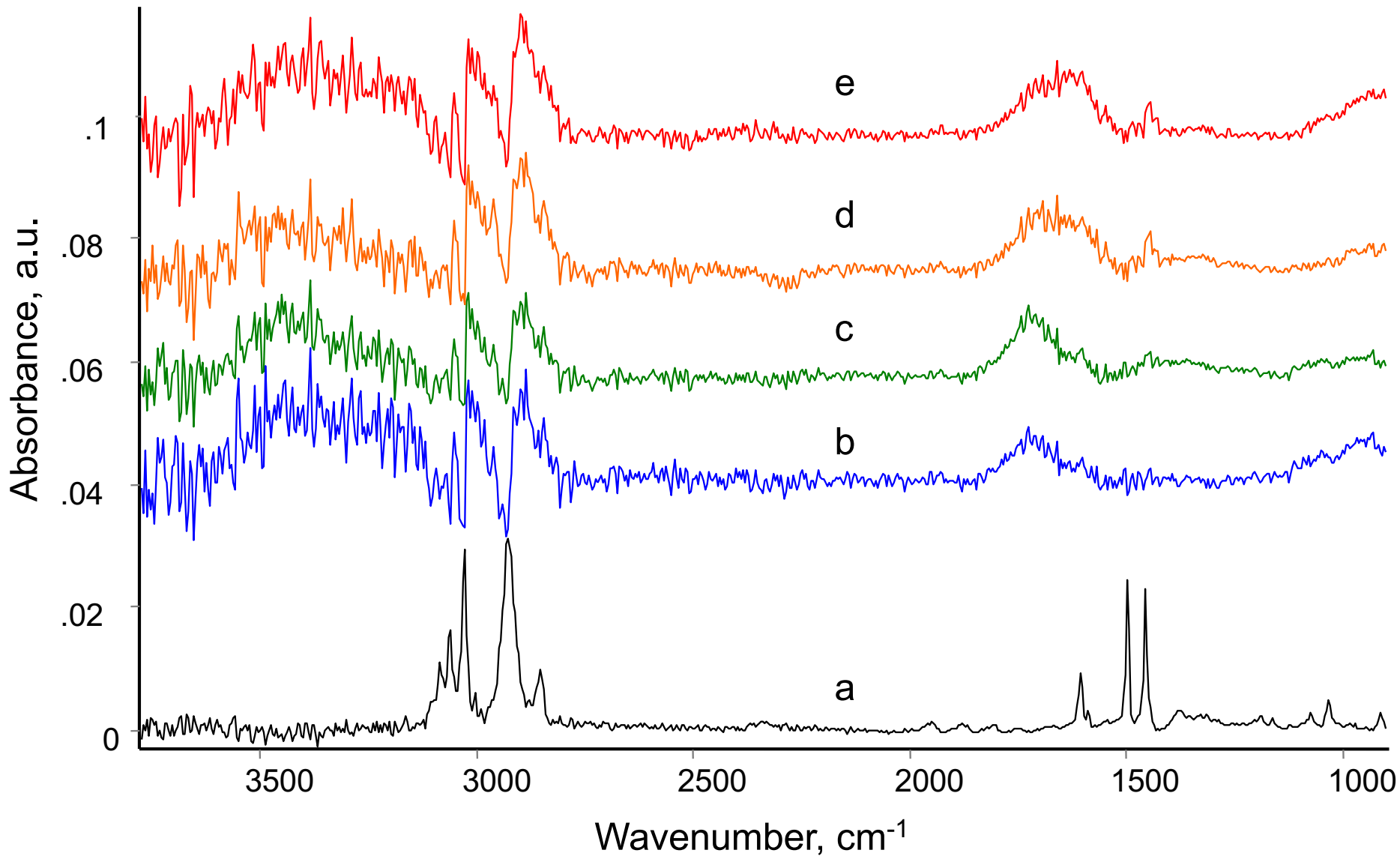

Fig. 30. FTIR transmission spectrum of untreated PS nanolayer with thickness of 580 nm (a) and differential spectra of PS nanolayer treated by ion-plasma treatment for (b) 40 seconds, (c) 80 seconds, (d) 800 seconds, (e) 1600 seconds. Spectra of silicon wafer are subtracted. Spectra are processed by baseline linearisation method and shifted along absorption axis for better observation. To obtain differential spectra (b-e) spectrum of untreated PS (a) is subtracted from corresponding spectra of treated samples and absorption scale is stretched for better observation.

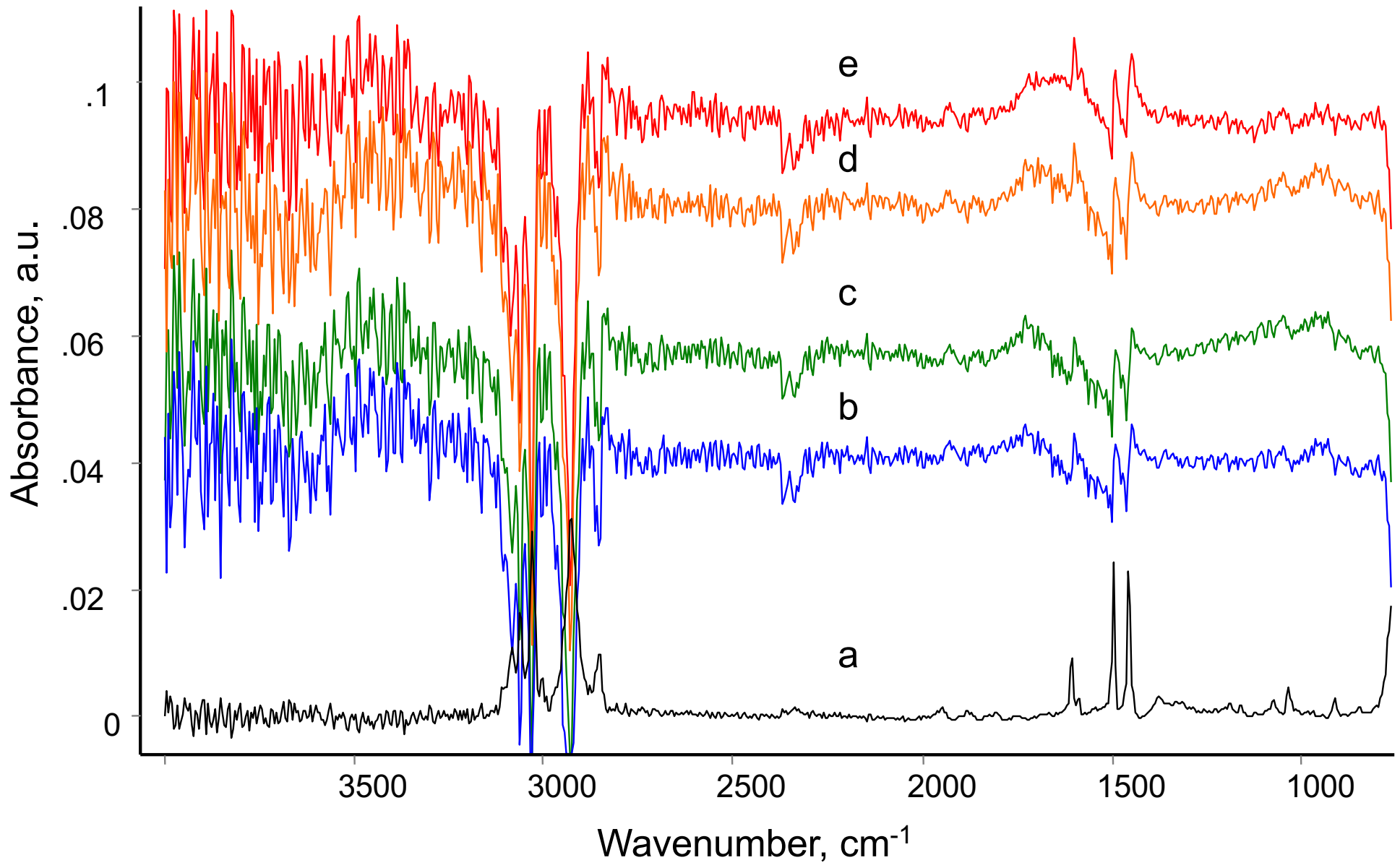

Fig. 31. FTIR transmission spectrum of untreated PS nanolayer with thickness of 930 nm (a) and differential spectra of PS nanolayer treated by ion-plasma treatment for (b) 40 seconds, (c) 80 seconds, (d) 800 seconds, (e) 1600 seconds. Spectra of silicon wafer are subtracted. Spectra are processed by baseline linearisation method and shifted along absorption axis for better observation. To obtain differential spectra (b-e) spectrum of untreated PS (a) is subtracted from corresponding spectra of treated samples and absorption scale is stretched for better observation.

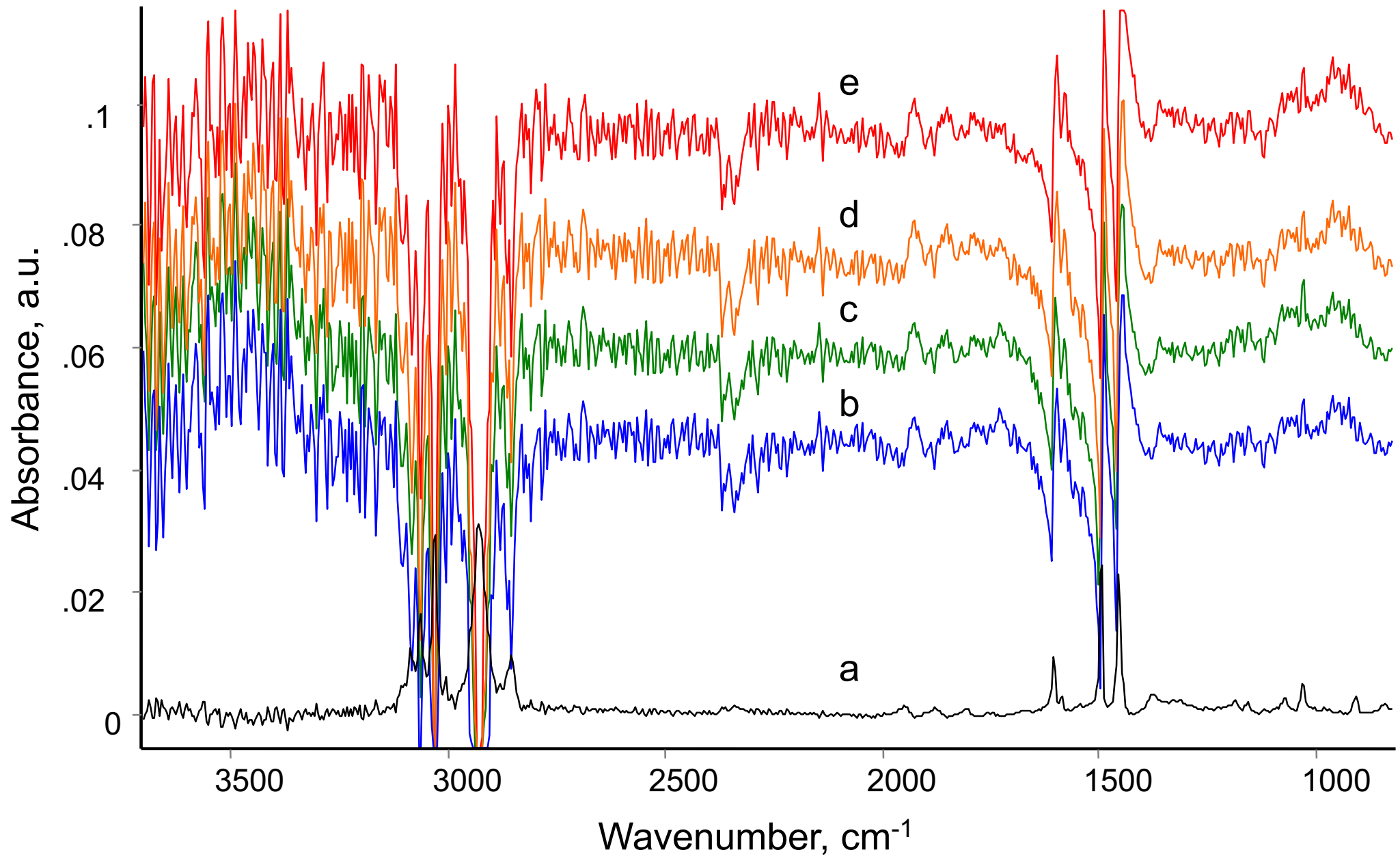

Fig. 32. FTIR transmission spectrum of untreated PS nanolayer with thickness of 1330 nm (a) and differential spectra of PS nanolayer treated by ion-plasma treatment for (b) 40 seconds, (c) 80 seconds, (d) 800 seconds, (e) 1600 seconds. Spectra of silicon wafer are subtracted. Spectra are processed by baseline linearisation method and shifted along absorption axis for better observation. To obtain differential spectra (b-e) spectrum of untreated PS (a) is subtracted from corresponding spectra of treated samples and absorption scale is stretched for better observation.

Differential spectra showed that the carbonized layer contains hydroxyl, carbonyl, and unsaturated carbon-carbon groups. The qualitatively carbonized layer shows the same tendency of changes for all samples of different thickness. Short time treatment is characterized by higher intensity in the region of oxygen-containing groups. With a long treatment time, the lines of unsaturated carbon-carbon groups appear more intensely.

Note that for spectra with a PS layer thickness of 170 nm or more, the subtraction of PS lines is a problem. In the differential spectrum, intensity spikes remain at the location of the most intense vibration lines of PS macromolecules at 3026, 2922, 1493, and 1452 $cm^{-1}$. This is especially evident in the spectra of samples with a nominal PS layer thickness of 930 and 1330 nm.

The obtained spectra were quantitatively analysed. For this purpose, the baseline method was used and the peak intensity of the corresponding line was measured. The results for the carbonyl group are shown in Fig. 33. For thin samples of 80 and 120 nm nominal thickness, the dependence of the line intensity on the processing time is practically not observed. For thicker samples, an increase in the intensity of the carbonyl line with the treatment time is observed, and then a decrease in the intensity with a long treatment time. However, the intensity of carbonyl line lays in the same region for samples of different thickness. Since the oxidation process is associated with the formation of vacancies of knocked-out atoms and the oxidation of the remaining macromolecules, therefore, the intensity of the carbonyl group line is associated with the thickness of the carbonized layer. The thickness of the carbonised layer is approximately the same for all samples, the similarity of the intensity of the carbonyl group for all samples of different initial thicknesses proves it.

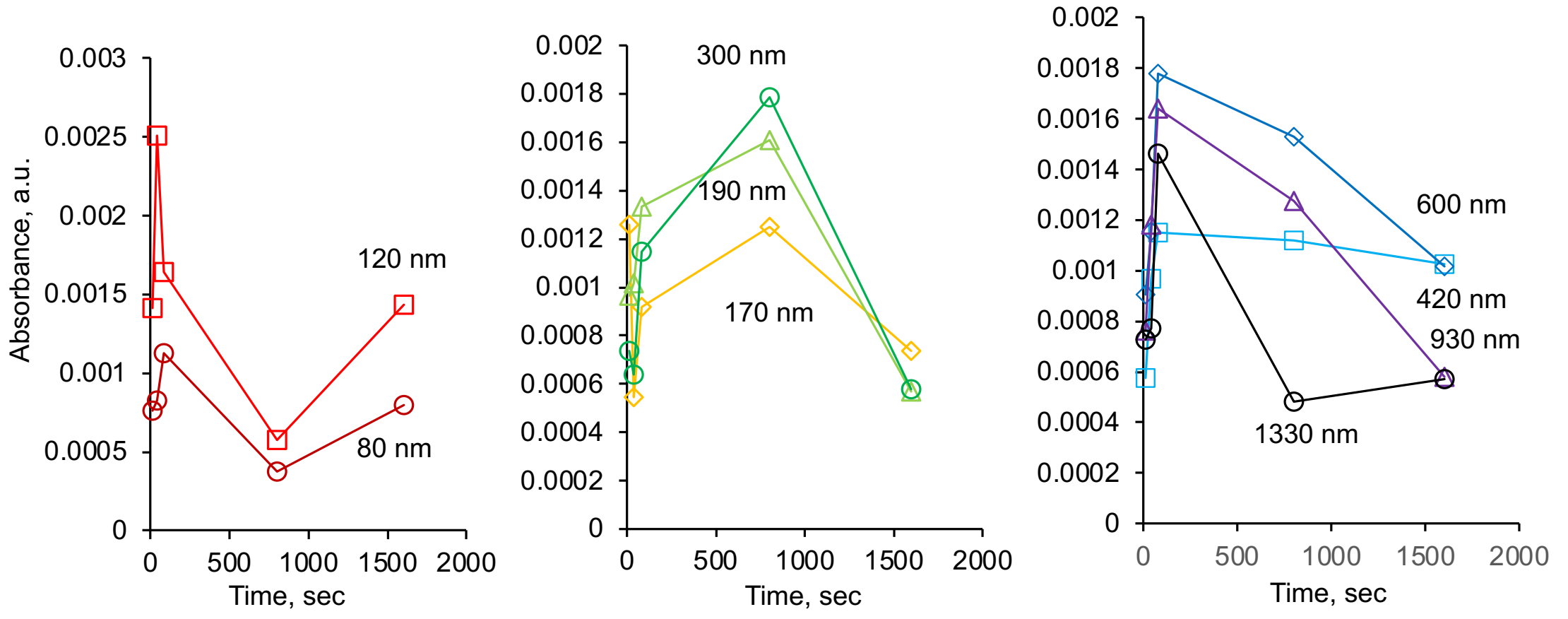

Fig. 33. Intensity of the carbonyl group line at 1700 $cm^{-1}$ in the spectrum of PS after ion beam treatment for samples of different thicknesses depending on the treatment time.

The analysis of PS macromolecules vibrations of was made using the aromatic ring line at 1493 $cm^{-1}$ and the methyl group line at 2922 $cm^{-1}$. The results are shown in Fig. 34.

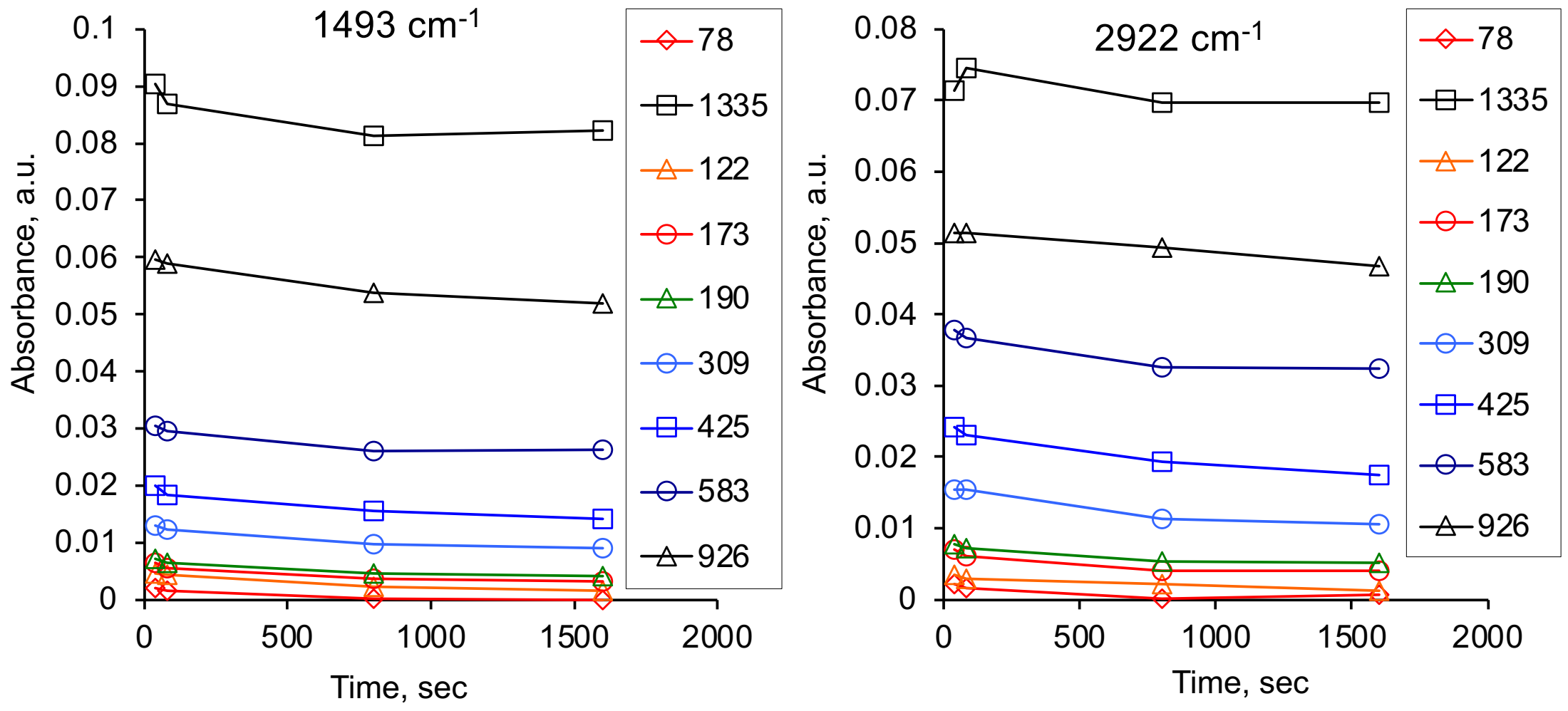

Fig.34. Intensity of aromatic ring vibrations line at 1493 $cm^{-1}$ and methyl group vibrations line at 2922 $cm^{-1}$ in the spectrum of PS after ion beam treatment for the samples of different thickness depending on the treatment time. Sample thickness is given in nm.

As can be seen from the graphs, the intensity of the aromatic ring and methyl group lines decreases with the treatment time. Moreover, the decrease continues even after the formation of the carbonized layer. However, the intensity of these lines is also proportional to the thickness of the PS layer and depends on the ratio of the wave vector length and the sample thickness.

To take these factors into account, the line intensity was normalized to the thickness of the unchanged underneath layer and to the ratio of the line intensities of the spectra of layers with a known thickness. As a standard, the theoretical line intensity was calculated based on the assumption that the line intensity is determined by the thickness of the unchanged underneath layer of PS, in which the concentration of absorbing groups does not change. The results of intensity normalization and theoretical dependences of intensities on sample thickness are presented in Figs. 35 and 36 for the ion beam treatment times of 40 and 1600 seconds.

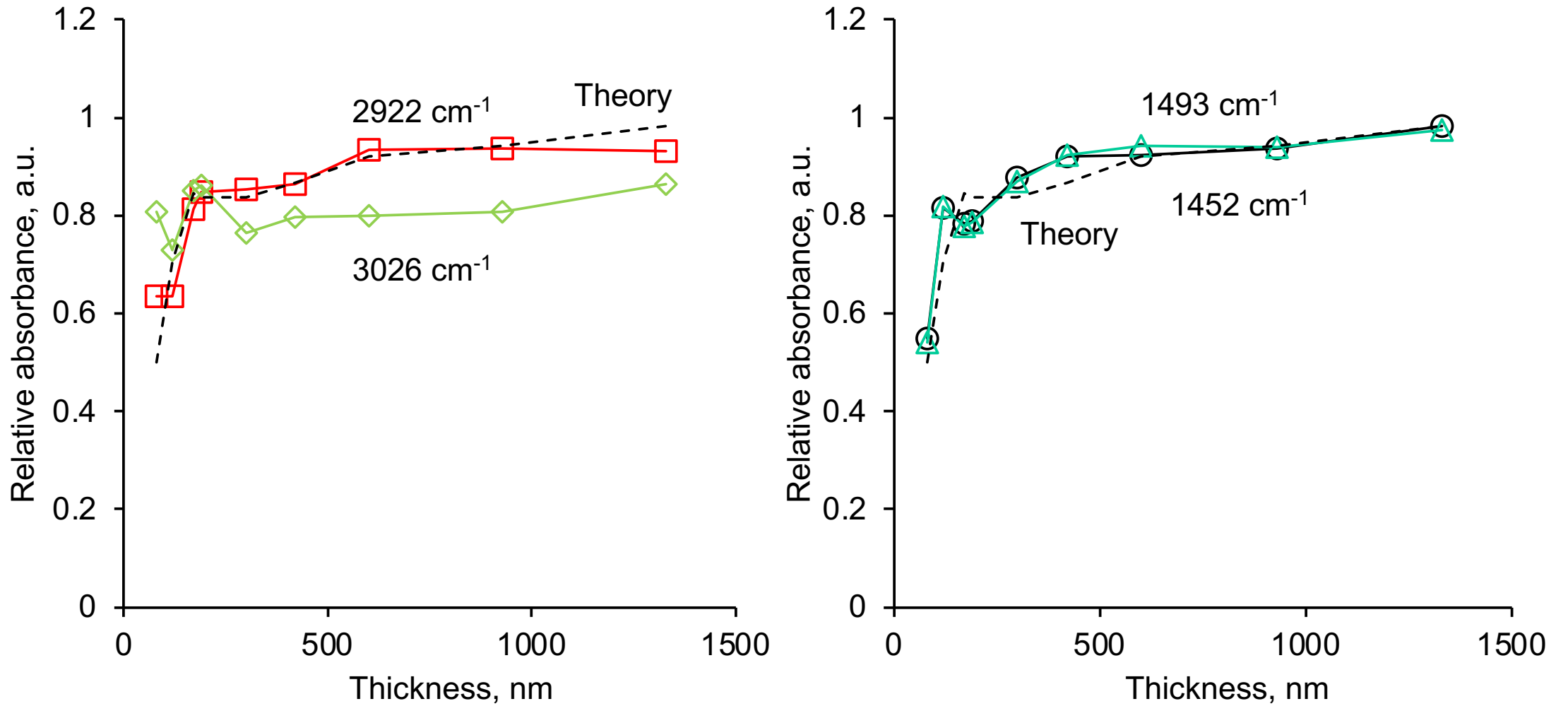

Fig.35. Normalized intensity of the aromatic ring and methyl group lines of PS treated with 40 seconds of ion beam depending on the initial thickness of the samples. The theoretical curve is shown by the dotted line.

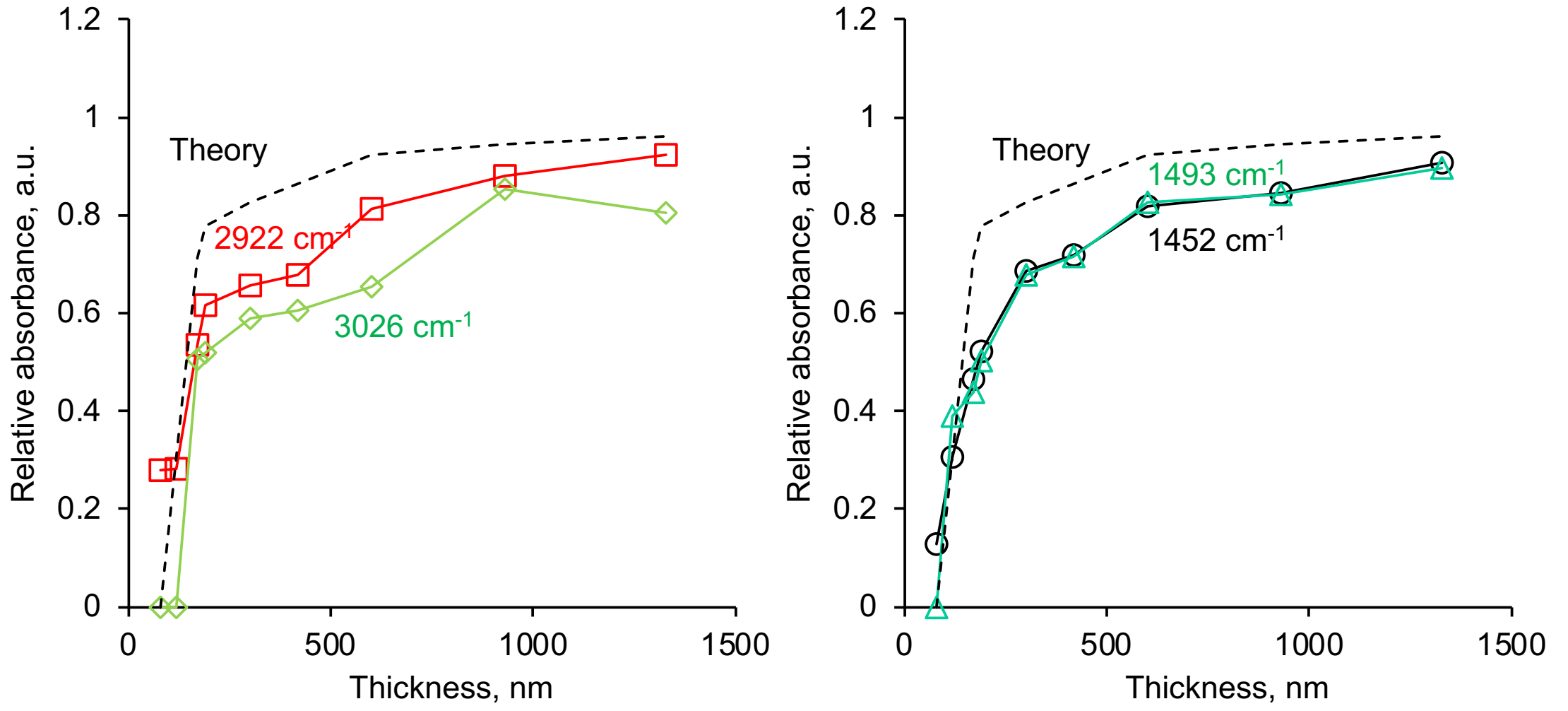

Fig.36. Normalized intensity of the aromatic ring and methyl group lines of PS treated with 1600 seconds of ion beam depending on the initial thickness of the samples. The theoretical curve is shown by the dotted line.

The graphs show that the intensity curves of the aromatic ring and methyl group vibration lines almost completely coincide with the theoretical curve for 40 seconds treatment time. This means that the unchanged underneath PS layer does not change as a result of ion beam treatment at this short treatment times. In contrast to these results, the intensity curves of these same lines lie below the theoretical curve at 1600 seconds treatment time. That is, during long treatment time, the underneath PS layer loses some aromatic rings and methyl groups. This agrees with the ellipsometry results on the change in the underneath layer under the carbonized PS layer.

### 3.3. Films after ion beam treatment and then washed in solvent

To analyze the underlying PS layer, the samples were washed in toluene after the ion beam treatment. After washing and drying, the ellipsometry and FTIR spectra of the samples were recorded again. Fig. 37 shows the thickness of the remaining part of the PS on the silicon wafer after washing. These samples were treated only in plasma without high voltage. The results (a) correspond to the all samples of 120 nm treated different time in the plasma. The results (b) correspond to the samples of different thickness treated 1600 seconds in the plasma.

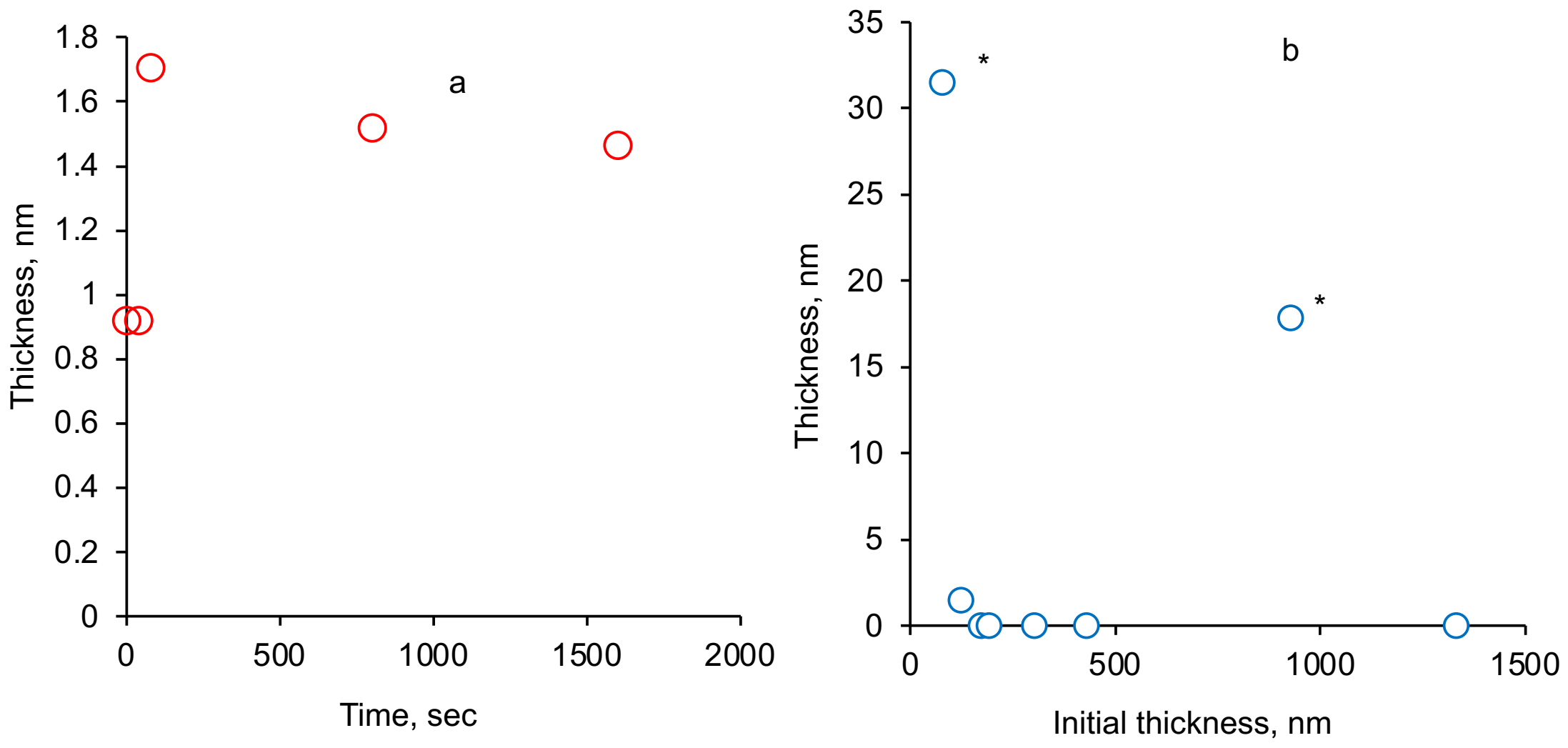

Fig. 37. Thickness of the insoluble part of PS (gel-fraction) on a silicon wafer according to ellipsometry data: (a) dependence on the plasma treatment time without applying high voltage to the electrode, initial thickness of the PS was 120 nm; (b) dependence on the initial thickness of the samples, plasma treatment time was 1600 seconds.

Untreated PS is washed off almost completely. Some amount of about 0.9 nm remains after washing off the 120 nm PS film, which may be an insoluble contaminant in the original PS or solvent. After plasma treatment, some amount of PS within 1.5-1.7 nm range remains. These may be macromolecules on the PS surface cross-linked under the low energy particles from plasma. When thick films are treated, in two cases the thickness of the residual layer was 32 and 18 nm, which may be due to accidental exposure during sample storage or the presence of contaminants. But a rest of the samples showed a practical absence of an insoluble part in PS treated with plasma for 1600 seconds. The ellipsometry spectra themselves are not shown, since they are uninformative in this case.

Ellipsometry spectra are shown for washed samples after ion beam treatment in Fig. 38-41. The curves of psi-functions of samples with thickness of 78 and 120 nm completely repeat the curves for unwashed samples. The coincidence of experimental and theoretical curves is good for the sample with thickness of 78 nm. For the sample with thickness of 120 nm, a noticeable deviation of experimental and theoretical curves is observed for samples treated for 40 and 80 seconds. But for samples treated for 800 and 1600 seconds, the coincidence of theoretical and experimental curves is good.

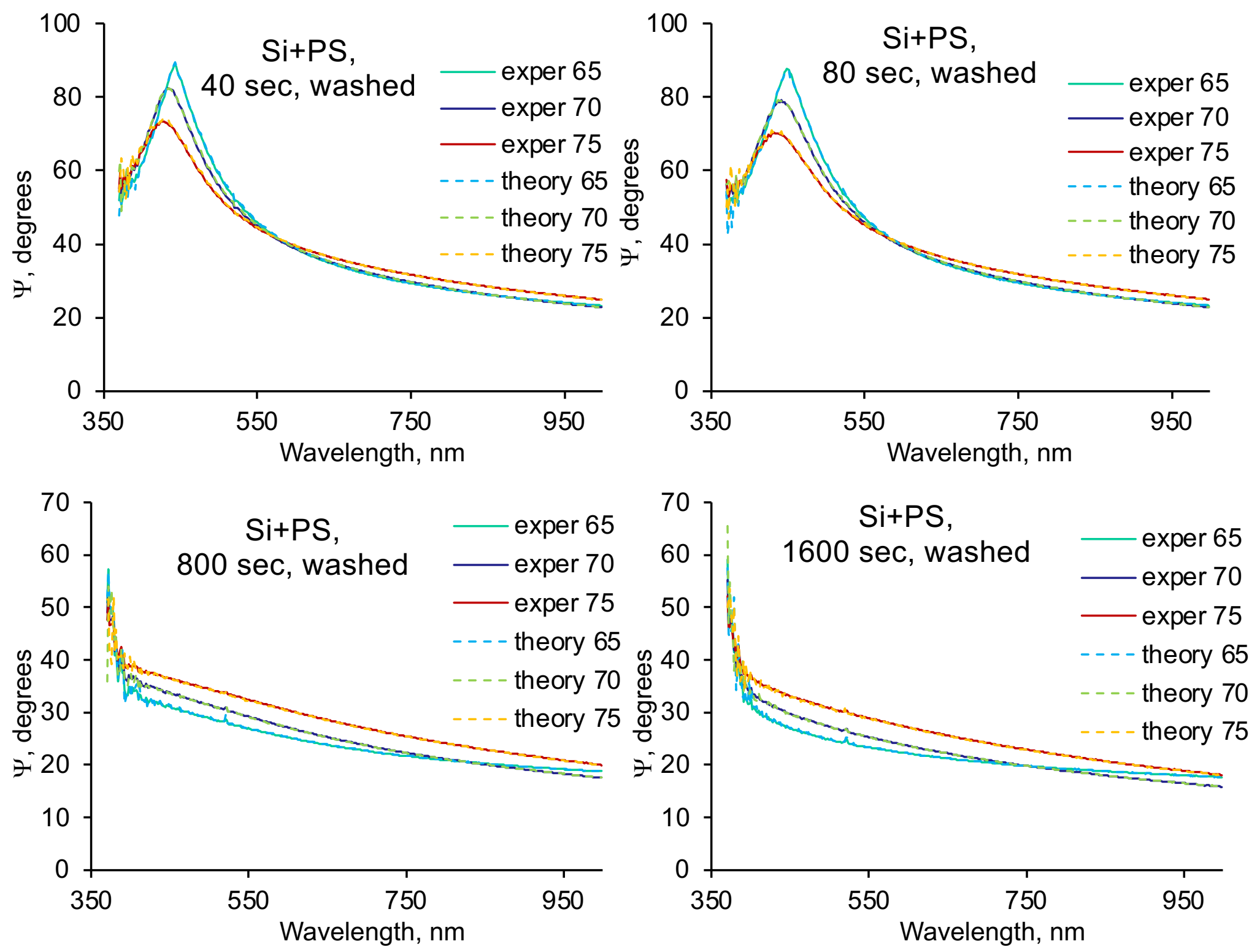

Fig.38. Ellipsometric Ψ function of a silicon wafer coated with polystyrene of 78 nm thickness and treated by nitrogen ions of 20 keV energy during 40, 80, 800 и 1600 seconds, and then washed in toluene. Solid lines are experimental. Dashed lines are theoretical calculations. Beam incidence angles are given in degrees.

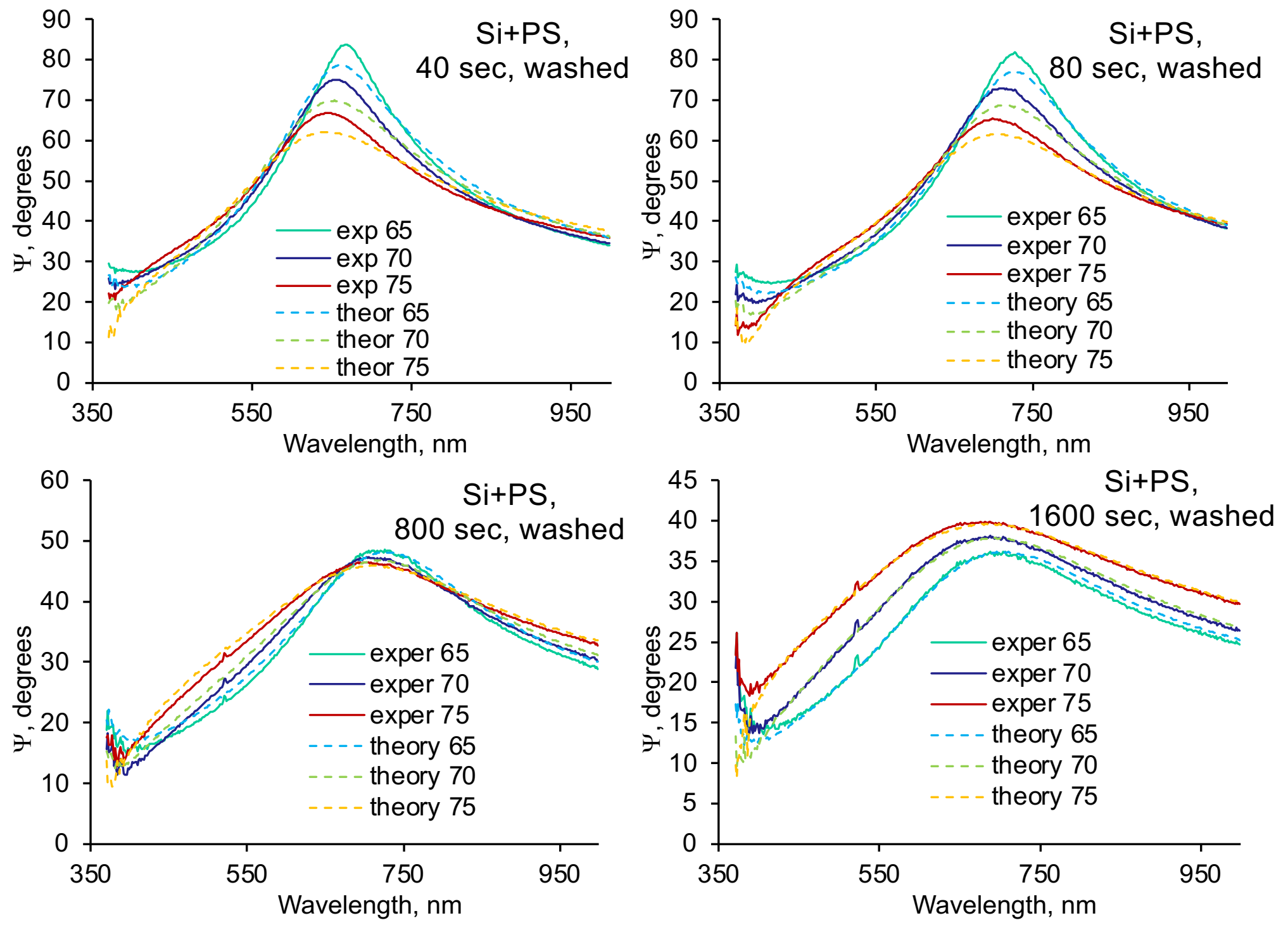

Fig.39. Ellipsometric Ψ function of a silicon wafer coated with polystyrene of 120 nm thickness and treated by nitrogen ions of 20 keV energy during 40, 80, 800 и 1600 seconds, and then washed in toluene. Solid lines are experimental. Dashed lines are theoretical calculations. Beam incidence angles are given in degrees.

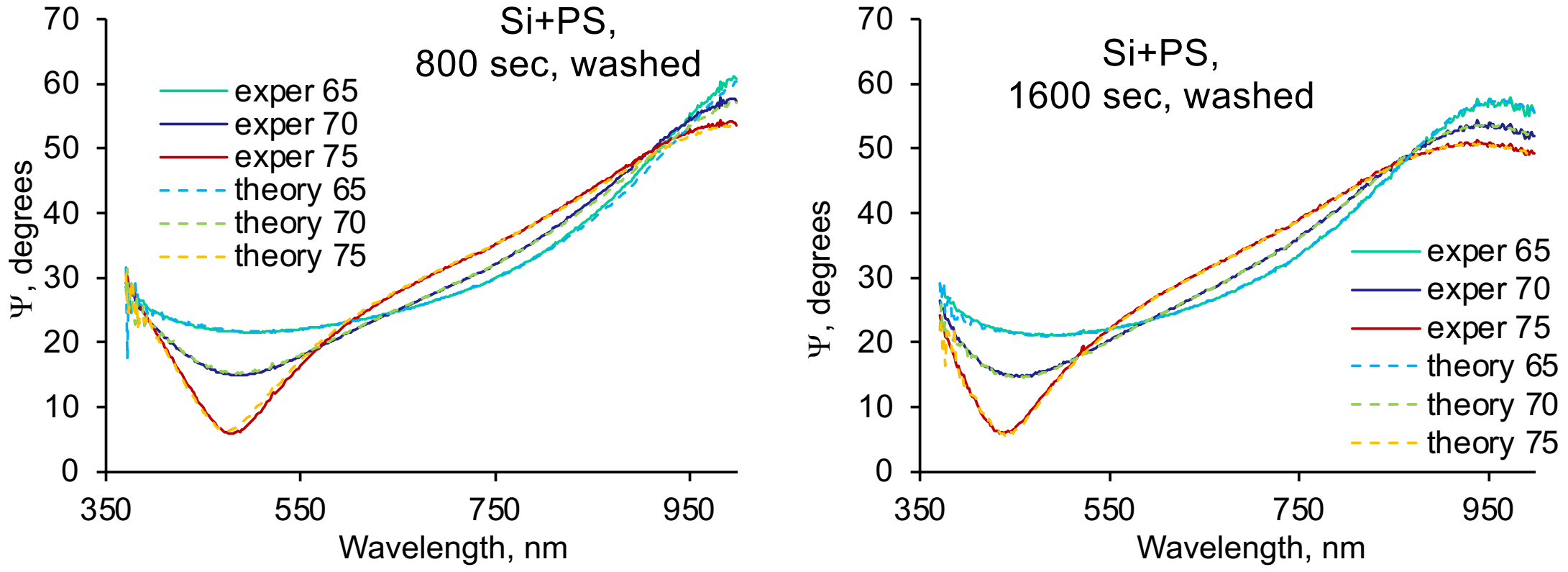

Fig.40. Ellipsometric Ψ function of a silicon wafer coated with polystyrene of 170 nm thickness and treated by nitrogen ions of 20 keV energy during 800 и 1600 seconds, and then washed in toluene. Solid lines are experimental. Dashed lines are theoretical calculations. Beam incidence angles are given in degrees.

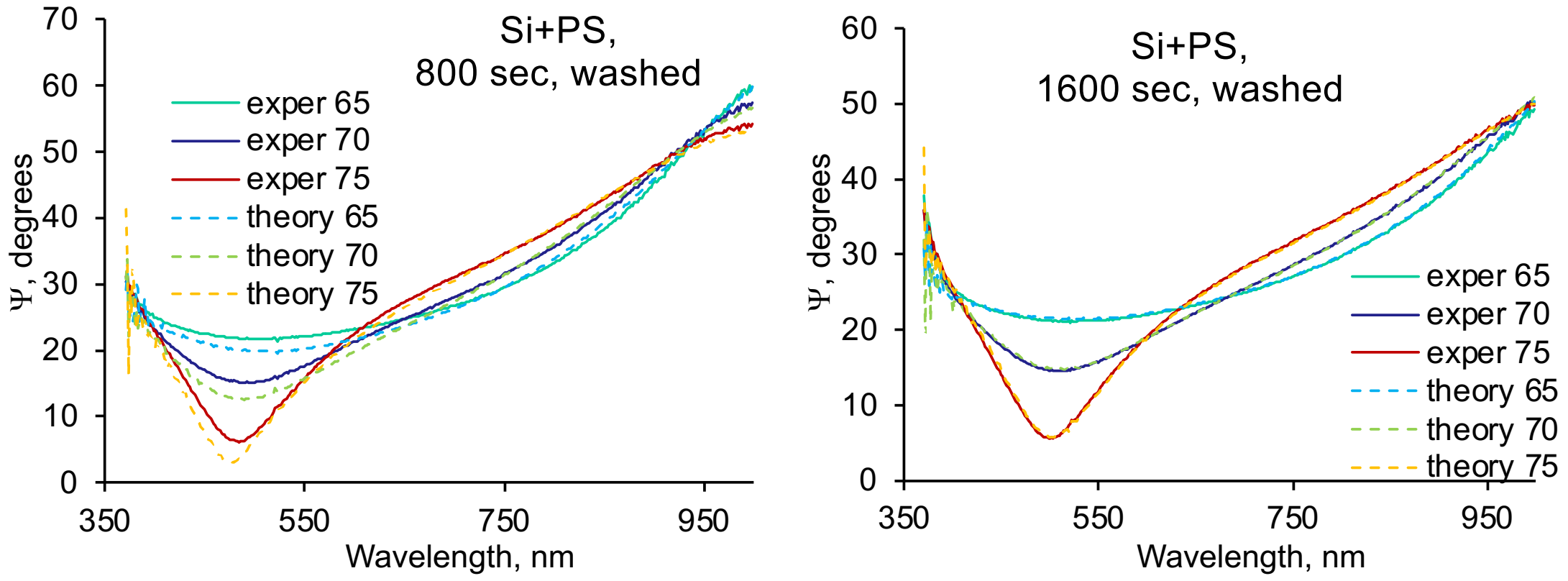

Fig.41. Ellipsometric Ψ function of a silicon wafer coated with polystyrene of 190 nm thickness and treated by nitrogen ions of 20 keV energy during 800 и 1600 seconds, and then washed in toluene. Solid lines are experimental. Dashed lines are theoretical calculations. Beam incidence angles are given in degrees.

For samples with thickness of 170 and 190 nm treated for 40 and 80 seconds, it was not possible to obtain a sufficiently stable spectrum. The beam focus on the detector slit is blurred. Observation of samples under a microscope shows that the surface is strongly deformed. The ellipsometer beam is scattered on the sample surface and cannot be focused on the detector. For the treatment times of 800 and 1600 seconds, the ellipsometric spectra are quite good and are approximated by theoretical curves with sufficient accuracy.

Thicker samples could not be measured by the ellipsometry method. Their surface after washing became folded. Such non-uniform surface of the samples does not allow obtaining the ellipsometric spectra. Therefore, the analysis of the PS thickness after washing was made only for 4 thicknesses of the samples. The results of the analysis are shown in Fig. 42.

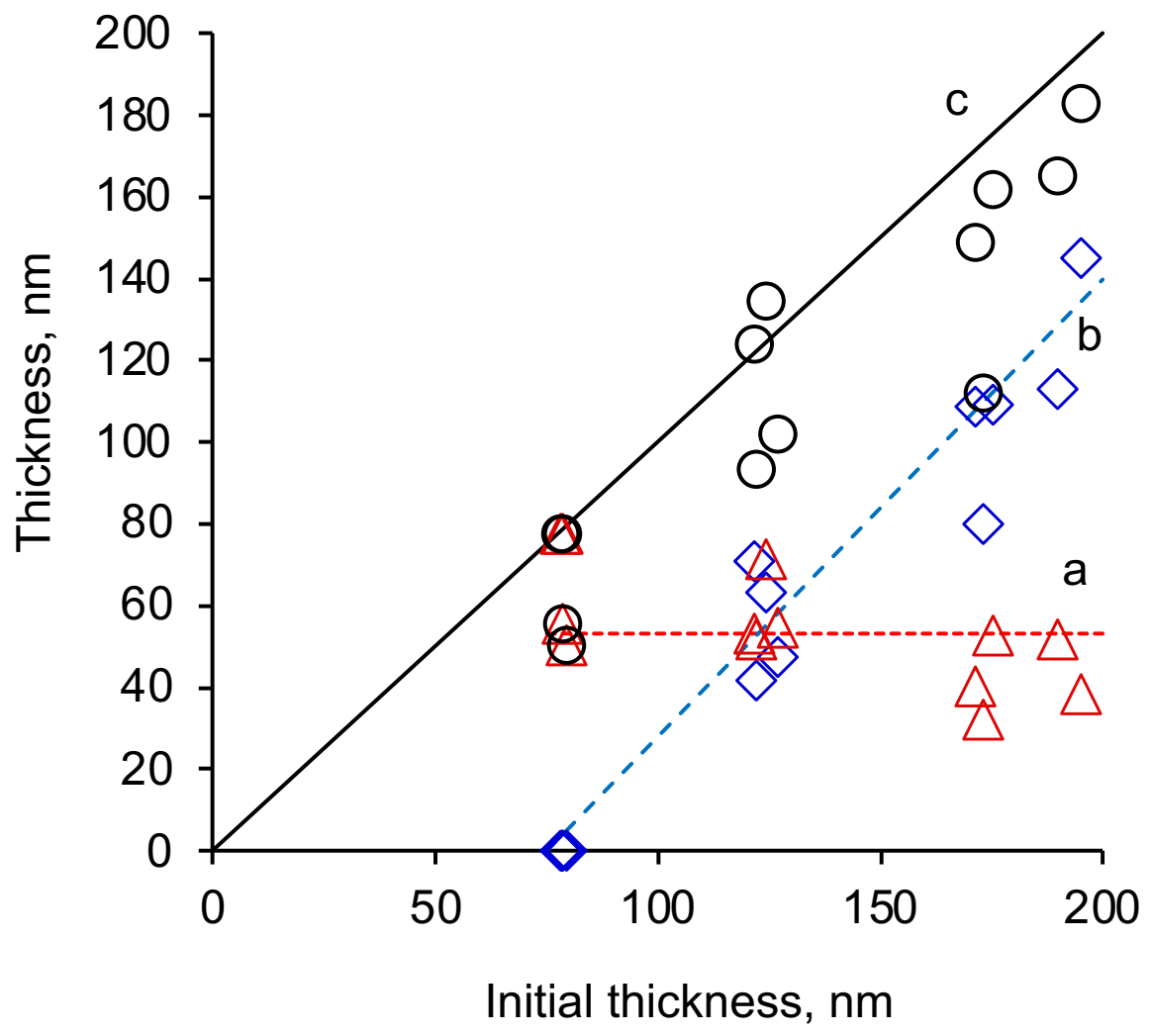

Fig.42. Dependence of the thickness of the insoluble fraction of PS after ion beam treatment according to ellipsometry data on the initial thickness of PS films. Red triangles and the red dashed line show the thickness of the carbonized layer, blue diamonds and the blue dashed line show the thickness of the underlying layer, black circles show the total thickness of the film. The straight line shows the bisector. The points are for all treatment times.

The results of the insoluble fraction measurements show that the carbonized and underlying PS layers are completely crosslinked in all the samples studied. This means that already with a 40 second treatment, the PS layer up to 190 nm is completely crosslinked, despite the fact that nitrogen ions do not penetrate to such a depth in PS. Longer treatment times also lead to crosslinking of all PS layers.

To analyze the insoluble part (gel fraction) of PS, FTIR spectra of PS samples were recorded after plasma treatment and ion beam treatment and then washing in toluene. Therefore, all these samples were all washed from the soluble part of PS. Spectra of samples of different thickness after plasma treatment without high voltage are shown in Fig. 43. The spectra show that the main part of PS is washed and does not remain on the silicon wafer. The intensities of the lines of the remaining part of PS are at the noise level. These results correspond to the data obtained using ellipsometry.

The FTIR spectra of PS samples of different thicknesses treated with ion beam for different times and then washed from the soluble sol fraction in toluene are shown in Figs. 44–52.

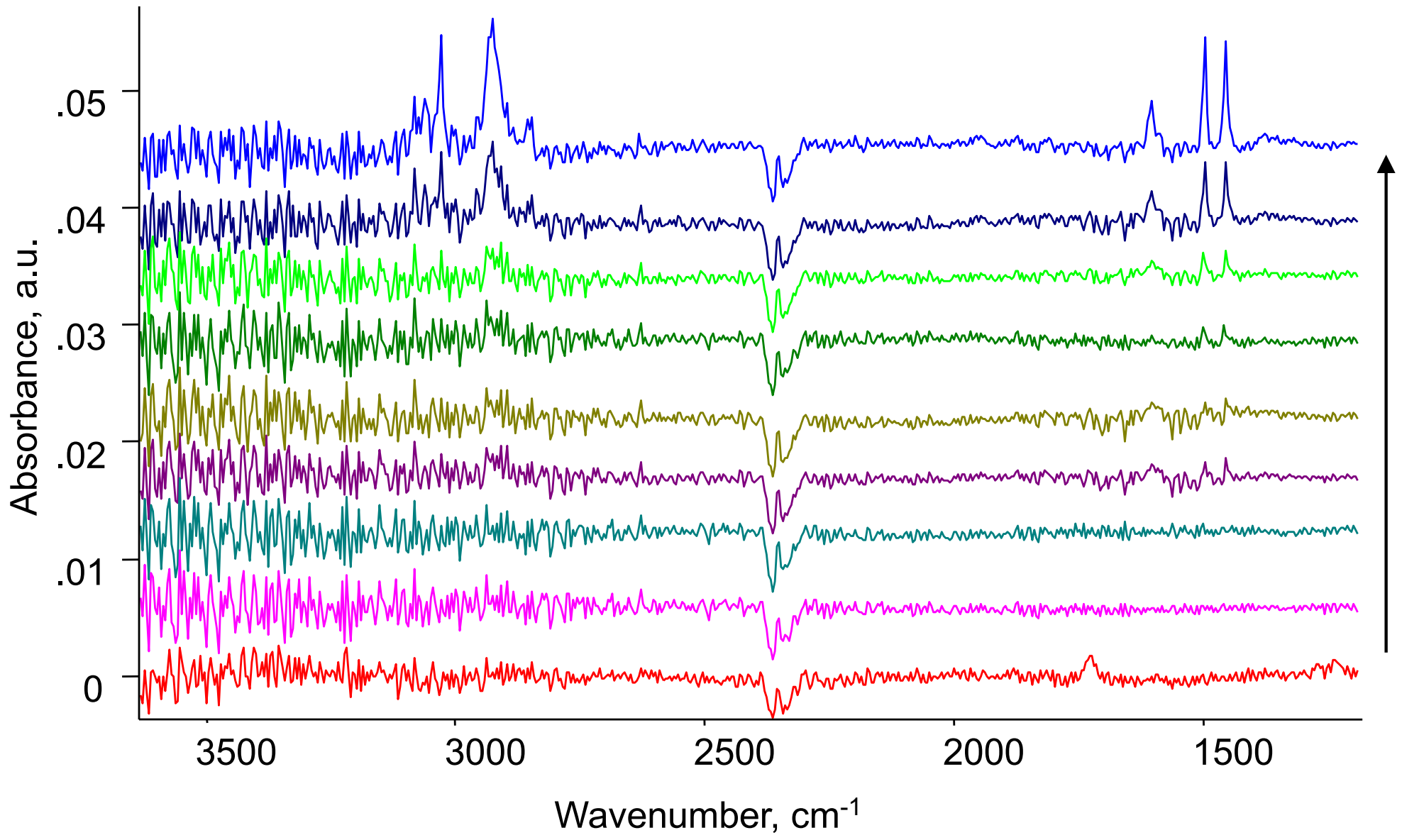

Fig.43. FTIR transmission spectra of PS nanolayers of different thicknesses treated with plasma for 1600 seconds and then washed in toluene. The order of spectra (by arrow) is as follows: 78 nm, 123 nm, 172 nm, 191 nm, 304 nm, 430 nm, 594 nm, 928 nm, 1331 nm. Spectra of silicon wafer are subtracted. Spectra are treated by baseline linearisation and shifted along the absorption axis for better observation.

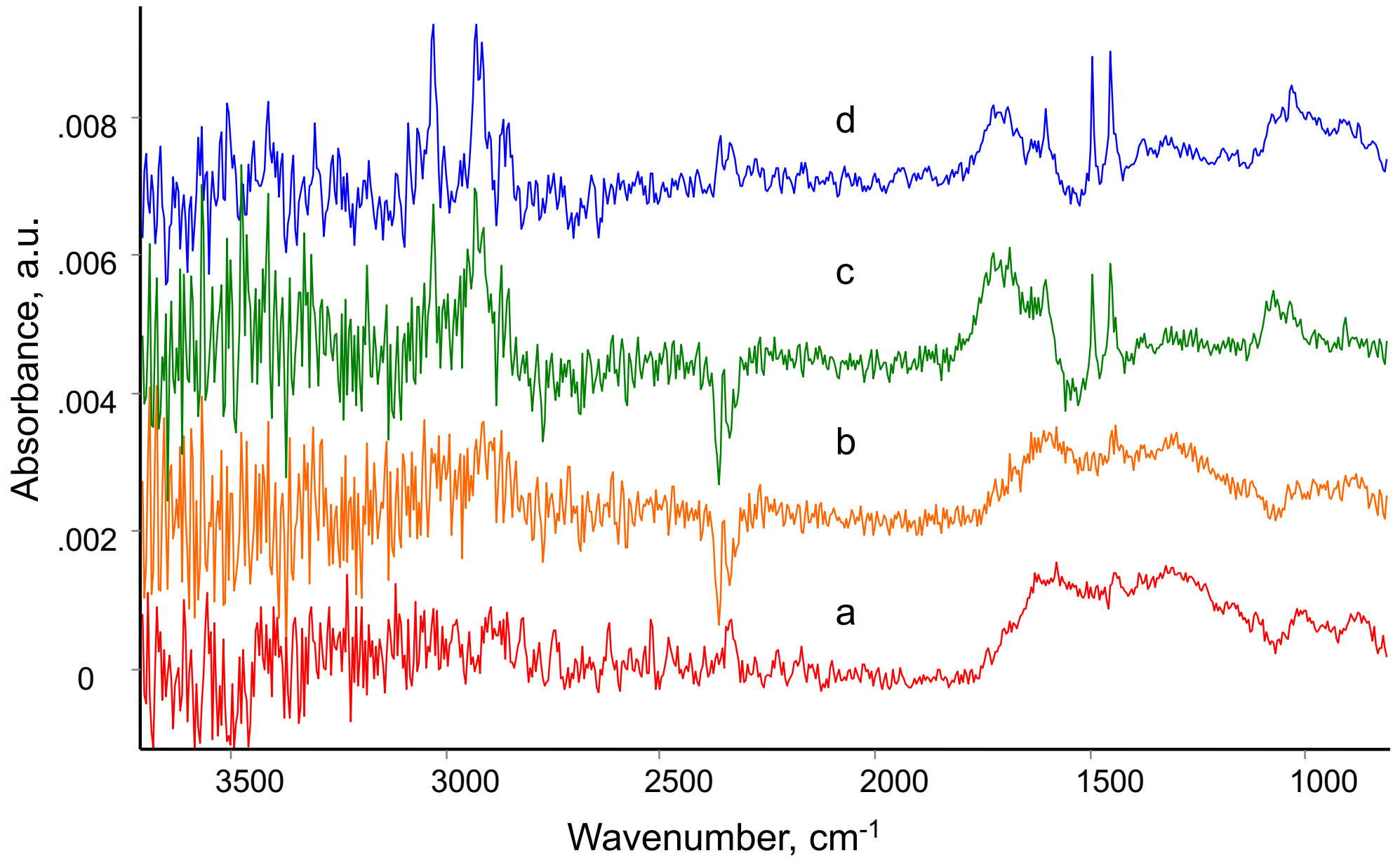

Fig. 44. FTIR transmission spectrum of PS nanolayer with thickness of 78 nm treated by ion-plasma treatment and washed in toluene: (a) 40 seconds, (b) 80 seconds, (c) 800 seconds, (d) 1600 seconds. Spectra of silicon wafer are subtracted. Spectra are processed by baseline linearisation method and shifted along absorption axis for better observation.

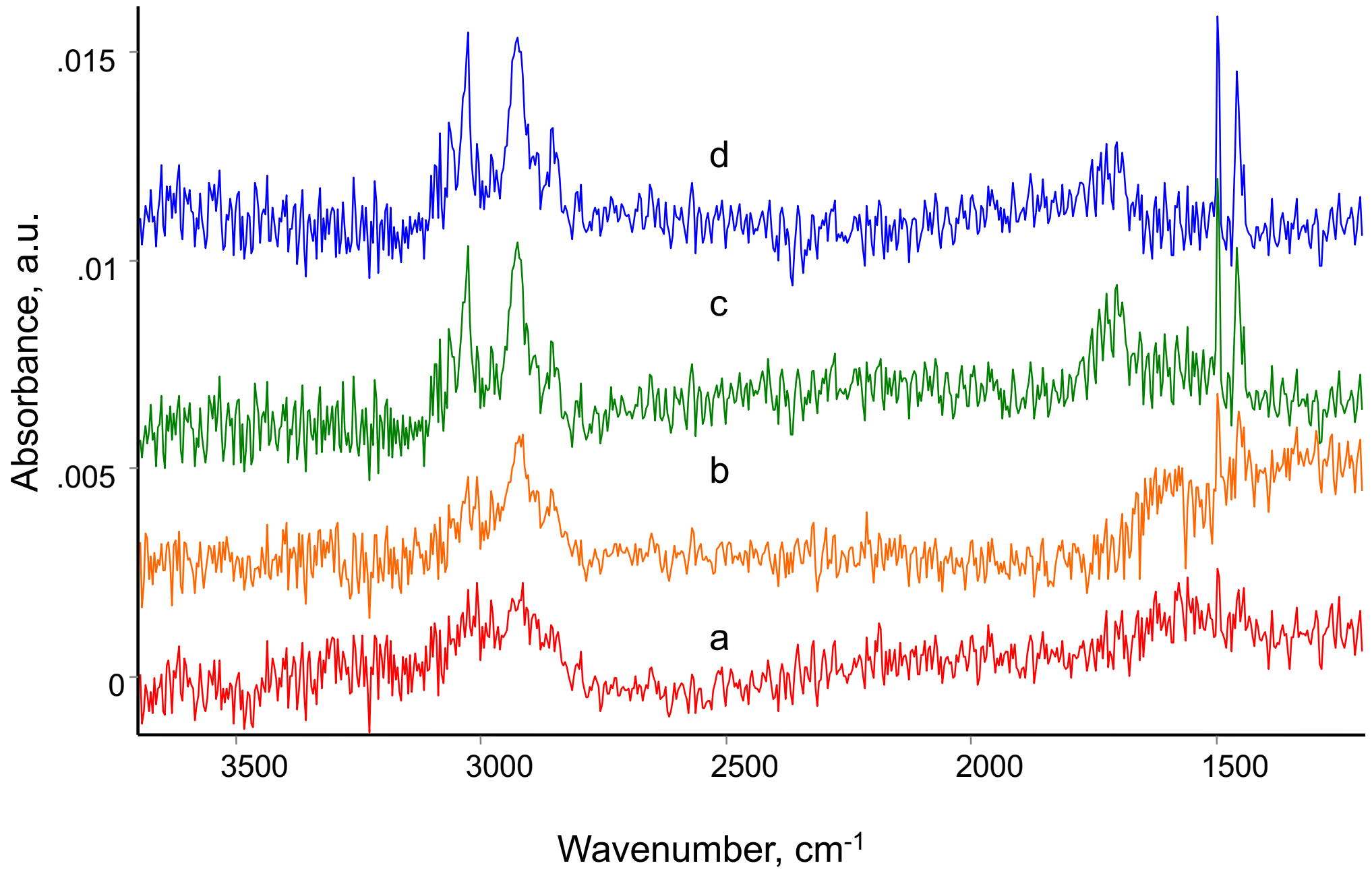

Fig. 45. FTIR transmission spectrum of PS nanolayer with thickness of 122 nm treated by ion-plasma treatment and washed in toluene: (a) 40 seconds, (b) 80 seconds, (c) 800 seconds, (d) 1600 seconds. Spectra of silicon wafer are subtracted. Spectra are processed by baseline linearisation method and shifted along absorption axis for better observation.

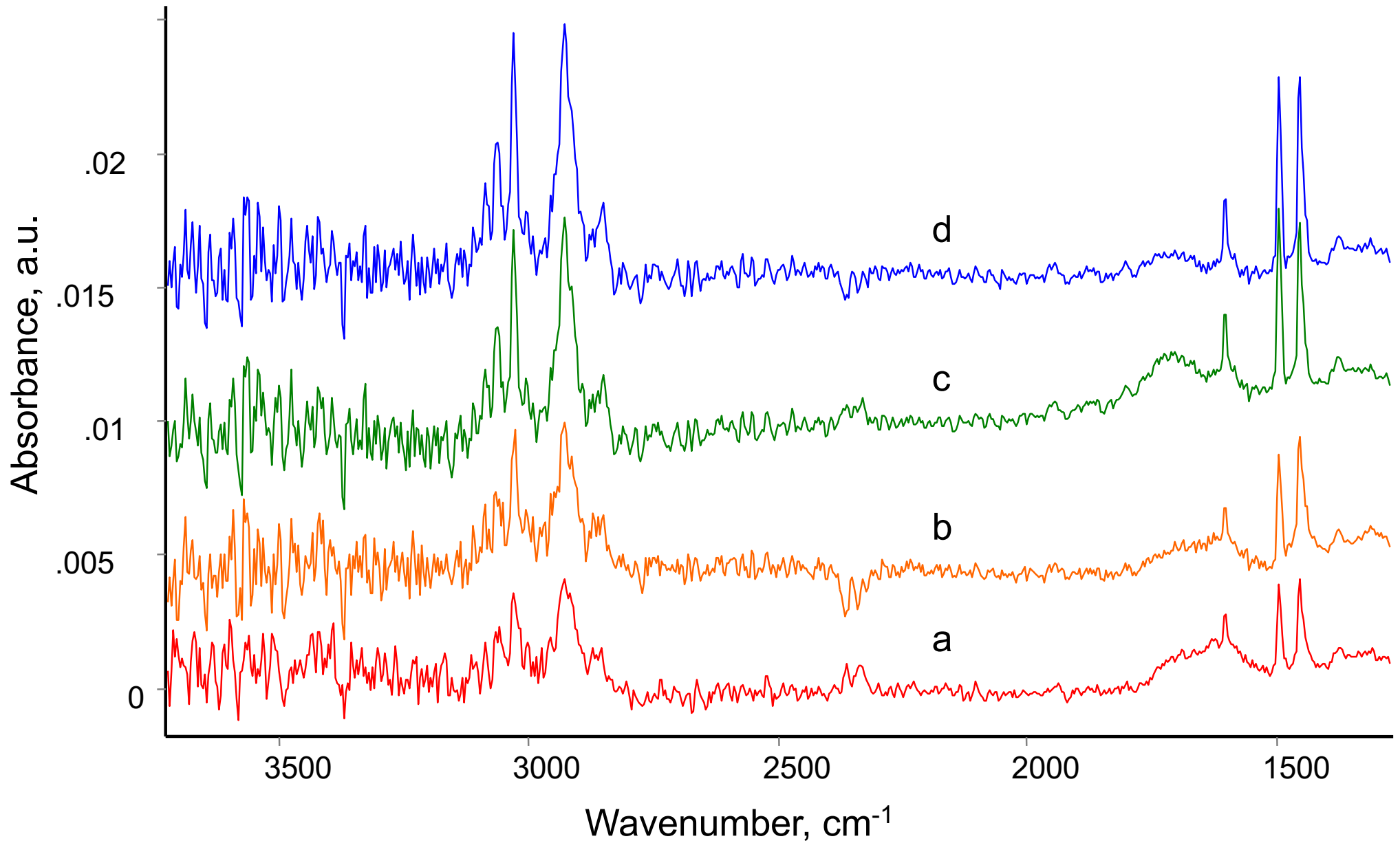

Fig. 46. FTIR transmission spectrum of PS nanolayer with thickness of 173 nm treated by ion-plasma treatment and washed in toluene: (a) 40 seconds, (b) 80 seconds, (c) 800 seconds, (d) 1600 seconds. Spectra of silicon wafer are subtracted. Spectra are processed by baseline linearisation method and shifted along absorption axis for better observation.

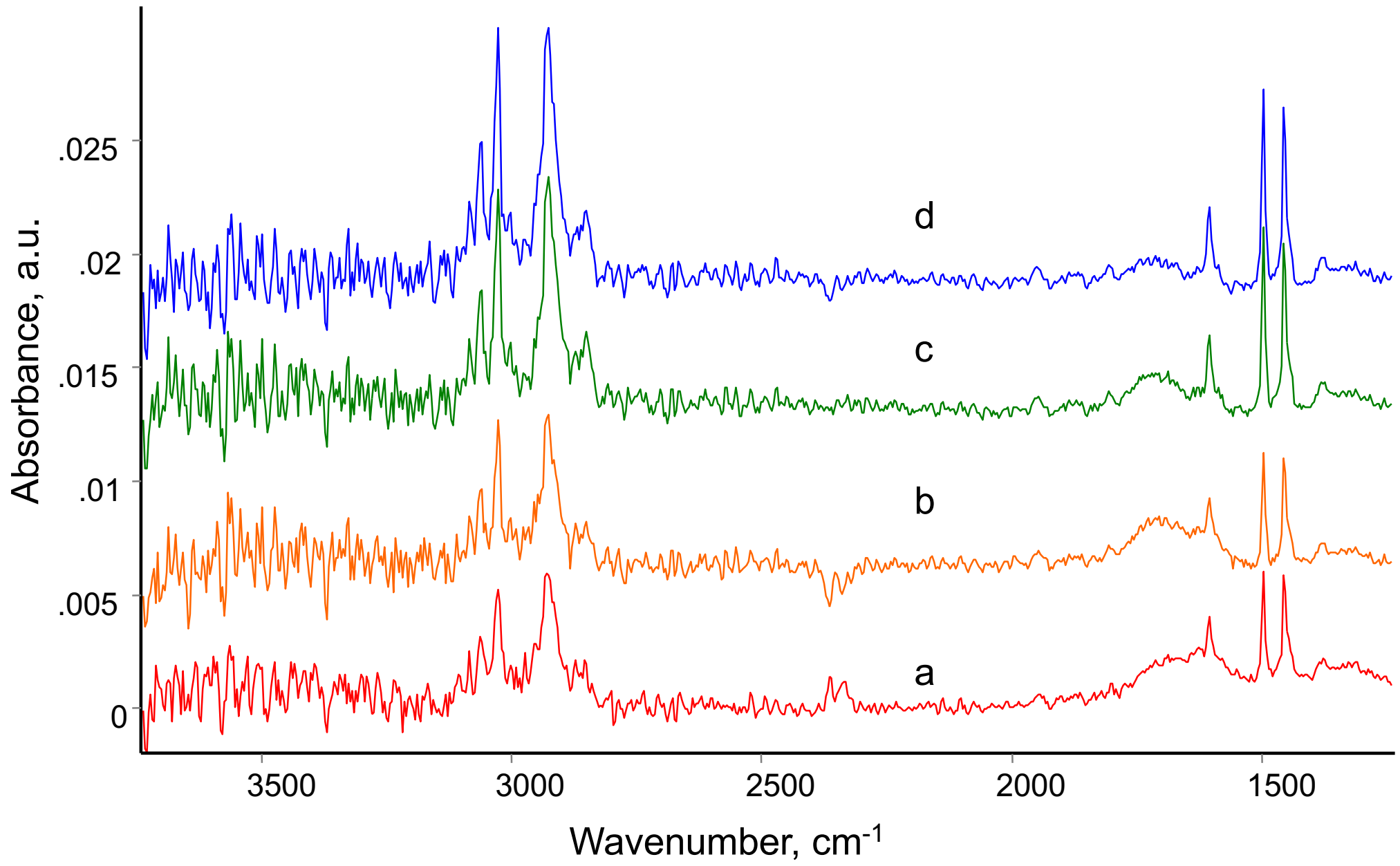

Fig. 47. FTIR transmission spectrum of PS nanolayer with thickness of 190 nm treated by ion-plasma treatment and washed in toluene: (a) 40 seconds, (b) 80 seconds, (c) 800 seconds, (d) 1600 seconds. Spectra of silicon wafer are subtracted. Spectra are processed by baseline linearisation method and shifted along absorption axis for better observation.

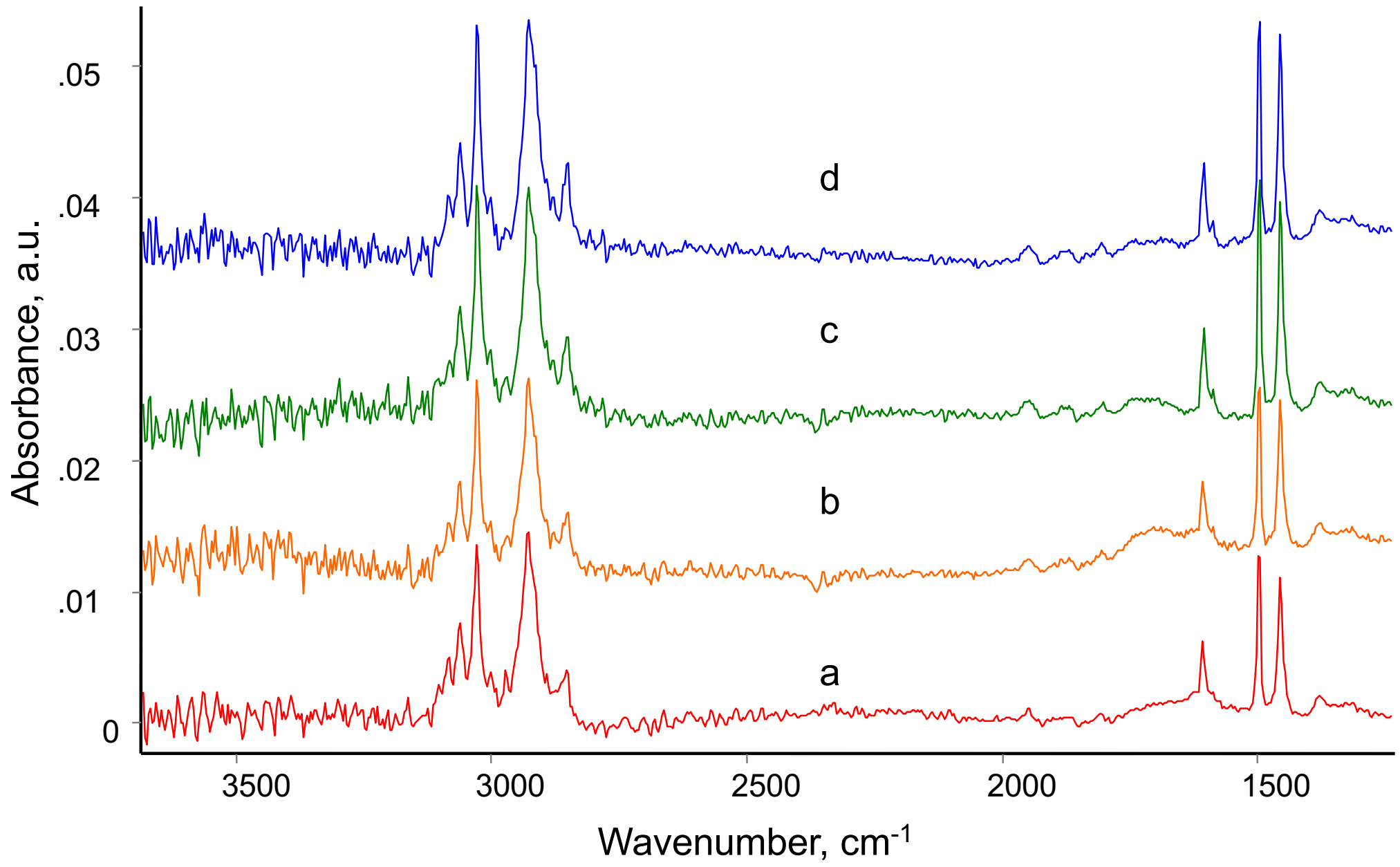

Fig. 48. FTIR transmission spectrum of PS nanolayer with thickness of 310 nm treated by ion-plasma treatment and washed in toluene: (a) 40 seconds, (b) 80 seconds, (c) 800 seconds, (d) 1600 seconds. Spectra of silicon wafer are subtracted. Spectra are processed by baseline linearisation method and shifted along absorption axis for better observation.

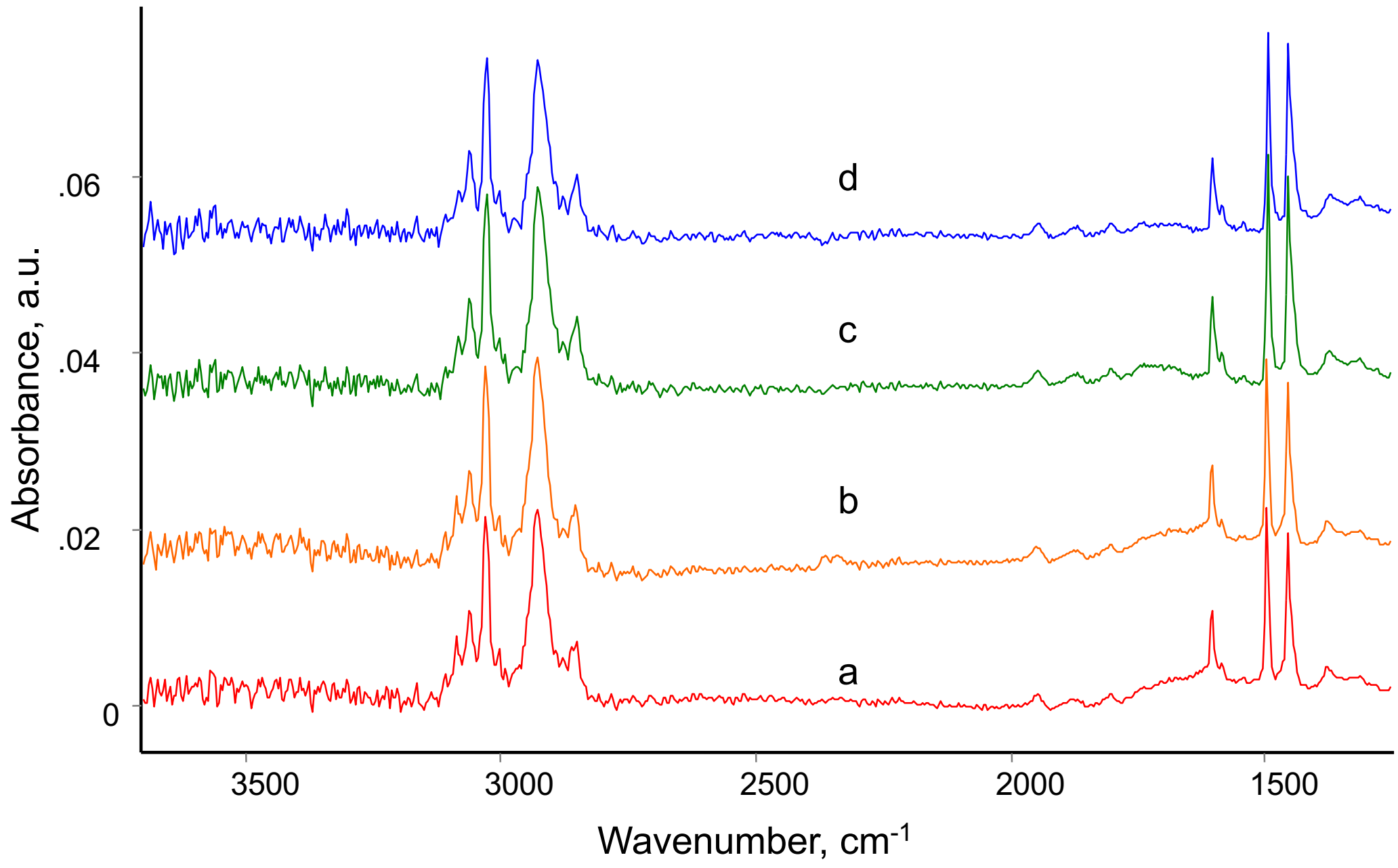

Fig. 49. FTIR transmission spectrum of PS nanolayer with thickness of 425 nm treated by ion-plasma treatment and washed in toluene: (a) 40 seconds, (b) 80 seconds, (c) 800 seconds, (d) 1600 seconds. Spectra of silicon wafer are subtracted. Spectra are processed by baseline linearisation method and shifted along absorption axis for better observation.

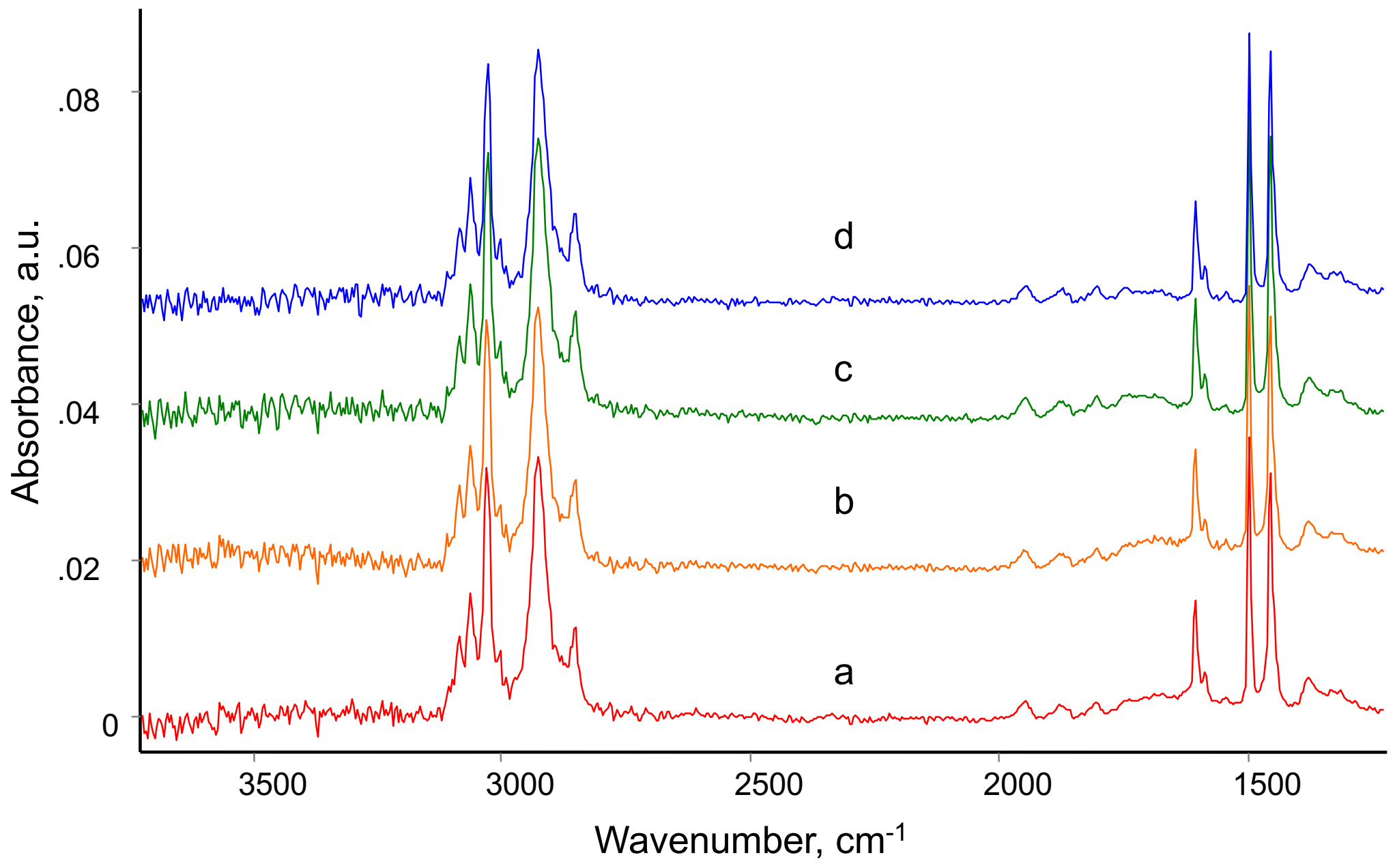

Fig. 50. FTIR transmission spectrum of PS nanolayer with thickness of 582 nm treated by ion-plasma treatment and washed in toluene: (a) 40 seconds, (b) 80 seconds, (c) 800 seconds, (d) 1600 seconds. Spectra of silicon wafer are subtracted. Spectra are processed by baseline linearisation method and shifted along absorption axis for better observation.

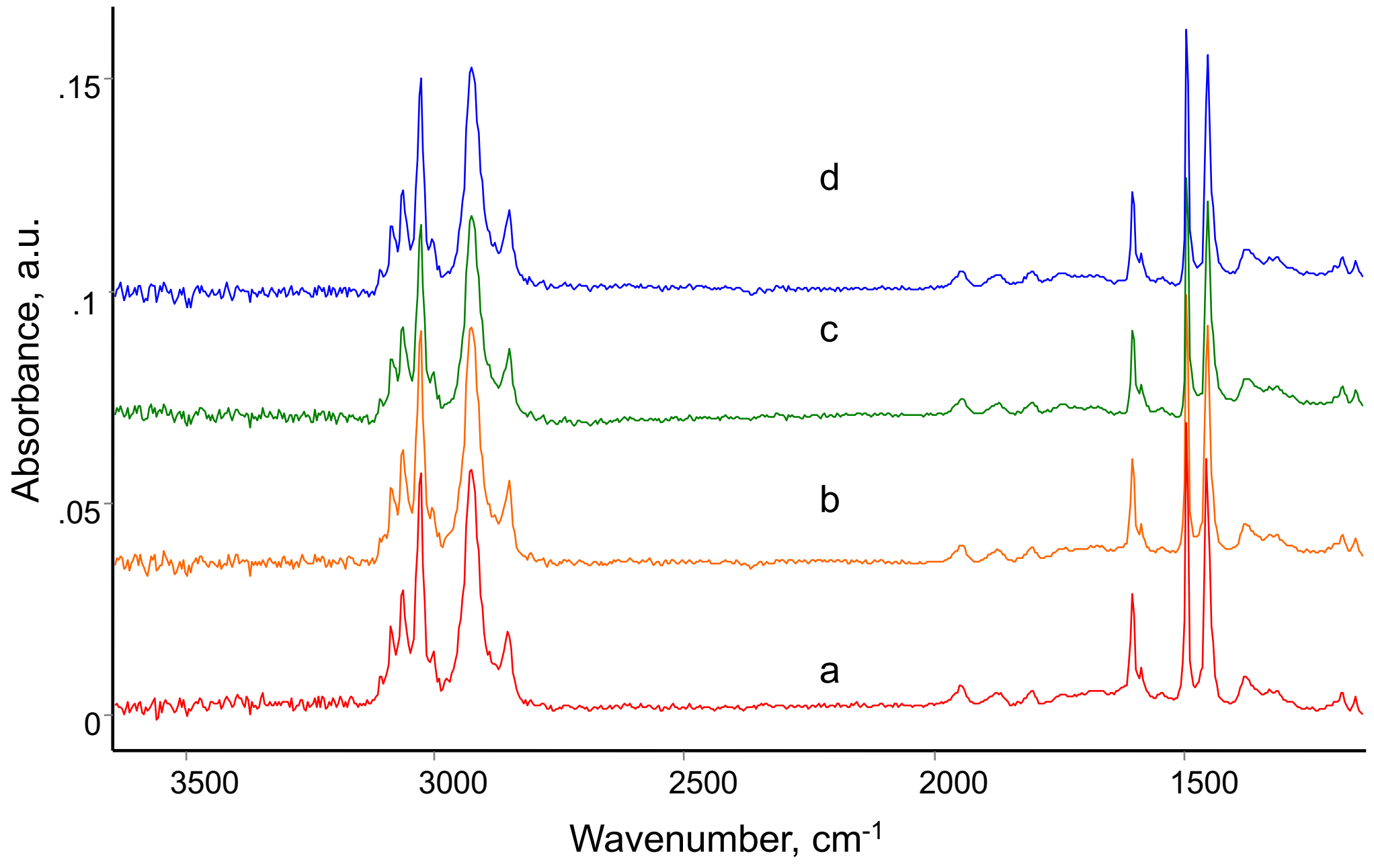

Fig. 51. FTIR transmission spectrum of PS nanolayer with thickness of 926 nm treated by ion-plasma treatment and washed in toluene: (a) 40 seconds, (b) 80 seconds, (c) 800 seconds, (d) 1600 seconds. Spectra of silicon wafer are subtracted. Spectra are processed by baseline linearisation method and shifted along absorption axis for better observation.

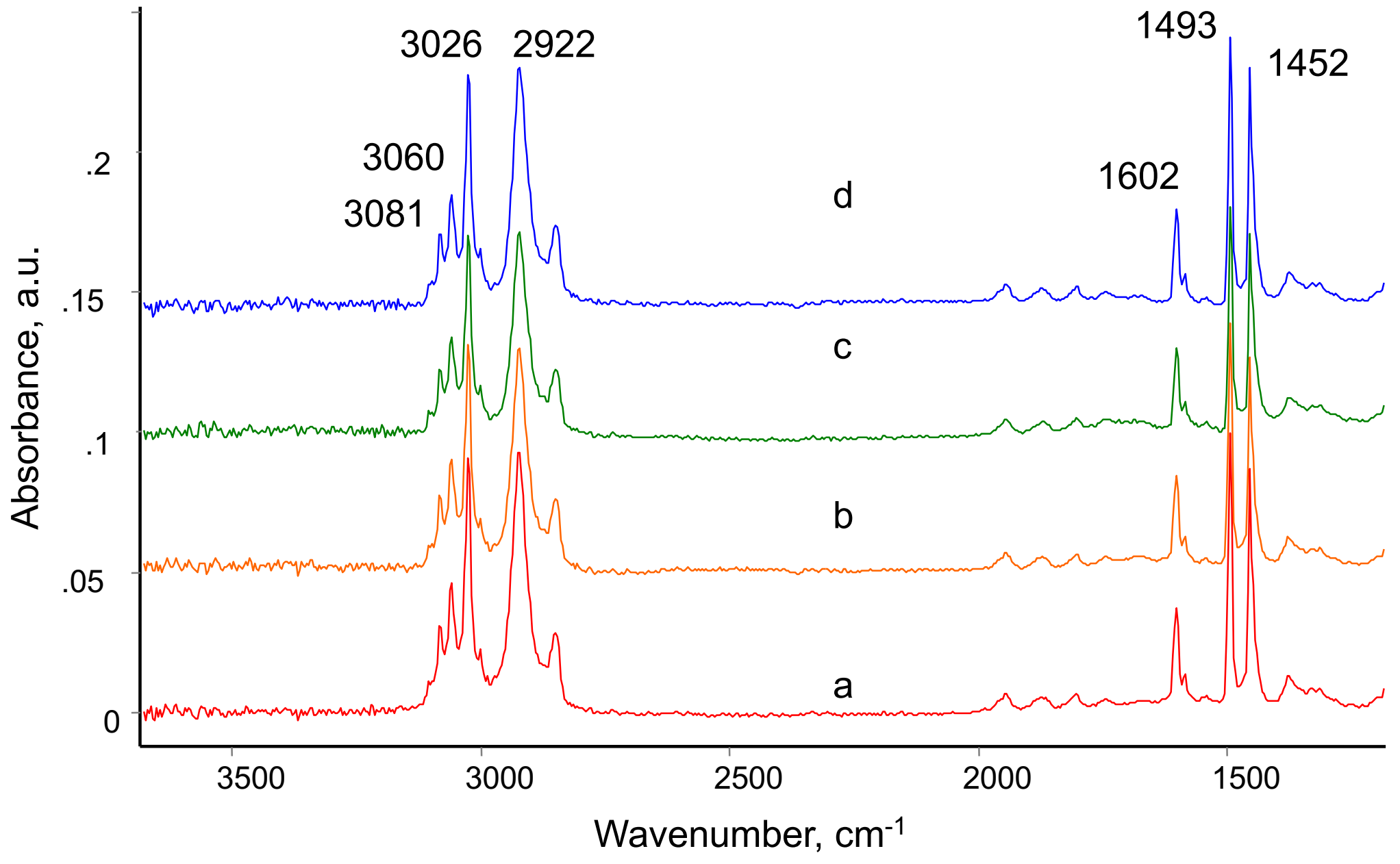

Fig. 52. FTIR transmission spectrum of PS nanolayer with thickness of 1330 nm treated by ion-plasma treatment and washed in toluene: (a) 40 seconds, (b) 80 seconds, (c) 800 seconds, (d) 1600 seconds. Spectra of silicon wafer are subtracted. Spectra are processed by baseline linearisation method and shifted along absorption axis for better observation.

The obtained spectra were used to calculate the intensity of the lines at 1452, 1493, 1602, 2922, 3026, 3060, 3081 cm$^{-1}$. As an example, the results of the analysis of the line 1452 $cm^{-1}$ of the spectra of PS treated with ions before and after washing in toluene are presented on Fig.53-57 for the samples of different treatment time and thickness.

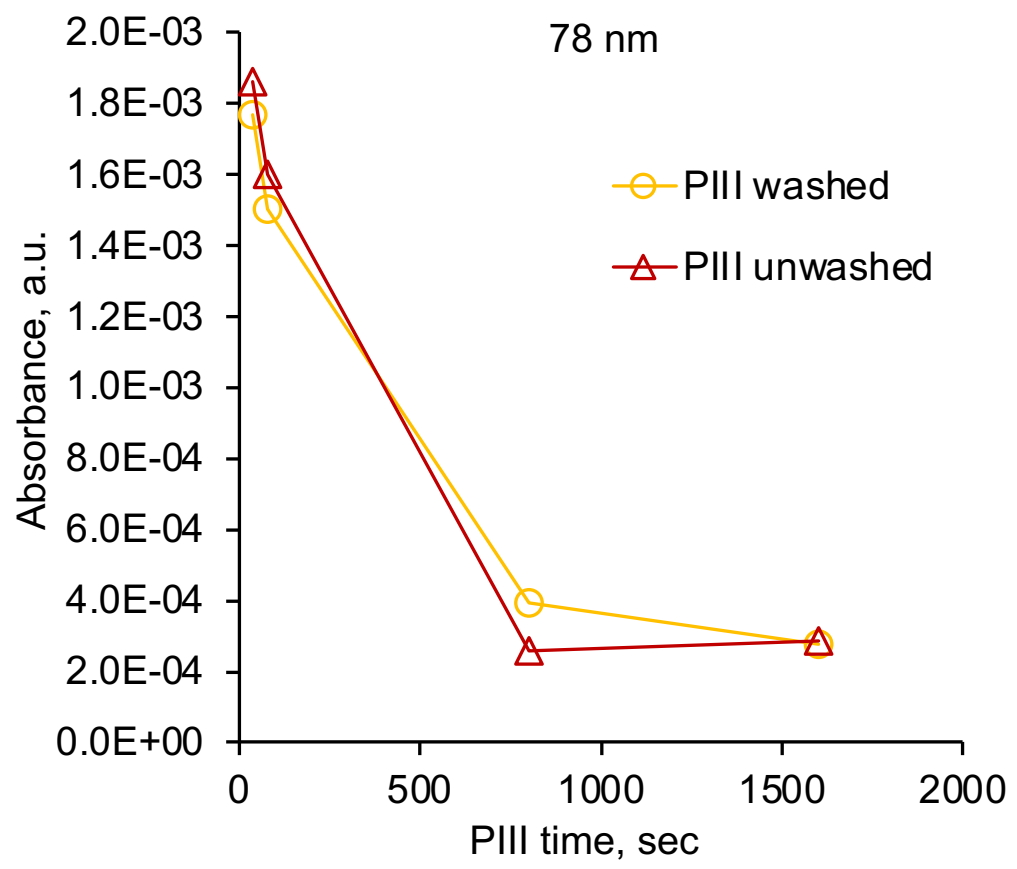

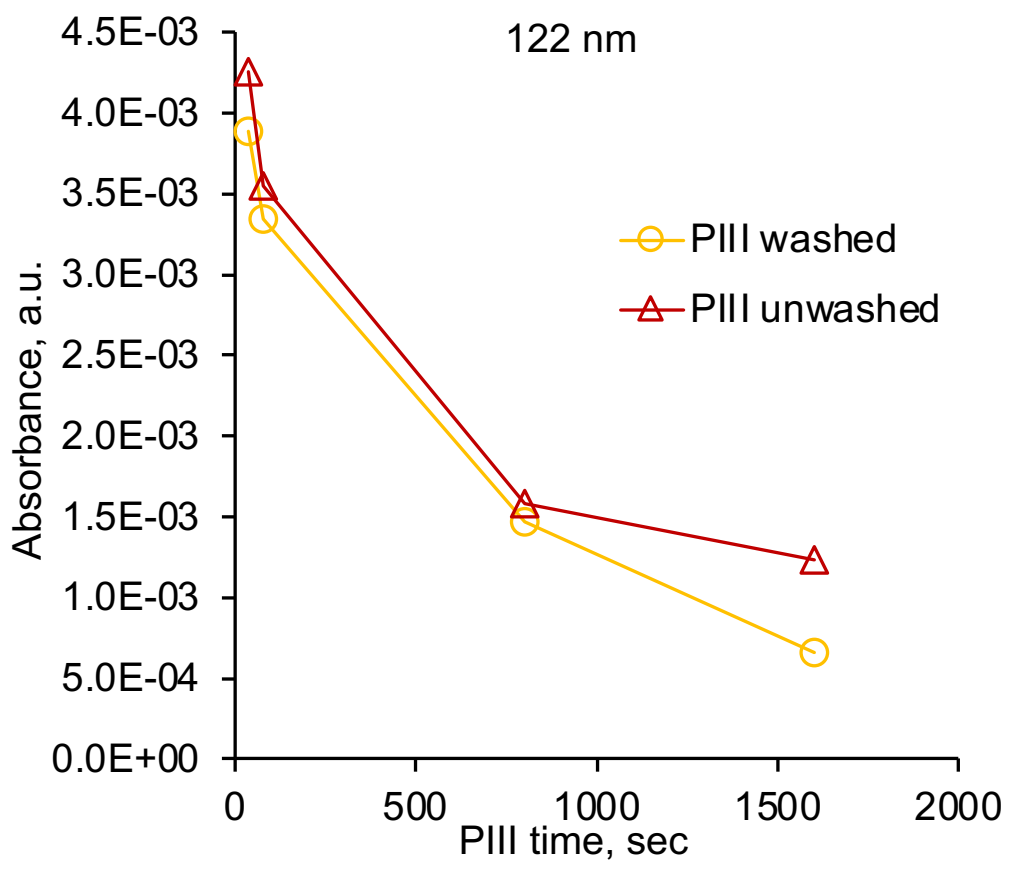

Fig. 53. Absorption of the 1452 $cm^{-1}$ line of the spectrum of PS with a thickness of 78 and 122 nm treated with ion beam and then washed in toluene depending on the treatment time.

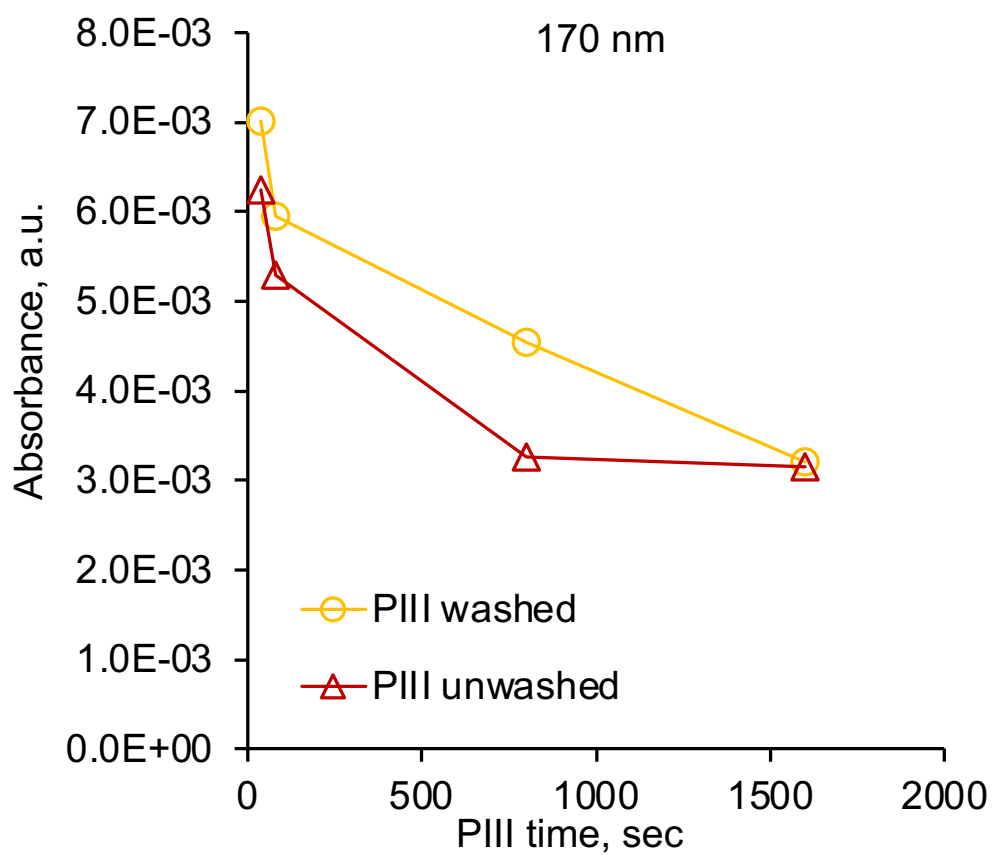

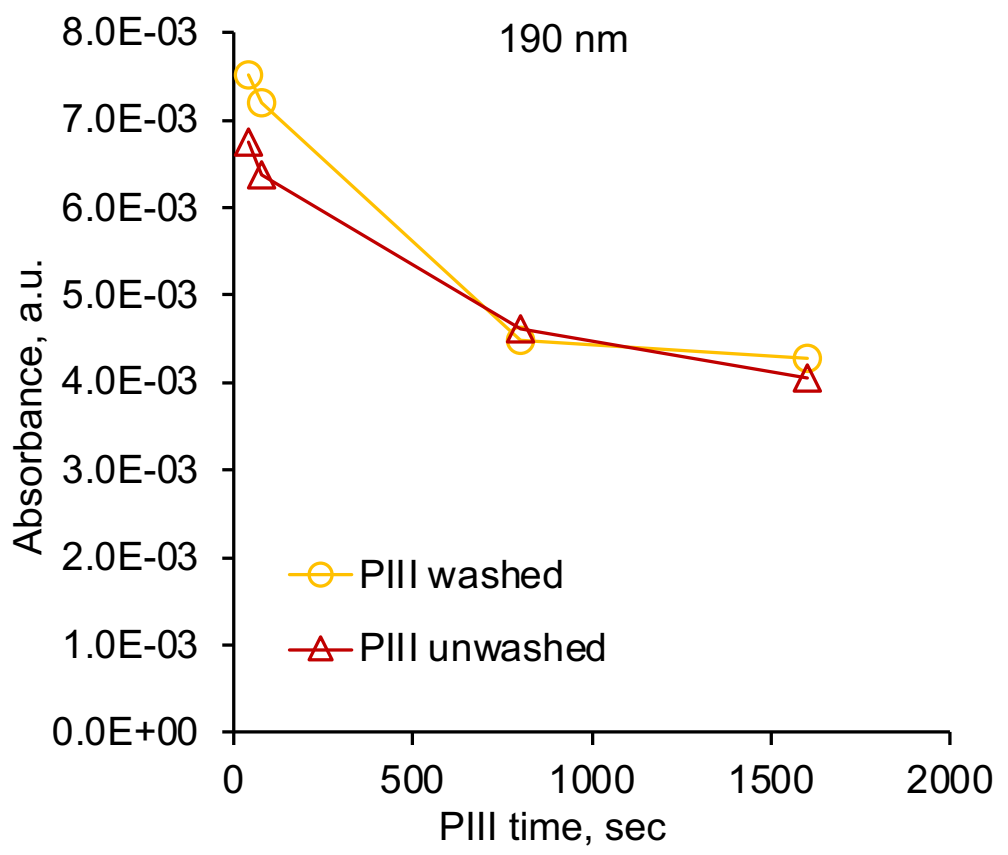

Fig. 54. Absorption of the 1452 $cm^{-1}$ line of the spectrum of PS with a thickness of 170 and 190 nm treated with ion beam and then washed in toluene depending on the treatment time.

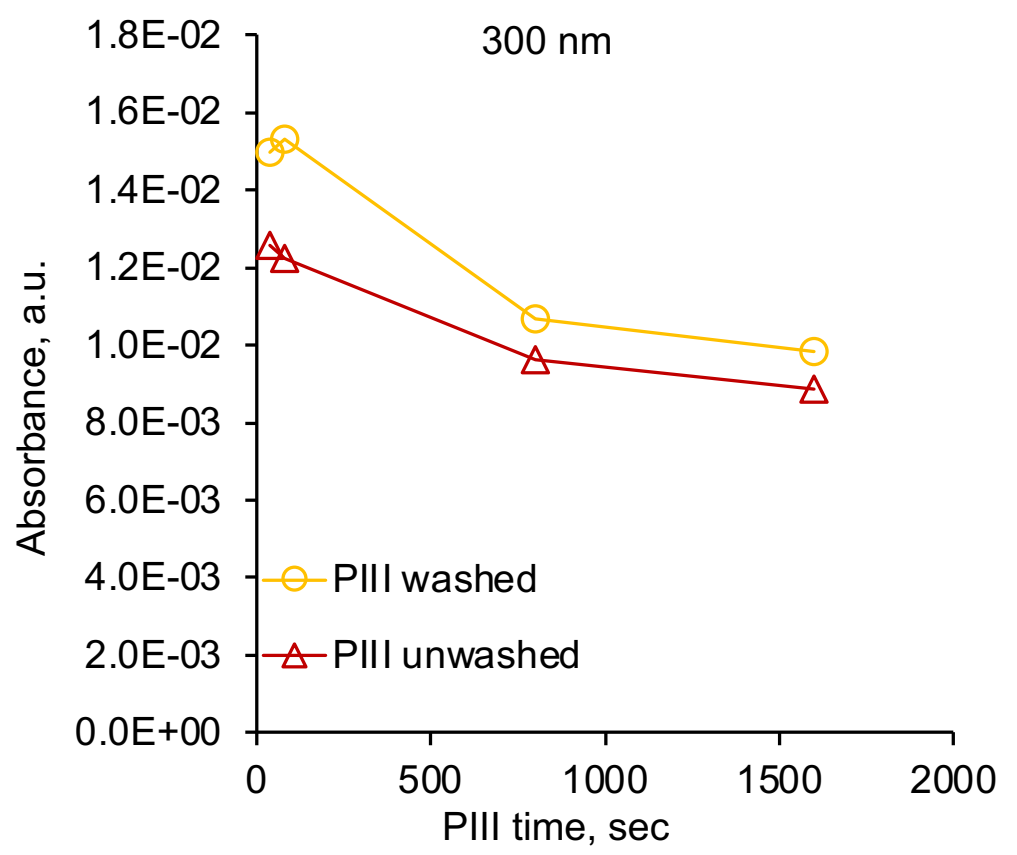

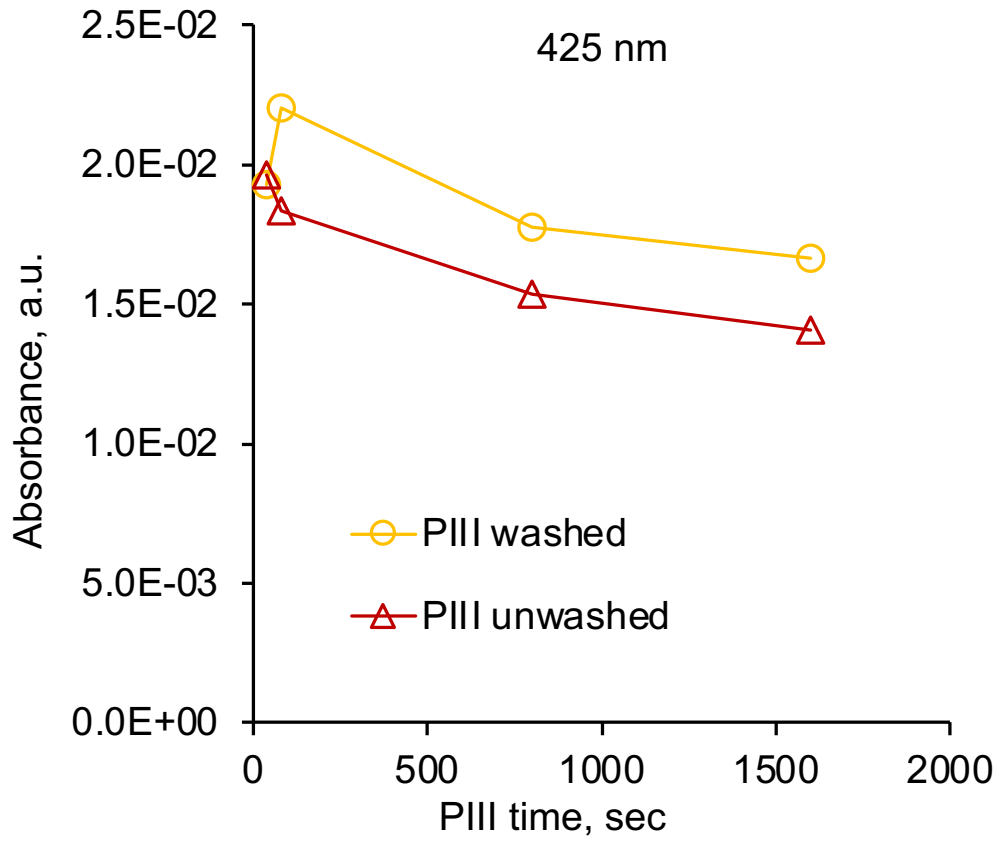

Fig. 55. Absorption of the 1452 $cm^{-1}$ line of the spectrum of PS with a thickness of 300 and 425 nm treated with ion beam and then washed in toluene depending on the treatment time.

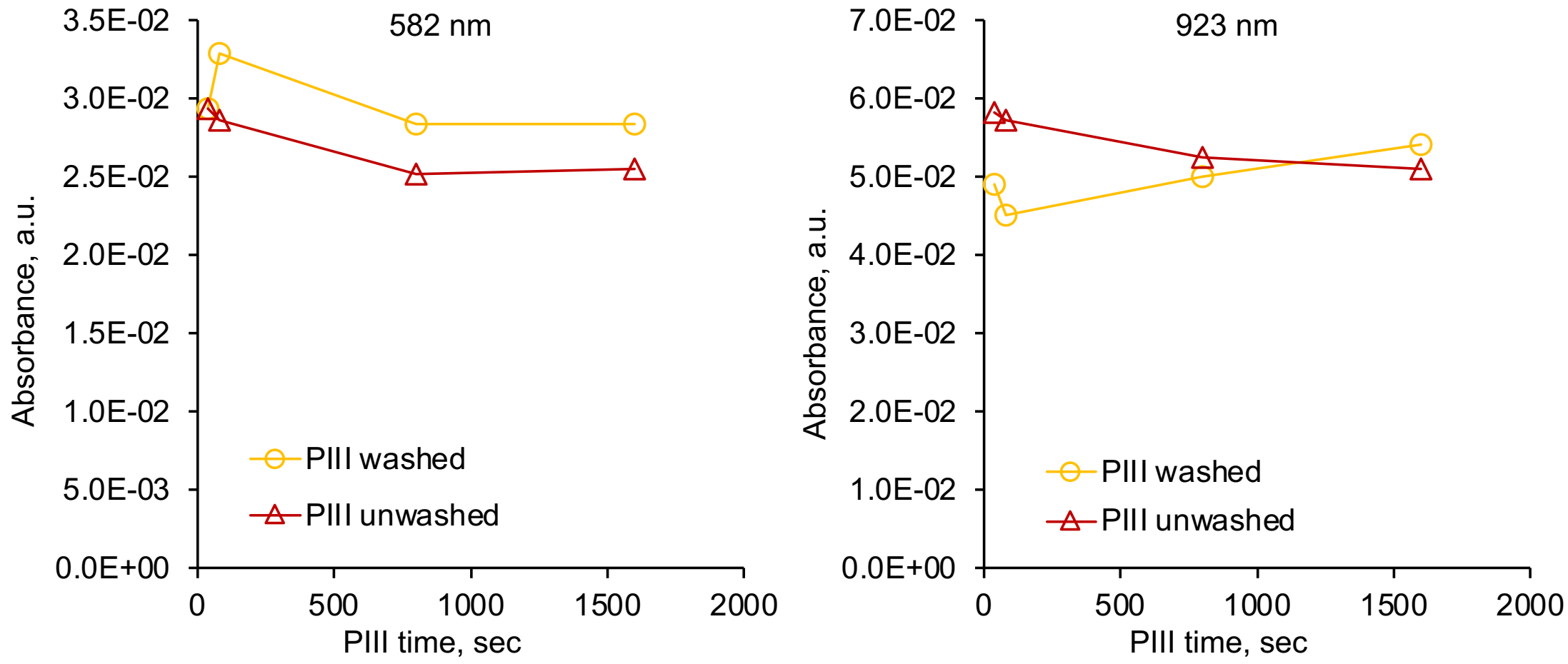

Fig. 56. Absorption of the 1452 $cm^{-1}$ line of the spectrum of PS with a thickness of 582 and 923 nm treated with ion beam and then washed in toluene depending on the treatment time.

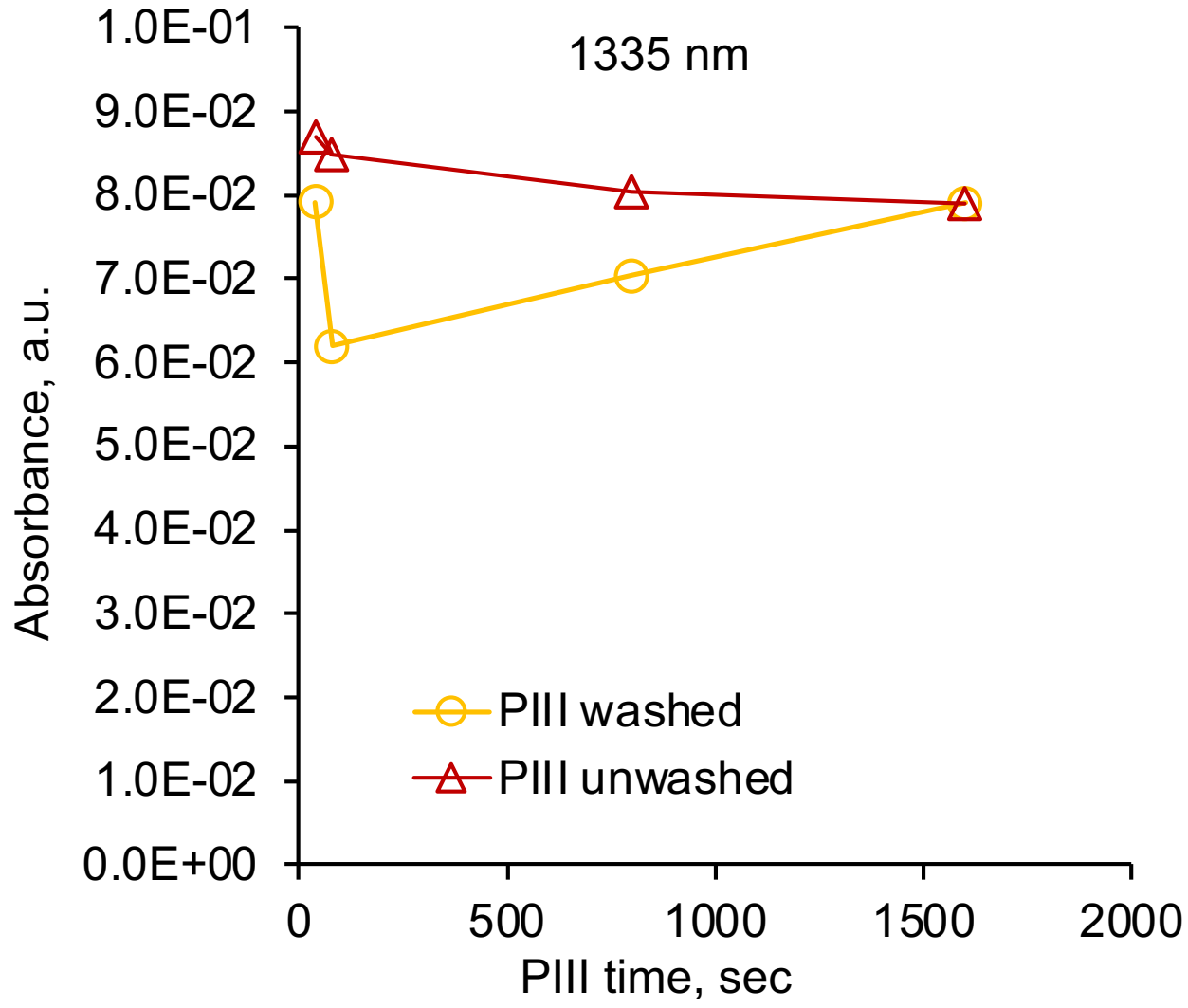

Fig. 57. Absorption of the 1452 $cm^{-1}$ line of the spectrum of PS with a thickness of 1335 nm treated with ion beam and then washed in toluene depending on the treatment time.

To determine the proportion of gel fraction in the treated PS, the intensity of all these lines was related to the intensity of the same lines in the spectra of treated but not washed samples. The results were then averaged over all lines. The results are shown in Fig. 58. The results show that the content of gel fraction is 100% for all samples of different thickness and treated with different times.

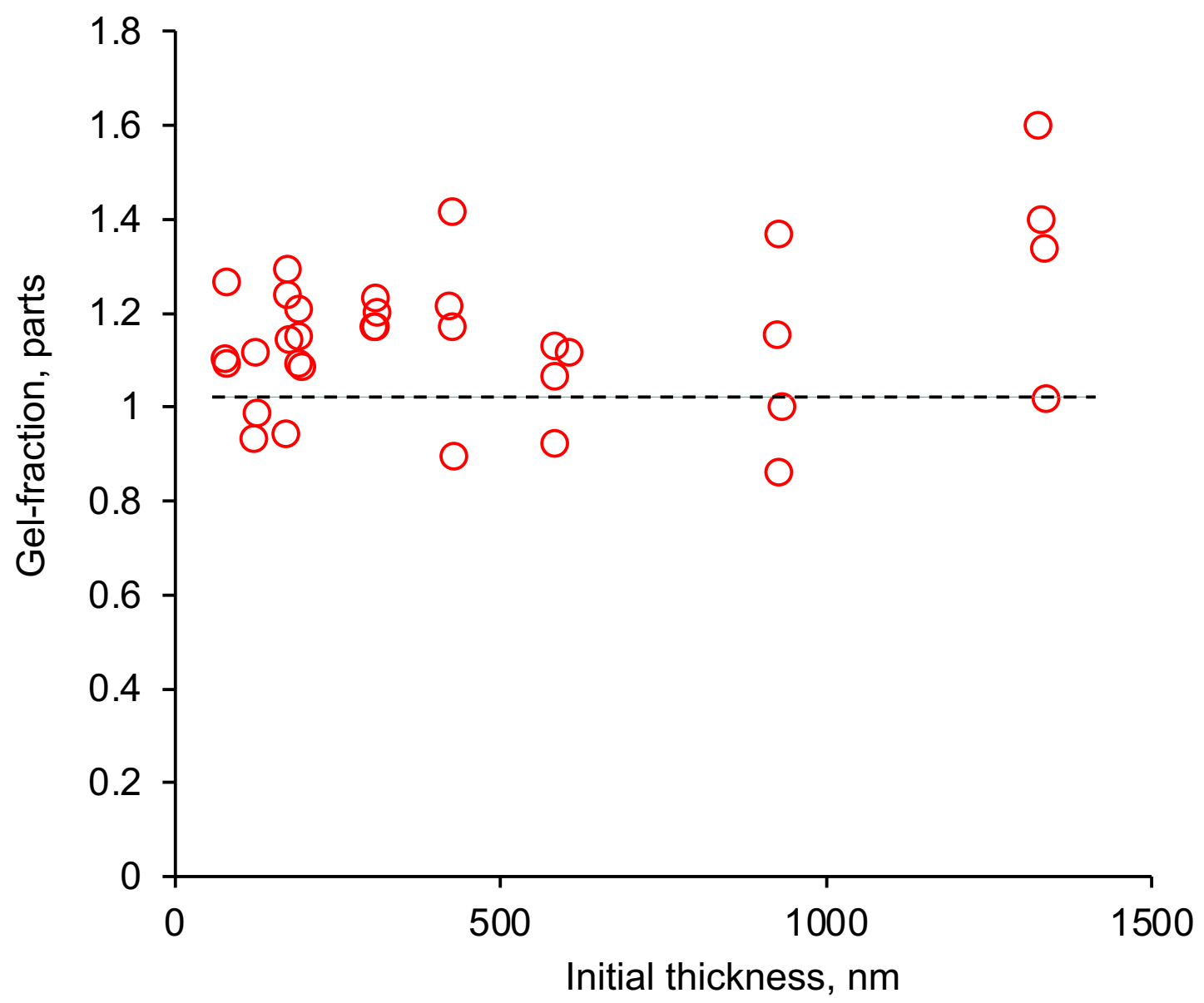

Fig.58. Gel fraction of PS samples treated with ion beam with different times in dependence on the initial thickness of PS samples. The gel fraction is defined as the ratio of the intensities of the FTIR spectrum lines of washed samples to the intensity of the FTIR spectrum lines of unwashed samples and averaged over the lines 1452, 1493, 1602, 2922, 3026, 3060, 3081 $cm^{-1}$.

Thus, the influence of ion penetration into the PS as a result of ion beam treatment extends into the PS to at least 1.3 µm, which is much greater than the depth of ion penetration into the PS.

## 4. Discussion

The ion beam treatment is a method of surface modification. This means that a thin surface layer of the material is modified. The thickness of the modified layer is determined by a depth of ion penetration into the material. Experimental studies of the thickness of the modified layer basically coincide with the depth of ion penetration determined both experimentally and theoretically based on the models of scattering of high-energy particles penetrating into a target. The results of both theoretical calculations and experimental observation are not questioned. This works for any kind of materials including polymers.

In thin layer of hydrocarbon polymer, carbon and hydrogen atoms are knocked out of the parent polymer macromolecule and form new structures. Hydrogen atoms mainly fly away from the polymer into the vacuum, and the remaining carbon forms unsaturated chemical structures with $sp^2$ hybridization of valence electrons of the type of condensed aromatic groups such as graphite or graphene, as well as diamond-like structures with $sp^3$ hybridization of valence electrons. This layer is characterized by high density, high refractive index and light absorption.

However, in the case of polymers, this physical model of the structure change as a result of ion-beam treatment is insufficient and cannot fully explain the properties of the treated polymer. The problem is that chemical processes in a polymer play a vital role after the treatment. Free radicals are formed in the polymer after the ion passes and stops, and the change in the chemical structure of the treated polymer cannot be explained without taking into account the chemical reactions of free radicals.
The free radicals initiate reactions of free valence migration at the carbon atom along the polymer chain. In this case, the hydrogen atom jumps to the neighboring carbon atom with free valence, creating an unbound valence on the parent carbon atom. Such a jump of the hydrogen atom occurs to the neighboring macromolecule, which creates migration of free valence between macromolecules. Migration of free valences along the polymer chain and between macromolecules occurs in all directions, including in the direction deep into the polymer. This process is described by the known kinetics of free radicals in hydrocarbons and occurs very quickly at room temperature.
If two free radicals in one macromolecule meet in their way, then an unsaturated group such as vinylene, vinyl, vinylidene and others may form. If two free radicals on different macromolecules meet, then a chemical bond between the macromolecules is formed, or in other words, cross-linking of macromolecules. Such a chemical bond cannot be broken by an inert solvent if the polymer is placed in a solvent after processing. Cross-linking of macromolecules leads to the formation of an insoluble part of the polymer, or in other words, a gel fraction of the polymer. When all macromolecules are cross-linked, the polymer becomes completely insoluble in the solvent. If the polymer is on a substrate, then such a cross-linked polymer cannot be washed off by a solvent.
If after the treatment a polymer is taken out into air, the residual free radicals react with oxygen and nitrogen of air, forming oxygen-containing and nitrogen-containing groups.
Analyzing the experimental results obtained above, it could be concluded that the thickness of the carbonized layer of PS corresponds to the penetration depth of nitrogen ions. In this case, the value of the penetration depth of nitrogen ions corresponds to the average value between the penetration depth of molecular and atomic ions.
The cross-linking of the treated polystyrene occurs throughout the entire thickness of the polystyrene layer up to 1330 nm, which greatly exceeds the penetration depth of the bombarding ions. Cross-linking occurs even for a low ion fluence of $5 \cdot 10^{14}$ ions $cm^{2}$.
Cross-linking of PS is also confirmed by a decrease in the concentration of the initial PS groups in the underneath layer, where nitrogen ions do not penetrate. In this case, a more noticeable decrease in the concentration of the initial PS groups is observed for a high fluence of processing, at which the concentration of macromolecular defects and the concentration of free radicals migrating into the bulk PS layer are higher.
Plasma treatment under the same conditions without high voltage acceleration for ions, as a control experiment, does not modify the deep layer. That is, changes in the deep layer are associated specifically with the effect of high-energy ions.
Oxidation of the treated PS is mainly associated with the carbonized layer. The concentration of oxygen-containing groups does not depend on the thickness of the original PS layer. That is, the migration of free radicals into the bulk layers of PS does not cause an increase in the total concentration of oxygen-containing groups. The oxidation process does not occur in the bulk layer.
Thus, the model of ion beam treatment of a polymer film on a silicon wafer looks like presented on Fig.59. As a result of ion bombardment, a thin carbonized layer appears in the polymer. Under the carbonized layer is a cross-linked layer with a slightly modified polymer structure. The entire polymer film is stable in the solvent.

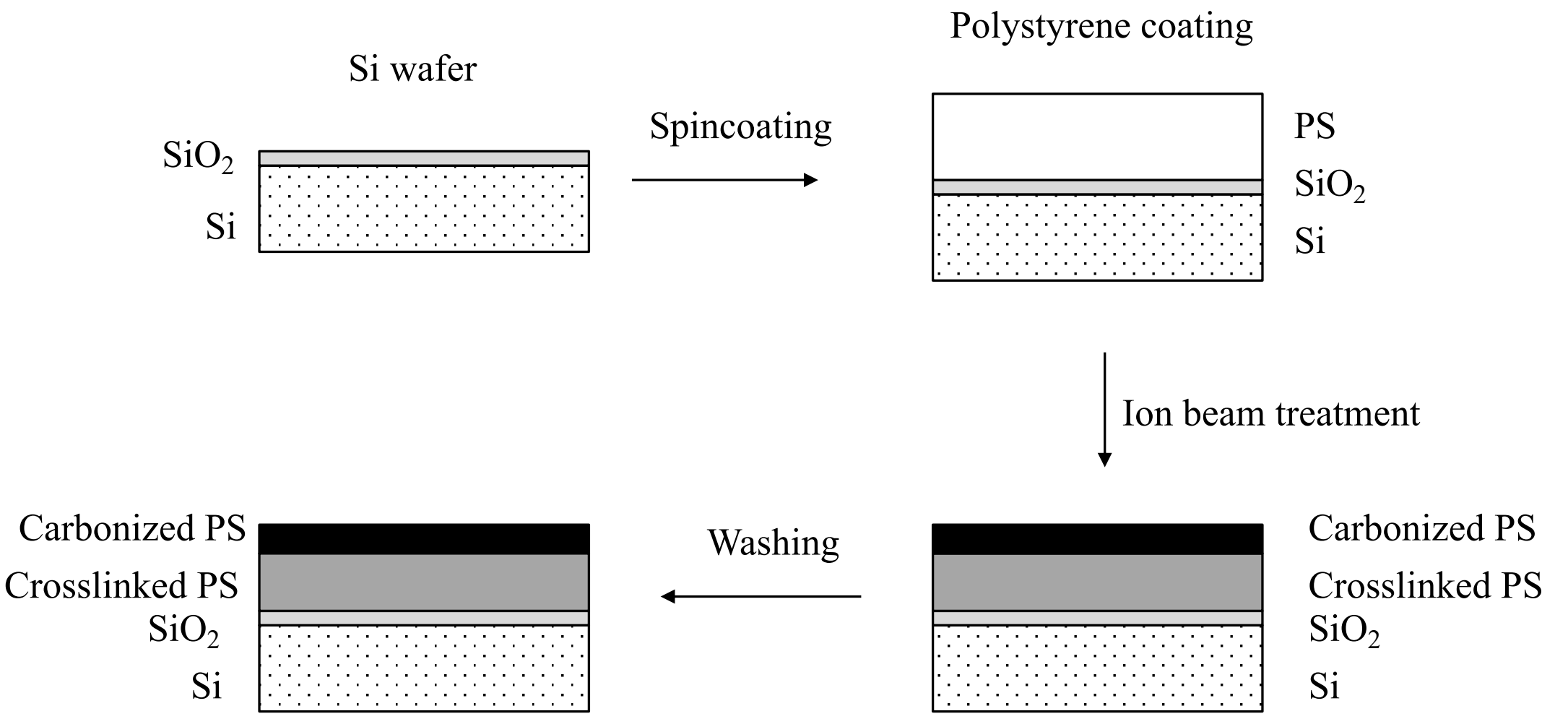

Fig.59. Experimental scheme corrected based on the results of the study. As a result of ion beam treatment, a stable polystyrene film is obtained that is insoluble in the solvent. The thickness of the stabilized film is much greater than the thickness of the carbonized layer determined by the ion penetration depth.

The question remains as to how deeply free radicals can penetrate into the bulk polymer layer. Previous results showed the thickness of the cross-linked polymer layer to be up to 1-3 μm. This was obtained for PS, polyisoprene and polyethylene films with an initial thickness of 20-100 μm using the gel-fraction measurements in a Soxhlet apparatus. This method does not provide results on the cross-linking gradient of the polymer layer and the migration depth of free radicals. On the other hand, the creation and analysis of spin-coated films more than 1 micron is also difficult. Therefore, the question of the migration depth of free radicals and the formation of a cross-linking gradient in the polymer remains open.

## 5. Conclusions

The presented studies on the example of polystyrene showed that as a result of ion beam treatment with nitrogen ions, a carbonized layer is formed with a thickness corresponding to the depth of ion penetration. In addition, the underlying layer of polystyrene also undergoes structural transformations consisting in partial destruction of the original chemical structure. The entire polymer layer up to 1330 nm is totally cross-linked with chemical bonds. Such deep formation of the polymer gel fraction throughout the entire depth can only be explained by the migration of free radicals from the carbonized surface layer deep into the underlying layer and the formation of cross-links of polymer macromolecules there along the path of free-radical reactions.

**Acknowledgments**: Author thanks Prof. Marcela Bilek and Prof. David McKenzie for providing employment in an equipped plasma laboratory.